\documentclass{aastex61}
\usepackage[utf8]{inputenc}
\usepackage{amsmath}
\usepackage{longtable}
\usepackage{graphicx}
\usepackage{listings}
\usepackage{hyperref}

\newcommand{\numbadtmag}{645}

\newcommand{\teff}{\ensuremath{T_{\rm eff}}}
\newcommand{\logg}{\ensuremath{\log g}}
\newcommand{\tmag}{\ensuremath{T}}
\newcommand{\feh}{[Fe/H]}
\newcommand{\msun}{\ensuremath{M_\odot}}
\newcommand{\rsun}{\ensuremath{R_\odot}}
\newcommand{\lbol}{\ensuremath{L_{\rm bol}}}
\newcommand{\rearth}{\ensuremath{R_{\oplus}}}

\definecolor{mygreen}{rgb}{0.0, 0.5, 0.0}
\definecolor{mylila}{rgb}{0.5, 0.0, 0.5}
\definecolor{myblue}{rgb}{0.0, 0.0, 0.5}
\definecolor{mypurple}{rgb}{0.58, 0.42, 0.87}

\newcommand{\subsubsubsection}[1]{\vspace{6pt}\noindent{\bf #1}\vspace{3pt}}

\begin{document}

\title{The {\it TESS\/} Input Catalog and Candidate Target List}
\author{Keivan G.\ Stassun}
\affiliation{Vanderbilt University}
\affiliation{Vanderbilt Initiative in Data-intensive Astrophysics (VIDA)}
\affiliation{Fisk University}

\author{Ryan J.\ Oelkers}
\affiliation{Vanderbilt University}
\affiliation{Vanderbilt Initiative in Data-intensive Astrophysics (VIDA)}
\affiliation{Vanderbilt Data Science Fellow}

\author{Joshua Pepper}
\affiliation{Lehigh University}
\affiliation{Vanderbilt Initiative in Data-intensive Astrophysics (VIDA)}

\author{Martin Paegert}
\affiliation{Harvard-Smithsonian Center for Astrophysics}
\affiliation{Vanderbilt Initiative in Data-intensive Astrophysics (VIDA)}

\author{Nathan De Lee} 
\affiliation{Northern Kentucky University}
\affiliation{Vanderbilt Initiative in Data-intensive Astrophysics (VIDA)}

\author{Guillermo Torres}
\affiliation{Harvard-Smithsonian Center for Astrophysics}

\author{David W. Latham}
\affiliation{Harvard-Smithsonian Center for Astrophysics}


\author{St\'ephane Charpinet}
\affiliation{Institut de Recherche en Astrophysique et Plan\'etologie, CNRS, Universit\'e de 
Toulouse}

\author{Courtney D.\ Dressing}
\affiliation{University of California at Berkeley}

\author{Daniel Huber}
\affiliation{University of Hawaii}

\author{Stephen R.\ Kane}
\affiliation{University of California, Riverside}

\author{S\'ebastien L\'epine}
\affiliation{Georgia State University}

\author{Andrew~Mann}
\affiliation{Columbia University}

\author{Philip S.\ Muirhead}
\affiliation{Boston University}

\author{B\'{a}rbara Rojas-Ayala}
\affiliation{Universidad Andr\'es Bello}

\author{Roberto Silvotti}
\affiliation{INAF-Osservatorio Astrofisico di Torino}


\author{Scott W.\ Fleming}
\affiliation{Space Telescope Science Institute}

\author{Al Levine}
\affiliation{Massachusetts Institute of Technology}

\author{Peter Plavchan}
\affiliation{George Mason University}


\begin{abstract}
The Transiting Exoplanet Survey Satellite (TESS) will be conducting a nearly all-sky photometric survey over two years, with a core mission goal to discover small transiting exoplanets orbiting nearby bright stars.  It will obtain 30-minute cadence observations of all objects in the TESS fields of view, along with 2-minute cadence observations of 200,000 to 400,000 selected stars.  The choice of which stars to observe at the 2-min cadence is driven by the need to detect small transiting planets, which leads to the selection of primarily bright, cool dwarfs.  We describe the catalogs assembled and the algorithms used to populate the TESS Input Catalog (TIC), 
including plans to update the TIC with the incorporation of the Gaia second data release in the near future.
We also describe a ranking system for prioritizing stars according to the smallest transiting planet detectable, and assemble a Candidate Target List (CTL) using that ranking. We discuss additional factors that affect the ability to photometrically detect and dynamically confirm small planets, and we note additional stellar populations of interest that may be added to the final target list.
The TIC is available on the STScI MAST server, and an enhanced CTL is available through the Filtergraph data visualization portal system at the URL \url{https://filtergraph.vanderbilt.edu/tess_ctl}.
\end{abstract}

\section*{ERRATUM}
Figure~5 in the originally published version of the paper incorrectly stated that stars with parallaxes available from the {\it Gaia\/} first data release (DR1) are included in the CTL only if their estimated surface gravity (\logg) is consistent with being a dwarf (i.e., \logg\ $\ge$ 4.1). The correct procedure, shown in the corrected Figure~5 below, instead uses the stellar {\it radius\/} as the criterion, and {\it includes both dwarfs and subgiants} according to the relation shown by the blue curve in Figure~13 ({\it right}). This is for the sake of consistency between the parallax-based method and the RPM$_J$-based method, and also to preserve potentially interesting subgiant targets as discussed in Section~4.3.3. 

Finally, the first sentence of the second paragraph of Section~3.1 should now read as follows: 
Next, we calculate the stellar radius for all stars with parallax measurements, using the stellar radius and mass estimation algorithms described in Section~3.3.2; whether they are included in or excluded from the CTL depends on whether the estimated stellar radius is, respectively, below or above the blue curve shown in Figure~13 ({\it right}).

\setcounter{figure}{4}
\begin{figure}[!ht]
    \centering
    \includegraphics[width=\linewidth]{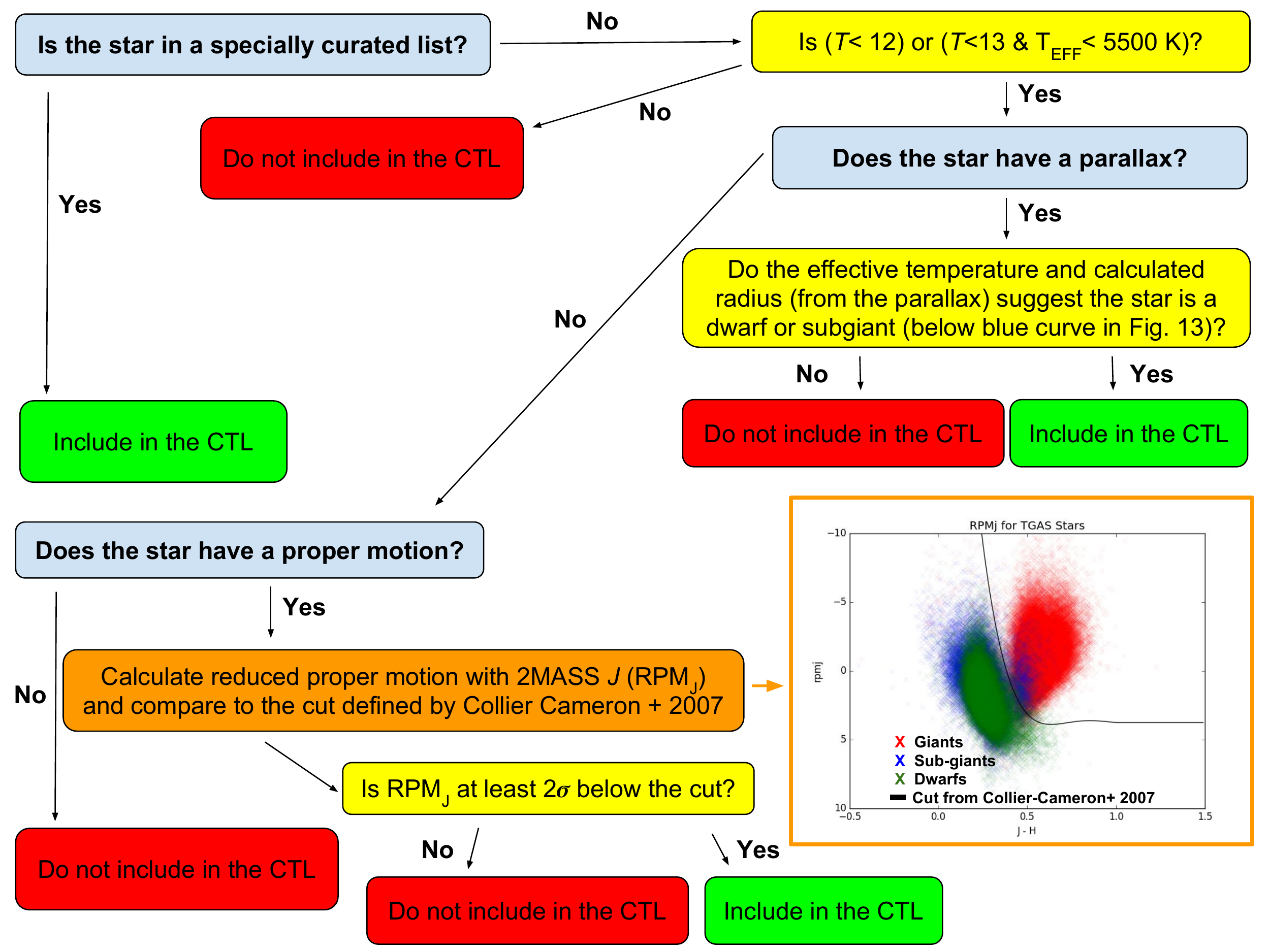}
    \caption{Overview of procedure for selecting stars for the CTL. The inset uses {\it Gaia\/} DR1 stars to show the effectiveness of separating red giants (red points) in the RPM$_J$ diagram using the cut defined by Collier Cameron et al.\ (2007). To the left/below the dividing line are dwarfs and subgiants; the RPM$_J$ cut removes $\gtrsim$95\% of red giants but removes only $\sim$10\% of subgiants (see Section 3.2).}
\end{figure}

\clearpage
\setcounter{figure}{0}

\section{Introduction\label{sec:intro}}

The TESS mission is designed to detect small transiting planets orbiting the brightest and nearest stars in the sky.  It will be conducting a sequential set of 27.4-day photometric surveys of $24^{\circ} \times 96^{\circ}$ sectors of the sky, first covering the southern ecliptic hemisphere over the course of one year, followed by the northern hemisphere.  All detectors on each TESS camera will be read out entirely every 30 minutes, while a set of roughly 200,000 postage stamp sections of the detectors will be downloaded at a 2-minute cadence for a set of pre-selected stars with high value as transit search targets.

Analogous in many ways to an expanded version of the NASA {\it Kepler\/} mission, TESS requires a source catalog comparable to the {\it Kepler\/} Input Catalog \citep[KIC;][]{Brown:2011}, the {\it K2\/} Ecliptic Plane Input Catalog \citep[EPIC;][]{Huber:2016}, and the {\it CoRoT\/} Exo-Dat Catalogue \citep{Deleuil:2009}.  The TESS Input Catalog (TIC) is a catalog of luminous sources on the sky, for use by the TESS mission to select target stars to observe, and to provide stellar parameters essential for the evaluation of transit signals.  The TIC will enable the selection of optimal targets for the planet transit search, enable calculation of flux contamination in the TESS aperture for each target, and will provide reliable stellar radii for calculating planetary radii,which will determine which targets receive mission-supported photometric and spectroscopic follow-up. The TIC is also essential for the community for the selection of targets through the Guest Investigator program. 

The area of the sky projected onto each TESS pixel is large (20$\times$20~arcseconds) and the point spread function is typically 1--2 pixels in radius (depending on stellar brightness and position on the image). Consequently it is expected that a given TESS target may include flux from multiple objects. Therefore,
we have created the TIC to contain every optically luminous, persistent object in the sky down to the limits of available wide-field photometric point source catalogs.  We have not included objects without significant persistent optical flux, and we have also not included, for technical and logistical reasons, objects that move rapidly enough that their celestial positions cannot be calculated with linear proper motions (i.e., we do not include solar system objects such as planets, asteroids, TNOs, etc).

The purpose of the TIC incorporates four basic needs. They are:

\begin{enumerate}

\item To provide in the TIC basic astronomical information for all sources in the TESS footprint (i.e., the entire sky), similar to what the KIC, the EPIC, and the NASA Exoplanet Archive (NEA) together provide for {\it Kepler\/} and {\it K2}. It is a catalog in which anyone can look up information about any object for which the TESS mission produces a light curve, barring moving or transient objects.

\item To enable selection of primary transit-search (i.e., 2-min cadence) targets for TESS. The incorporation of all luminous objects in the sky for the full TIC will allow a calculation of the flux contamination for all potential TESS targets. In practice, that involves a flux contamination value for every star in the Candidate Target List (CTL; see Sec.~\ref{sec:CTL}) subset of the TIC.

\item To provide stellar parameter information for the TESS Science Processing Operations Center (SPOC) to evaluate exoplanet transit candidates.  As TESS light curves are searched for transit candidates, information about the target star, such as effective temperature (\teff), surface gravity (\logg), mass ($M_\star$), radius ($R_\star$), and other parameters are used to calculate the planet properties. 

\item To enable false positive determination.  Both the SPOC and the public will need to be able to identify all known astrophysical sources within some angular radius of any object in the TIC, down to a certain faintness limit, with their accompanying information. That will allow users to decide how likely it is that signals in the light curve of a target are due to an astrophysical source that is not the target.

\end{enumerate}

The CTL represents a distillation of the full TIC to enable more detailed calculations of stellar parameters and a ranking of the highest-priority stars.  It is a list of several million stars that meet certain simple criteria that serve as the first cut for selection of 2-minute targets.  Since many stellar properties relevant for consideration as targets require non-trivial calculation, we found it necessary to winnow the initial TIC of half a billion objects to the more manageable CTL.  That step allowed us to conduct repeated tests, adjustments, and variations to a number of the procedures described in this paper, in a time frame that would have been prohibitive when applied to the full TIC.

All stars in the CTL are then prioritized for their desirability as TESS transit search targets.  Priorities are established via a scheme that emphasizes detectability of small planets. In the following sections, we detail the construction of the TIC (Section \ref{sec:tic}) and of the CTL (Section~\ref{sec:CTL}). For both the TIC and the CTL we describe the algorithms used to determine associated stellar parameters. Because some stellar parameters are calculated for all TIC stars, and other parameters are just computed for the stars in the CTL, the discussion of stellar properties is spread between those sections. We also describe the target prioritization scheme for the CTL in Section~\ref{sec:CTL}. We conclude in Section~\ref{sec:discussion} with a discussion of known limitations in the current TIC and CTL, as well as planned future improvements. 

The TIC and CTL are also accompanied by official release notes, which are provided in Appendix~\ref{sec:relnotes}. Public access to the TIC is provided via the MAST server, and access to an enhanced CTL is provided via the Filtergraph data visualization service at the URL \url{https://filtergraph.vanderbilt.edu/tess_ctl}.

\section{The TESS Input Catalog (TIC)\label{sec:tic}} 

In this section we detail the procedures by which we have constructed the TIC, with particular detail provided regarding the algorithms, relations, and rules adopted for populating the TIC. 
The TIC includes a number of columns, each with a specified format and a permitted range of values. These are summarized in Appendix~\ref{sec:appendix_ticranges}. The provenance flags associated with various TIC quantities are listed in Appendix~\ref{sec:flags}. Steps taken to ensure internal consistency among various TIC quantities are given in Appendix~\ref{sec:internal_consistency}. 
\textit{It is important to understand that, as described below, the TIC deliberately includes both point sources (stars) and extended sources (e.g., galaxies); positional searches of the TIC will in general return some extended sources as well as stars. These can be separated by use of the {\tt objtype} flag (see Appendix~\ref{sec:appendix_ticranges}).}

\subsection{Assembly of the TIC\label{sec:overview}}

\begin{figure}[ht]
    \centering
    \includegraphics[width=\linewidth,trim=5 5 5 5,clip]{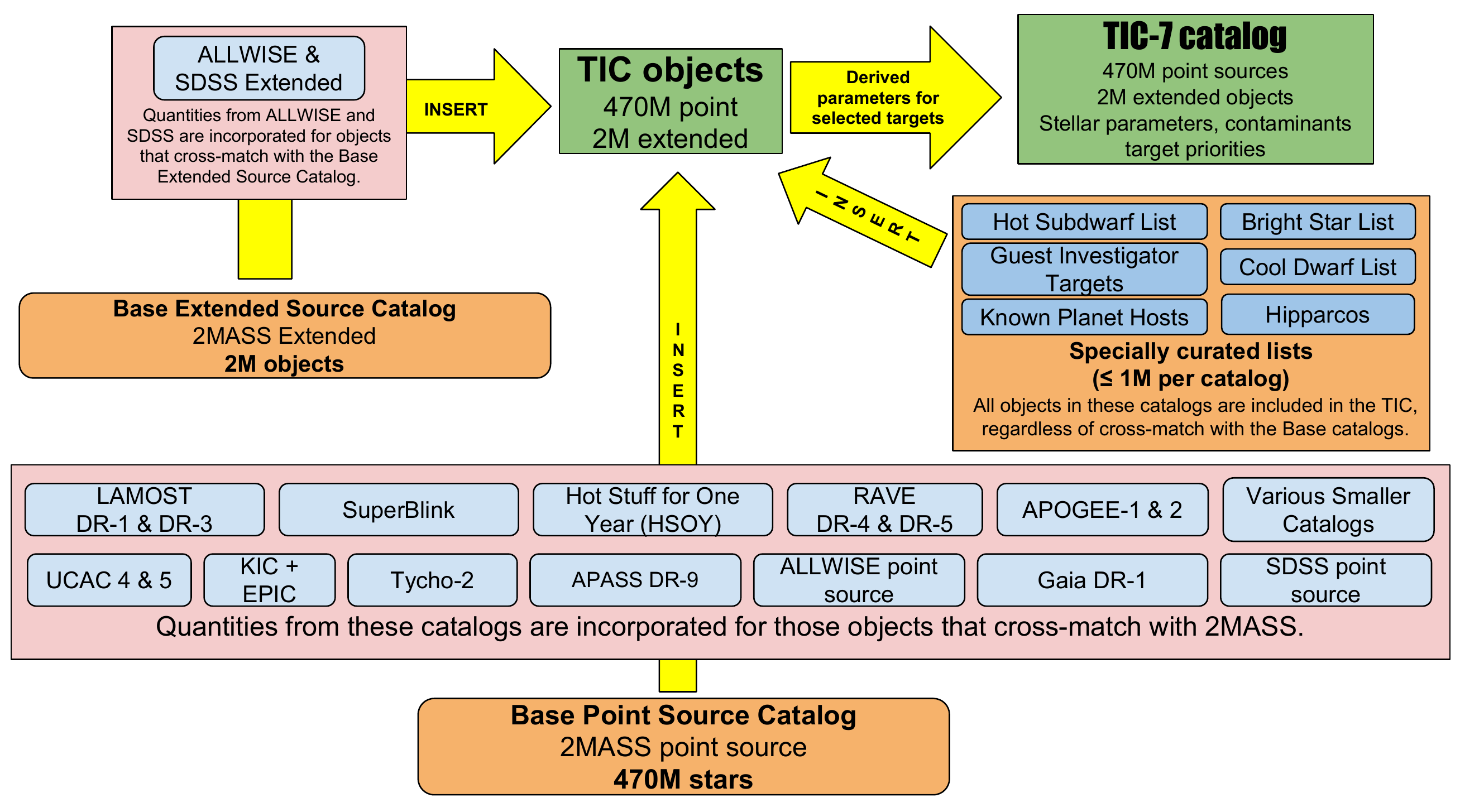}
    \caption{Visual overview of the photometric catalogs used to construct the overall TESS Input Catalog (TIC). Yellow arrows depict the order in which catalogs are cross-matched and/or merged. The final TIC (current version indicated by the integer number) is represented by the green box at the upper right. }
    \label{fig:tic_overview}
\end{figure}

The process by which the various photometric catalogs have been assembled for the construction of the overall TIC (currently version 7) is summarized graphically in Figure \ref{fig:tic_overview}.  The compilation of the TIC is the product of merging three base catalogs to create a full list of point sources, and prominent extended sources. We provide the quality flags from the base catalogs in the TIC but do not include the quality flags from the other assorted catalogs. This means the TIC inherits structure and biases from these catalogs as we describe in Section~\ref{subsec:structure}.
In principle, Gaia would serve as an ideal base catalog, and indeed that was our original intent. Unfortunately the Gaia second data release (DR2) was not available sufficiently in advance of TESS launch to be incorporated into the TIC for TESS Year~1 operations. However, we do plan a substantial update of the TIC with Gaia DR2 in time for the TESS Year~2 operations, as discussed further in Sec.~\ref{sec:gaiadr2}.

The full list of TIC objects is subjected to a set of algorithms to determine the astronomical and physical parameters for each object as outlined in \S~\ref{subsec:parms} and prioritized within the CTL as described in \S~\ref{subsec:priority}. 
The full TIC includes $\sim473\times10^6$~objects ($\sim470\times10^6$ point sources, $\sim2\times10^6$ extended sources, and $\sim1\times10^6$ objects from the specially curated lists). The TIC is staged for public use on the MAST portal system at the URL \url{https://mast.stsci.edu}. The current release notes for the TIC are provided in Appendix~\ref{sec:relnotes}.

\subsubsection{Point sources}

The TIC point source base catalog is constructed from the full 2MASS point source catalog \citep{Skrutskie:2006} of $\sim470\times10^6$ objects. Next, this catalog is crossed-matched against the following catalogs: LAMOST-DR1 and DR3 \citep{Luo:2015}, KIC+EPIC \citep{Brown:2011,Huber:2016}, RAVE DR4 and DR5 \citep{Boeche2011,Kunder:2017}, APOGEE-1 and APOGEE-2 \citep{Majewski:2015,Zasowski:2017}, UCAC4 and UCAC5 \citep{Zacharias:2013,Zacharias:2017}, {\it Tycho-2\/} \citep{Hog:2000}, APASS-DR9 \citep{Henden:2009}, Hot-Stuff-for-One-Year \citep[hereafter, HSOY;][]{Altmann:2017}, Superblink \citep{Lepine:2017}, HERMES-TESS DR1 \citep{Sharma:2018}, SPOCS \citep{Brewer:2016}, Geneva-Copenhagen \citep{Holmberg:2009}, Gaia-ESO \citep{Gilmore:2012}, ALLWISE point sources \citep{Cutri:2013}, and Sloan Digital Sky Survey point sources \citep[SDSS;][]{Alam:2015, Blanton:2017,Abolfathi:2017}. 

All coordinates are initially drawn from 2MASS. Those objects are cross-matched with {\it Gaia\/} DR1 and the coordinates are updated to the {\it Gaia\/} values if a single match, or updated to the flux-weighted {\it Gaia\/} values if multiple match (see \S~\ref{subsubsec:gaiadupes}). We keep the 2MASS coordinates otherwise. Additional sources originating from a few select small catalogs (e.g., the specially curated cool dwarf list) have their coordinates taken directly from those catalogs.

The proper motions and parallaxes for {\it Tycho-2\/} stars from the {\it Gaia\/} DR1 catalog \citep{deBruijne:2012,Gaia:2016} have also been cross-matched using the 2MASS IDs provided in the {\it Gaia\/} source catalog, and the {\it Gaia\/} DR2 catalog will be cross-matched for future TIC releases. Cross matches are done directly with 2MASS IDs when possible, otherwise they are done via cone search typically with a 10~arcsec search radius and, when possible, a comparison of magnitudes with a tolerance of 0.1~mag.

If entries in multiple proper motion catalogs are available for a given star, we prioritize the proper motion measurements giving preference to those from the {\it Tycho-Gaia\/} Astrometric Solution \citep[hereafter TGAS;][]{Gaia:2016}, then SUPERBLINK \citep{Lepine:2011}, then {\it Tycho-2}, then {\it Hipparcos}. We have chosen to rank {\it Tycho-2\/} above {\it Hipparcos\/} because it has a significantly longer time baseline that can help to identify and remove binary motion contamination. SUPERBLINK was ranked above {\it Tycho-2\/} and {\it Hipparcos\/} because it incorporates a similar ranking for the preference of proper motions using {\it Tycho-2\/} and {\it Hipparcos}.

However, only a small subset of all stars in the TIC have proper motions available in the catalogs described above. To greatly increase the number of stars with proper motions we include the catalogs UCAC4, UCAC5, and HSOY. We select from either UCAC4, UCAC5, or HSOY depending on the value of the star's total proper motion. For stars with total proper motions less than 200~mas~yr$^{-1}$ we use HSOY. For stars with total proper motions between 200--1800~mas~yr$^{-1}$ we use UCAC5. For stars with total proper motions greater than 1800~mas~yr$^{-1}$ we use UCAC4. This approach was adopted to: 1) use proper motions that incorporate \textit{Gaia} DR1 for a large number of distant stars \citep[HSOY,][]{Altmann:2017}; 2) eliminate the large number of false-positive, high proper-motion stars in HSOY, which are not found in UCAC5; and 3) provide accurate proper motions for the high proper-motion stars, through the UCAC-provided correction catalog.\citep{Zacharias:2013}.

Specially curated lists of objects (see Appendix~\ref{sec:spec_cat}), such as stars from {\it Hipparcos\/} \citep{Perryman:1997}, known cool dwarf stars \citep{Muirhead:2018}, known planet hosts, hot subdwarfs, bright stars, and guest investigator targets, are added to the TIC and CTL. The stellar properties supplied by those specially curated lists supersede the default values in the TIC and CTL (see Appendix~\ref{sec:spec_cat} for details), but their prioritizations for the transit search are calculated in the same manner as other targets. 

\subsubsection{Blended Objects in 2MASS that are Resolved by \textit{Gaia} \label{subsubsec:gaiadupes}}

Whenever a single match between a \textit{Gaia} source and 2MASS source exists, we accept the match from the official \textit{Gaia} DR-1 cross-match table \citep{Gaia:2016}. However, in many cases, multiple \textit{Gaia} targets will match to a single 2MASS star. The procedure for identifying the most appropriate \textit{Gaia} source associated with a single 2MASS source is as follows.

First, we store the flux weighted mean of right ascension and declination for all \textit{Gaia} sources matching to the parent star, and accept the weighted means as the stellar coordinates. We then store the flux weighted mean of the angular distances between each of the matching \textit{Gaia} sources, and adopt this as the error for the star's right ascension and declination. We do not simply adopt the flux weighted positional error of an individual \textit{Gaia} source because that would be unrealistically small. For example, if two equally bright sources are 1\arcsec~from a given 2MASS source, the positional error is not $\ll 1$\arcsec\ (as a flux weighted combination of \textit{Gaia} positional errors would imply), but rather 1\arcsec. We update the provenance flag for the positions identified in this way from \textit{tmass} to \textit{tmmgaia}. The \textit{Gaia} ID assigned to the TIC star is chosen the be the brightest of the \textit{Gaia} stars. 

We update the \textit{Gaia} magnitude by combining the fluxes of all matching \textit{Gaia} stars. We use two provenance flags to differentiate between the original \textit{Gaia} magnitude and the combined \textit{Gaia} magnitude, \textit{tmgai} and \textit{tmmga} respectively.

\subsubsection{Extended sources}

The extended source base catalog is the full 2MASS extended source catalog \citep{Skrutskie:2006}. This base catalog is then positionally cross-matched with the ALLWISE extended source catalog \citep{Cutri:2013}, and the SDSS extended source catalog \citep{Alam:2015}, and then merged with the point source catalog. 

Early versions of the TIC (up to and including TIC-5) included the full SDSS extended source catalog as a part of the basis of the extended source catalog. However, we found that including the entire SDSS catalog also introduced a large number of artifacts around bright stars caused by diffraction spikes in the SDSS images. These artifacts contributed to source confusion, and led to duplicated bright objects in the flux contamination estimates. Therefore we added SDSS-colors to 2MASS extended objects, but did not include any extended object from SDSS, unless it is also in 2MASS.

\subsection{Algorithms for calculated stellar parameters \label{subsec:parms}}

In this section, we describe the algorithmic procedures we adopted for calculating various stellar parameters. 
We begin with the procedure for calculating an apparent magnitude in the {\it TESS\/} bandpass, $T$, as this is the most basic quantity required of any object in the TIC.
In order to provide maximum flexibility, where possible we developed multiple relations involving the various possible combinations of available colors. 
Next, since certain parameters such as reddening can at present only be computed with relations involving a $V$ magnitude, we separately describe the procedure for calculating a $V$ magnitude from other colors when a native $V$ magnitude is not available. The {\it Gaia\/} $G$ magnitude is especially useful, since it is available for the vast majority of point sources in the TIC.
Finally, we describe the procedures for calculating physical parameters of the stars, such as effective temperature. 

\subsubsection{TESS magnitudes}\label{sec:tmag_relations}

The most basic quantity required for every TIC object, aside from its position, is its apparent magnitude in the TESS bandpass \citep[see][for a definition of the TESS bandpass and of the TESS photometric system]{Sullivan:2015}, which we represent as \tmag.
We estimate \tmag\ using a set of empirical relations developed from PHOENIX stellar atmosphere models \citep{Husser:2016} to convert the available catalog magnitudes and colors to \tmag\ magnitudes. 
The relationships between \tmag\ and other magnitudes for stars with low surface gravities ($\log g < 3.0$) and low metallicities (${\rm [Fe/H]} < -0.5$) can be quite different from those of near solar-metallicity dwarfs. Since dwarfs are the targets of greatest interest for TESS,
we adopt a single set of relations, which are strictly valid for \logg\ $>$ 3 and \feh\ $> -0.5$. They also are also strictly valid for stars with $\teff > 2800$~K. We do apply the relations outside these ranges in order to ensure that every star in the TIC has an estimated \tmag, but we note that in such cases the estimated \tmag\ are subject to larger errors. 

For each of the color-\tmag\ relations, we have opted to define a single polynomial relation. As a result the calibrated relationships have somewhat larger scatter than might otherwise be possible with a more complex relation, but the scatter is still usually smaller than the errors from the original photometric catalogs, and thus we believe it is a worthwhile tradeoff for the sake of simplicity. The effects of reddening on \tmag\ are not currently included---many of the highest priority TESS targets will be relatively nearby and should not experience much reddening---but are expected to be implemented in future versions of the TIC (see \S~\ref{sec:redT}).

We first present the relations for point sources, next for extended objects, and finally we provide ancillary relations we developed for calculating other magnitudes such as $B$ and $V$. We note that while most of the magnitude calculations are based on Gaia and 2MASS magnitudes, we exclude values from our calculations with the following 2MASS quality flags: {\tt X, U, F, E, D}. For the point sources we report a set of preferred relations, in order of preference, followed by a set of fallback relations that we use for a small number of stars for which it is not possible to apply any of the preferred relations. Column~63 in the TIC specifies which relation was used for any given star (see Appendix~\ref{sec:relnotes}); these flag names are listed below with their associated relations.

Regarding the uncertainties in the TESS magnitudes, we have investigated the use of the full covariance matrix but we find that those errors are typically small compared to the scatter of the calibrations. Thus we believe it is more conservative to simply use the scatter provided with each of the relations below, added in quadrature to the errors propagated from the photometric uncertainties.

Finally, in a small number of cases where we are unable to calculate \tmag\ we arbitrarily assign a value of $\tmag = 25$. This is the case for only \numbadtmag\ objects in the TIC.

\subsubsubsection{Point sources, preferred relations}


\noindent\textbf{{\tmag~from \textit{Gaia} photometry}}

\noindent{\textit{\textbf{gaiaj}}: $T$ from $G$ and $J$, valid for all \feh\ and \logg. }
$$T = G + 0.00106(G-J)^3 + 0.01278(G-J)^2 - 0.46022(G-J) + 0.0211$$
with a scatter of 0.015 mag. 

\noindent{\textit{\textbf{gaiah}}: $T$ from $G$ and $H$ valid for $\feh \ge -0.5$ and $\logg \ge 3.0$.}
$$T = G + 0.00510(G-H)^3 + 0.02230(G-H)^2 - 0.38134(G-H) - 0.0058$$
valid for $-0.3 \le G-H \le 2.3$, with a scatter of 0.010 mag.
$$T = G + 0.01029(G-H)^3 - 0.06163(G-H)^2 - 0.31892(G-H) + 0.2113$$
valid for $2.3 < G-H \le 5.0$, with a scatter of 0.014 mag. 

\noindent{\textit{\textbf{gaiak}}: $T$ from $G$ and $K_S$ valid for $\feh \ge -0.5$ and $\logg \ge 3.0$.}
$$T = G + 0.00942(G-K_S)^3 + 0.00288*(G-K_S)^2 - 0.35664(G-K_S) + 0.0125$$
valid for $-0.3 \le G-K_S \le 2.5$, with a scatter of 0.010 mag.
$$T = G + 0.02925(G-K_S)^3 - 0.29659(G-K_S)^2 + 0.60877(G-K_S) - 0.8676$$
valid for $2.5 < G-K_S \le 5.2$, with a scatter of 0.019 mag. 

\noindent{\textit{\textbf{gaiav}}: $T$ from $G$ and $V$ valid for all \feh\ and $\logg \le 5.0$.}
$$T = G - 6.5(V-G) + 0.031$$
valid for $0.0 \le V-G < 0.02$, with a scatter of 0.054 mag.
$$T = G + 6.04460(V-G)^2 - 3.64200(V-G) - 0.0336$$
valid for $0.02 \le V-G \le 0.2$, with a scatter of 0.018 mag.
$$T = G - 0.11179(V-G)^3 + 0.57748(V-G)^2 - 1.28108(V-G) - 0.3173$$
valid for $0.2 < V-G \le 2.8$, with a scatter of 0.015 mag.

\medskip 

\noindent{\textbf{$T$ from 2MASS photometry}}

The following relations are valid for stars with $-0.1 < J-K_S < 1.0$.  In all cases the $J$ and $K_S$ magnitudes are taken from 2MASS. 

\noindent{\textit{\textbf{vjk}}:~\tmag\ from $V$, $J$, and $K_S$:}
$$T = J + 0.00152\:(V-K_S)^3 - 0.01862\:(V-K_S)^2 + 0.38523\:(V-K_S) + 0.0293,$$
where the scatter of the calibration is 0.021 mag.  In this relation, the $V$ magnitude might be taken from APASS, or calculated from the Tycho-2 $B_T$ magnitude (see below).


\noindent{\textit{\textbf{bpjk}}:~\tmag\ from $B_{ph}$, $J$, and $K_S$, where $B_{ph}$ is a photographic $B$ magnitude (from 2MASS via USNO A2.0):} 
$$T = J + 0.00178\:(B_{ph}-K_S)^3 - 0.01780\:(B_{ph}-K_S)^2 + 0.31926\:(B_{ph}-K_S) + 0.0381,$$
where the scatter is 0.020 mag.

\noindent{\textit{\textbf{bjk}}:~\tmag\ from $B$, $J$, and $K_S$, where $B$ is a Johnson magnitude:}
$$T = J + 0.00226\:(B-K_S)^3 - 0.02313\:(B-K_S)^2 + 0.29688\:(B-K_S) + 0.0407,$$
where the scatter is 0.031 mag.  In this relation, the $B$ magnitude might be taken from any of several catalogs.

The following relations are used when there are no Johnson $V$, Johnson $B$, or photographic $B_{ph}$ magnitudes available. In all cases the $J$ and $K_S$ magnitudes are taken from 2MASS.

\noindent{\textit{\textbf{jk}}:~\tmag\ from $J$ and $K_S$.}
$$T = J + 1.22163\:(J-K_S)^3 - 1.74299\:(J-K_S)^2 + 1.89115\:(J-K_S) + 0.0563,$$
valid for $J-K_S \le 0.70$, where the scatter is 0.080 mag. 
$$T = J - 269.372\:(J-K_S)^3 + 668.453\:(J-K_S)^2 - 545.64\:(J-K_S) + 147.811,$$
valid for $J-K_S > 0.70$, where the scatter is 0.17 mag.

There are some stars for which the $J-K_S$ colors are too blue or too red to determine a reliable \tmag\ from the above relations. For these stars we use:

\noindent{\textit{\textbf{joffset}}:~\tmag\ for $J-K_S < -0.1$:}
$$ T = J + 0.5~~(\pm 0.8~{\rm mag}).$$
\noindent{\textit{\textbf{joffset2}}:~\tmag\ for $J-K_S > 1.0$:}
$$ T = J + 1.75~~(\pm 1.0~{\rm mag}).$$

\subsubsubsection{Point sources, fallback relations}

For a small subset of point sources in the TIC we have only a few magnitudes listed with passing quality flags from the available catalogs. For these, we adopt the following relations, if needed. 

\noindent{\textit{\textbf{vjh}}:~\tmag\ from $V$, $J$, and $H$ (no $K_S$ available):}
$$T = V - 0.28408\:(J-H)^3 + 0.75955\:(J-H)^2 - 1.96827\:(J-H) - 0.1140$$
where the scatter is 0.063 mag. 

\noindent{\textit{\textbf{jh}}:~\tmag\ from $J$, and $H$ (no $V$ or $K_S$ available):}
$$T = J - 0.99995\:(J-H)^3 - 1.49220\:(J-H)^2 + 1.93384\:(J-H) + 0.1561$$
where the scatter is 0.040 mag.

Finally, for faint stars for which only one magnitude is valid, the best we can do to compute \tmag\ is to apply a simple offset from the one available magnitude. We wish to estimate a \tmag\ magnitude in these cases, even if only crudely, because we wish to include all known objects in the sky as part of our flux contamination calculations (see Section~\ref{sec:flux_contam}). 
Because we rely on a specially curated cool-dwarf catalog to identify faint, cool objects that are known to be {\it bona fide} M dwarfs, here we use models to compute \tmag\ only for $\teff > 3840$~K. The uncertainties are representative of the spread for a range of \teff, and the numerical value is an average.

\medskip
\noindent{\textit{\textbf{gaiaoffset}}}:
$$T = G - 0.5~~(\pm 0.6~{\rm mag}),$$ 
\noindent{\textit{\textbf{voffset}}}:
$$T = V - 0.6~~(\pm 0.9~{\rm mag}),$$
\noindent{\textit{\textbf{joffset}}}:
$$T = J + 0.5~~(\pm 0.8~{\rm mag}),$$
\noindent{\textit{\textbf{hoffset}}}:
$$T = H + 0.7~~(\pm 1.3~{\rm mag}),$$
\noindent{\textit{\textbf{koffset}}}:
$$T = K_S + 0.8~~(\pm 1.4~{\rm mag}).$$

\subsubsubsection{Extended objects}

\noindent{\textit{\textbf{jk}}:~\tmag\ from $J$ and $K_S$.}
$$T = J + 1.22163\:(J-K_S)^3 - 1.74299\:(J-K_S)^2 + 1.89115\:(J-K_S) + 0.0563,$$
valid for $J-K_S \le 0.70$, where the scatter is 0.080 mag. 
$$T = J - 269.372\:(J-K_S)^3 + 668.453\:(J-K_S)^2 - 545.64\:(J-K_S) + 147.811,$$
valid for $J-K_S > 0.70$, where the scatter is 0.17 mag. It should be noted that since these relations were derived from models for dwarf stars, when applied to extended objects the errors will typically be larger than the formal scatter.

\noindent{\textit{\textbf{SDSS}}~\tmag\ from SDSS $g$ and $i$ (for 2MASS extended sources lacking good 2MASS photometry):}
$$T=i-0.00206\:(g-i)^3-0.02370\:(g-i)^2+0.00573\:(g-i)-0.3078,$$
where the scatter is 0.030 mag.
If an SDSS extended source was found to have an unreasonable \tmag\ magnitude
($T \notin [-5,25]$),
then if $g = 30$, signifying a failure in measuring an SDSS magnitude, and {\it not} $i = 30$, we adopt $T = i-0.5$~($\pm 1.0$ mag). If $g \neq 30$ and $i = 30$, then we adopt $T = g-1$~($\pm 1.0$ mag).

\subsubsection{$V$ magnitudes}

In order of preference we adopt high-confidence $V$ magnitudes from Mermilliod we used for deriving the relation $V(G - K)$, then Hipparcos $V$, then third-order and linear fits for $V(V_T, B_T)$, then the CCD-based magnitudes listed in the UCAC4 catalog (which are not far from Johnson $V$; see~Figure~\ref{fig:vmag_from_ucac}), unless they were flagged as unreliable (i.e., if the `number of images' used is reported as zero), then from APASS DR9, and finally secondary sources, such as calculated from $G - K_s$ or 2MASS colors only. For $V$ from Hipparcos, Tycho2, UCAC4 and APASS we compute the Gaia-based $V(G - K_s)$ and compare it to the $V$ reported in the catalog. We accept the catalog value if the difference is less than $0.5$ mag, otherwise we assume a bad measurement and use $V(G - K_s)$ as a valid $V$ magnitude. In all cases, whether the $V$ magnitude is observed or calculated from one of the following relations, we convert the adopted $V$ magnitudes to Johnson $V$ using standard conversion relations. Catalog flags that define the source of each magnitude are currently not provided by the TIC schema but may be provided in future versions.

\subsubsubsection{$V$ magnitude from $G$}

\begin{figure}[ht]
    \centering
    \includegraphics[width=\linewidth,trim=10 10 0 10,clip]{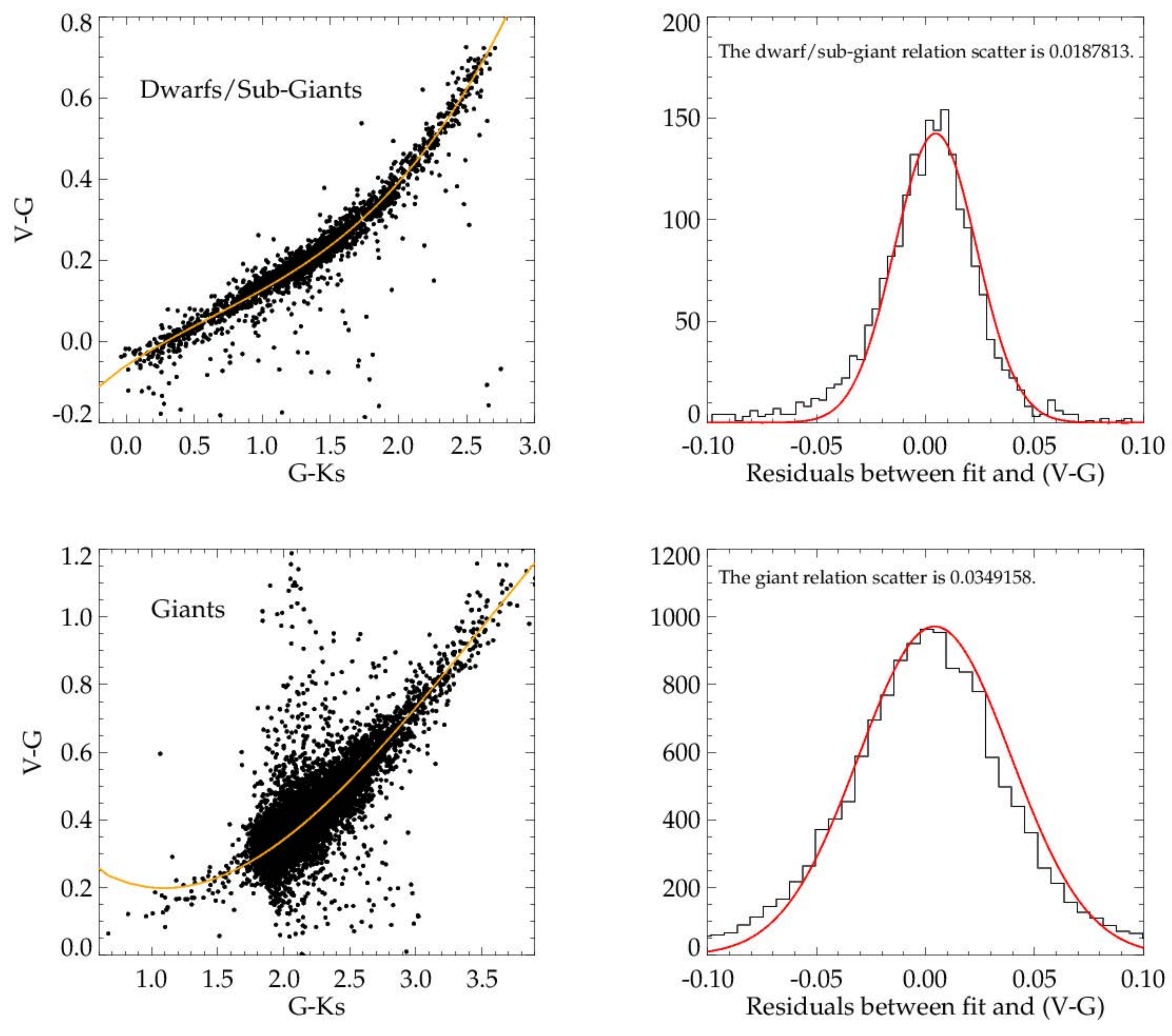}
    \caption{The fits describing the translation from \textit{Gaia} G magnitude to Johnson$V$ magnitude. (\textit{top}) The dwarf relation defined with \citet{Mermilliod:1987} values, with a scatter of 0.019~mag; (\textit{bottom}) The giant relation defined with APASS DR-9 values, with a scatter of 0.035~mag.
    }
    \label{fig:v2g}
\end{figure}

\begin{figure}[!ht]
    \centering
    \includegraphics[width=0.9\linewidth,clip,trim=0 75 0 75]{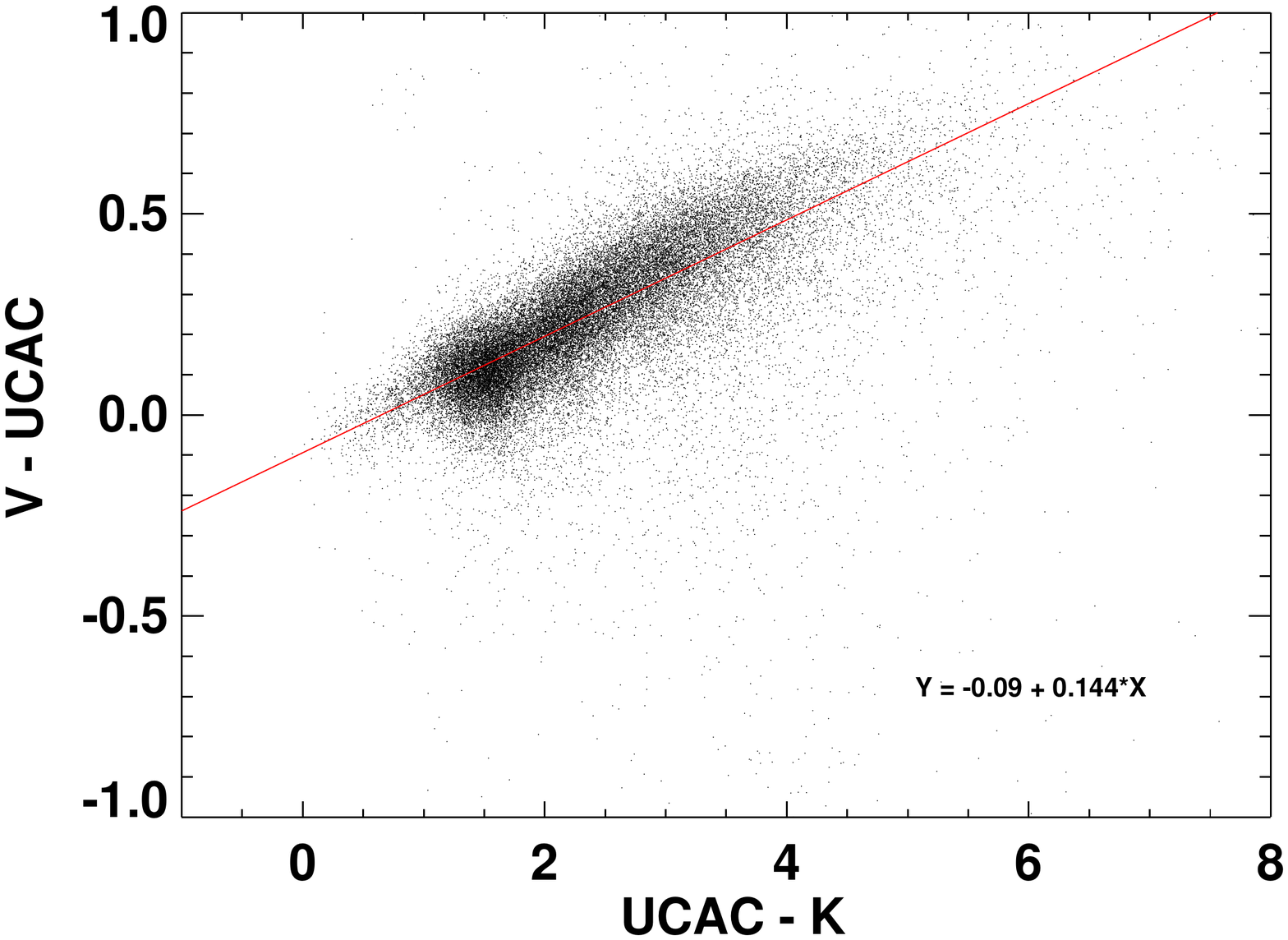}
    \caption{Determination of Johnson $V$ from UCAC4 (near-$V$) aperture magnitudes as a function of the measured $V-K_S$ color from UCAC4 (using near-$V$ for the color). Shown is the fit relation (red) to $\sim$40,000 stars from a cross-match between UCAC4, APASS, and 2MASS.}
    \label{fig:vmag_from_ucac}
\end{figure}

In many cases, we found the observed $V$ magnitudes for bright stars to be discrepant between catalogs by $>2\%$. Therefore, we elected to calculate a Johnson $V$ magnitude using magnitudes from the two largest catalogs in the TIC, \textit{Gaia} DR1 and 2MASS (see Figure~\ref{fig:v2g}). The relations for dwarfs and sub-giants were created comparing Johnson $V$ magnitudes measured in \citet{Mermilliod:1987} with $G-K_S$ color, using 2015 dwarfs and sub-giants within 100~pc and with photometric errors in $V$ of $\sigma_{V} < 0.1$. The relation for giants was created comparing APASS DR9 Johnson $V$ magnitudes with $G-K_S$ color, using 13,580 giants within 400~pc, and with photometric errors in $V$ of $\sigma_{V} < 0.1$. For stars that were used as part of determining these relations, we adopted the $V$ magnitudes from \citet{Mermilliod:1987} or APASS DR9 regardless of the prioritization above, for consistency. In both cases, 2.5 sigma-clipping was used to create the fit.

For dwarfs the following is adopted for $0.0 < G-K_S < 2.75$, with scatter of 0.019:
$$V = G - 0.0598208 + 0.2212224 (G-K_S) - 0.0705395 (G-K_S)^2 + 0.0363978 (G-K_S)^3$$

For giants the relation for dwarfs is adopted when $0.0 < (G-K_S) < 1.45$; for $1.45 < G-K_S < 3.7$ the following is adopted, with scatter of 0.035:
$$V = G + 0.4710164 - 0.5313155 (G-K_S) + 0.2883704 (G-K_S)^2 - 0.0274133 (G-K_S)^3$$

\subsubsubsection{$V$ magnitude from UCAC4}

In the case of the UCAC4 aperture magnitudes there appears to be no published conversion to Johnson $V$, so we developed our own, shown in Figure~\ref{fig:vmag_from_ucac}, based on $\sim$40,000 stars that we cross-matched between UCAC4, APASS, and 2MASS. The relation is defined as:
$$V  = A - 0.09 + 0.144 (A - K_S)$$
where $A$ is the UCAC4 aperture magnitude and $K_S$ is from 2MASS.

\subsubsubsection{$V$ magnitudes from secondary sources}

Finally, we calculate a $V$ magnitude for stars that do not have one in an existing catalog and for which there is not a reliable $G-K_S$. We use the relations described in the following subsections depending on what magnitude information is available for each star. The following alternate relations are mostly useful for dwarf stars, and are strictly valid for stars with $\logg > 3.0$. They are expected to give larger errors for metal-poor stars with $\feh < -0.5$ and for very cool dwarfs with $\teff < 2800$~K.

\noindent{$\bullet$~$V$ magnitude from $B_{ph}$ and $K_S$, where $B_{ph}$ is photographic (from 2MASS via USNO A2.0):}
$$V=J-0.00814\:(B_{ph}-K_S)^3-0.03725\:(B_{ph}-K_S)^2+0.63921\:(B_{ph}-K_S)+0.0323,$$
where the scatter is 0.023 mag.  In this relation, the $B$ magnitude might be taken from any of several catalogs.

\noindent{$\bullet$~$V$ from $B$ and $K_S$, where $B$ is a Johnson magnitude:}
$$V=J+0.00740\:(B-K_S)^3-0.03897\:(B-K_S)^2+0.57069\:(B-K_S)+0.0355,$$
where the scatter is 0.041 mag.  In this relation, the $B$ magnitude might be taken from any of several catalogs.

\noindent{$\bullet$~$V$ from $J$ and $K_S$:}
$$V=J+1.28609\:(J-K_S)^3-2.35587\:(J-K_S)^2+3.70190\:(J-K_S)+0.0766,$$
where the range of validity is $J-K_S \leq 0.70$, and the scatter is 0.027 mag. For the complementary color range $J-K_S > 0.70$ the relation is
$$V=J+63.3104\:(J-K_S)^2-86.2252\:(J-K_S)+31.1658.$$
The scatter of this relation is large (0.49 mag) so its usefulness is limited, but provided for completeness.

\subsubsection{$B$ magnitudes}

Johnson $B$ magnitudes are calculated for populating the $B$ magnitude column of the TIC and for ensuring all magnitudes provided in the TIC are on the same photometric system, similar to what is done for the $V$ magnitude calculations above.

\noindent{$\bullet$~Johnson $B$ magnitude from photographic $B_{ph}$ and $J$}:
$$J = B_{ph} - 0.00526\:X^3 + 0.03256\:X^2 + 0.14101\:X - 0.0149,$$
where $X = B_{ph}-J$. While the overall scatter of the calibration is 0.060 mag, it is better for small $X$ and much worse for large $X$. Therefore, when propagating errors and adding the scatter of the calibration in quadrature, 0.035 mag should be added to the scatter if $X < 3.5$, and 0.16 mag if $X > 3.5$.

The photometric errors for photographic magnitudes in 2MASS come from the USNO-A2.0 catalog. Based on that catalog's description\footnote{\url{http://vizier.u-strasbg.fr/vizier/VizieR/pmm/usno2.htx}} the errors in the photographic $B$ magnitudes are expected to be $\sim 0.3$ mag in the equatorial north and $\sim 0.5$ mag in the equatorial south.

\subsubsection{Effective Temperature\label{sec:teff}}

In this section, we describe the methods used to determine \teff\ for stars in the TIC. Some stars in the TIC have spectroscopically derived \teff\ values in the literature, and we use these where available. Specifically, we have ingested several large spectroscopic catalogs, and we adopt the spectroscopic \teff\ if the reported error is less than 300\;K, giving preference to the catalogs in the priority order listed in Table~\ref{tab:spectra_cats}. Additionally, the proper calculation of parameters such as \teff\ for cool dwarf stars ($\teff <3840$~K) is typically very difficult when using ensemble dwarf relations, like we do in this section. Therefore, a specially curated cool dwarf list was created to calculate these parameters; a brief explanation of the techniques used is in \S~\ref{subsubsec:cool} but we direct the reader to \citet{Muirhead:2018} for more details. 

\begin{center}
\vspace{-0.1in}
\begin{longtable}[c]{|c|c|c|c|c|}
 \hline
 Name & Data Release & Approximate Num. of Stars & Priority & Reference \\
 \hline
 SPOCS & & 1.6~k & 1 & \citet{Brewer:2016}\\
 PASTEL & & 93~k & 2 & \citet{Soubiran:2016}\\
 {\it Gaia}-ESO & & 29~k & 3 & \citet{Gilmore:2012}\\
 HERMES-{\it TESS} & DR-1 & 25~k & 4 & \citet{Sharma:2018}\\
 GALAH & & 10~k & 5 & \citet{Kordopatis:2013}\\
 APOGEE-2 & DR-14 & 277~k & 6 & \citet{Abolfathi:2017}\\
 APOGEE-1 & DR-12 & 160~k & 7 & \citet{Majewski:2015}\\
 LAMOST & DR-3 & 2.9~M & 8 & \citet{Luo:2015}\\
 LAMOST & DR-1 & 1.0~M & 9 & \citet{Luo:2015}\\
 RAVE & DR-5 & 484~k & 10 & \citet{Kunder:2017}\\
 RAVE & DR-4 & 482~k & 11 & \citet{DeSilva:2015}\\
 Geneva-Copenhagen & DR-3 & 16~k & 12 & \citet{Holmberg:2009}\\
 \hline
 \caption{Spectroscopic Catalogs in the TIC.}
 \label{tab:spectra_cats}
\end{longtable}
\vspace{-0.5in}
\end{center}

However, the majority of TIC stars do not have spectroscopic \teff\ values available. Therefore we have developed a procedure to estimate \teff\ based on empirical relationships of stellar $V-K_S$ color. Because 2MASS is the base catalog for the TIC, we have a $K_S$ magnitude for nearly every object (though to reiterate, we do not use 2MASS photometry whose quality flags are any of {\tt X, U, F, E, D}). 

The relation {\it for stars that are not identified as giants} (see Section~\ref{sec:giant_removal}) is in two pieces:

\begin{enumerate}
\item For $V-K_S$ in the range $[-0.10, 5.05]$ and [Fe/H] in the range $[-0.9, +0.4]$, we use the AFGKM relation from \citet{Huang:2015}:
\begin{align*}
X &= V-K_S \, {\rm (de-reddened)} \\
Y &= {\rm [Fe/H]} \\
\theta &= 0.54042 + 0.23676X - 0.00796X^2 - 0.03798XY + 0.05413Y - 0.00448Y^2 \\
T_{\rm eff} &= 5040/\theta
\end{align*}
The scatter of this calibration is 2\% in \teff, which should be added in quadrature to whatever errors come from photometric uncertainties propagated through the above equation.

\item For redder $V-K_S$ values in the range $[5.05, 8.468]$, use the relation from \citet{Casagrande:2008}, shifted by +205.26 K to meet with the \citet{Huang:2015} relation at $V-K_S = 5.05$:
\begin{align*}
X &= V-K_S \, {\rm (de-reddened)} \\
\theta &= -0.4809 + 0.8009X - 0.1039X^2 + 0.0056X^3 \\
T_{\rm eff} &= 5040/\theta + 205.26
\end{align*}
The scatter of this calibration is only 19 K, according to the paper, which should be added in quadrature to whatever errors come from photometric uncertainties propagated through the above equation. Note that this \citet{Casagrande:2008} relation does not include metallicity terms, but those authors claim that the dependence on metallicity is weak for M stars with \teff $>2800$ K.
\end{enumerate}

A number of previous studies have compared spectroscopic determinations of \teff\ to photometric determinations \citep[e.g.,][]{Damiani:2016,Gazzano:2010,Guenther:2012,Sebastian:2012}, and generally find that the error of the photometric \teff\ determination depends on the spectral type of the star. The scatter in the color-\teff\ relations noted above include this \teff\ dependence implicitly; the reported scatter is a percentage of \teff. For example, at 4000~K the formal \teff\ error is 80~K and at 8000~K it is 160~K.

\begin{figure}[!ht]
    \centering
    \includegraphics[width=\linewidth,clip,trim=0 5 0 60]{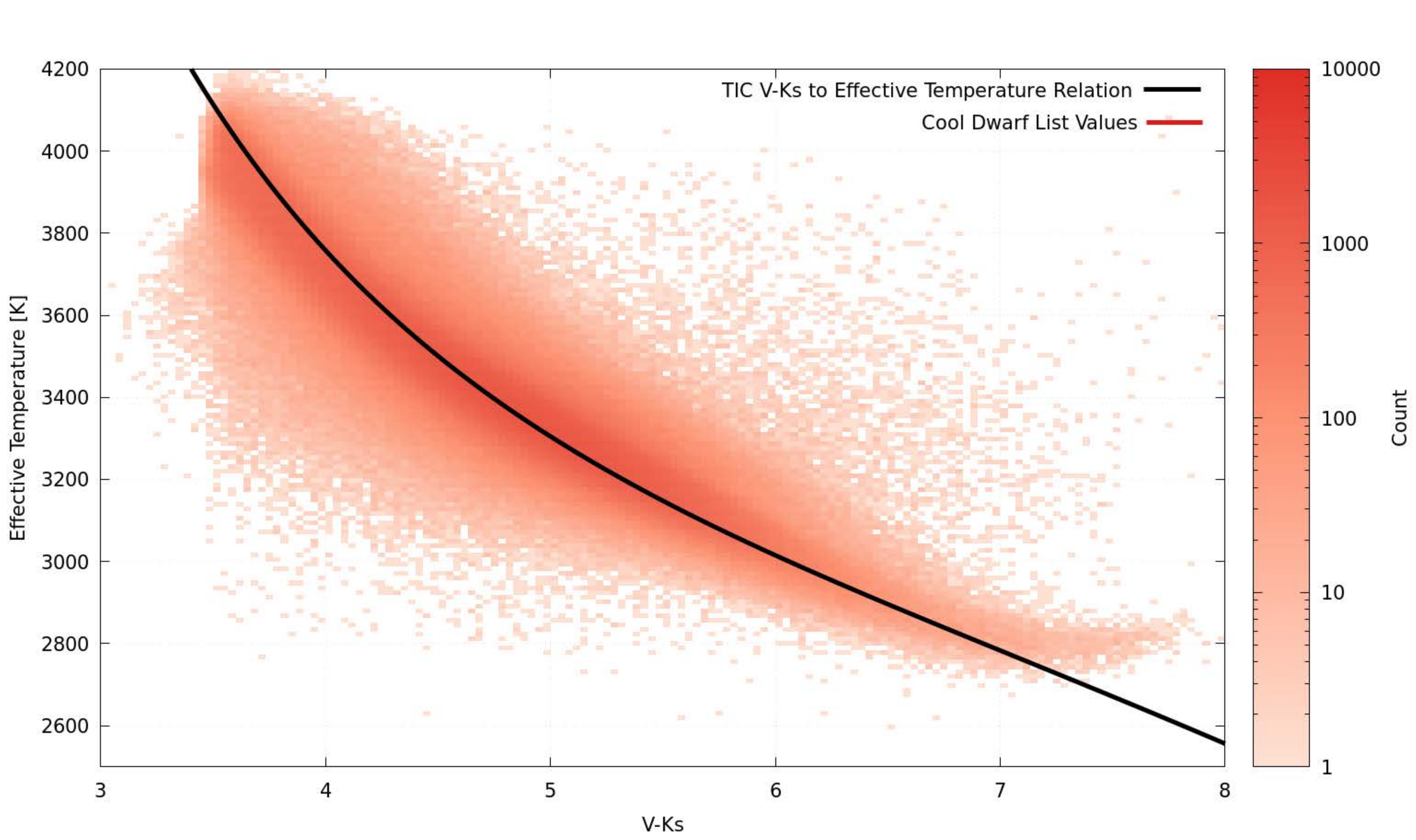}
    \caption{Plot of $V-K_S$ and \teff\ of the independently determined \teff\ for stars in the cool dwarf list (red dots) with the over plotted color-\teff\ relation for dwarfs (black curve).}
    \label{fig:new_teff}
\end{figure}

Color-temperature relations can be especially challenging at cool \teff\ because of the complexities of the spectral energy distributions of very cool stars. Determining reliable \teff\ is crucial to estimating the stellar radii and thus for sensitivity of small planets (see below). To validate the above relations at cool \teff, we compared the \teff\ predicted by the relation to the \teff\ supplied independently by the specially curated cool-dwarf list in the CTL (see Section~\ref{sec:spec_cat}). The result of this comparison (Figure~\ref{fig:new_teff}) indicates good agreement with no obvious systematics, except perhaps at the very coolest \teff\ ($V-K_S \approx 7$, $\teff \lesssim 2600$~K). 
To further verify the \teff\ that we determine for M-dwarf stars, we compared our calculated \teff\ with the values in the specially curated cool-dwarf list \citep[see Section~\ref{sec:spec_cat}, and][]{Muirhead:2018}. As shown in Figure~\ref{fig:new_teff}, the independently determined \teff\ from the cool-dwarf list closely follows the color-based \teff\ relation described above and we find a mean difference of only $-19 \pm 63$~K.

Finally, for completeness, we provide a similar relation for red giants, also from \citet{Huang:2015}:
\begin{align*}
X &= V-K_S \, , \, Y = \rm {[Fe/H]} \\
\theta &= 0.46447 + 0.30156\:X - 0.01918\:X^2 - 0.02526\:X\:Y + 0.06132\:Y - 0.04036\:Y^2 \\
T_{\rm eff} {\rm [K]} &= 5040/\theta
\end{align*}
which is valid for $V-K_S$ in the range $[1.99, 6.09]$ and [Fe/H] in the range $[-0.6, +0.3]$. The scatter in \teff\ is 1.7\% \citep{Huang:2015}.

The above expressions provide a continuous color-\teff\ relation from 2444~K to 9755~K. For stars with $V-K_S$ outside of the above ranges of validity, the TIC reports \teff\ = {\tt Null}. 
For stars that are deemed to be likely non-giants according to our reduced-proper-motion procedure (see Section~\ref{sec:giant_removal}), the \teff\ is calculated with the above relations but using the {\it de-reddened\/} $V-K_S$ color (see Section~\ref{sec:dereddening} for the de-reddening procedure). In general we \textit{do not} calculate the \teff\ for giants using a \textit{de-reddened} color because we only apply our de-reddening procedure for stars identified as dwarfs which are the stars most likely to be included in the transit candidate targeting list (see Section~\ref{sec:CTL}). We note, finally, that in order to be conservative the final \teff\ errors from the above polynomial relations include a 150~K contribution added in quadrature to the uncertainties from the photometric errors and the scatter of the calibrations.

\section{The Candidate Target List (CTL)\label{sec:CTL}}

The purpose of the CTL is to provide a subset of TIC objects that can be used to select the 
target stars for TESS 2-min cadence observations
in service of the TESS mission's primary science requirements, which are: 

\begin{enumerate}
\item To search $>$200,000 stars for planets with orbital periods less than 10~d and radii smaller than $2.5\;\rearth$.
\item To search for transiting planets with radii smaller than $2.5\;\rearth$ and with orbital periods up to 120~d among 10,000 stars in the ecliptic pole regions.
\item To determine masses for at least 50 planets with radii smaller than $4\;\rearth$. 
\end{enumerate}

Given the limited number of stars for which TESS will be able to acquire 2-min cadence light curves, it is crucial that the set of targets for TESS be optimized for small planet detection. To that end, we have compiled a catalog of bright stars that are likely to be dwarfs across the sky, from which a final target list for TESS can be drawn, based on in-flight observation constraints yet to be determined. This list of high-priority candidate 2-min cadence targets is the Candidate Target List (CTL). Our basic consideration is to assemble a list of dwarf stars all over the sky in the temperature range of interest to TESS, bright enough for TESS to observe, and taking extra steps to include the scientifically valuable M dwarfs. Our overall approach is to start with the 470 million stars in the TIC, and then apply cuts to select stars of the desired ranges in apparent magnitude and spectral type, and to eliminate evolved stars. At this stage we also compute additional stellar parameters that are relevant for target selection, which for logistical reasons or computational limitations, we do not compute for all other stars in the TIC.

First, we give a brief overview describing the assembly of the CTL from the TIC, including specifically the process by which we identify likely dwarf stars for inclusion in the CTL and identify likely red giants for exclusion from the CTL.
Next we describe the algorithms by which a number of stellar properties---such as stellar mass and radius---are computed for the CTL (Section~\ref{sec:ctl_algorithms}). Finally, we present the prioritization scheme used for identifying the top priority targets from the CTL for targeting (Section \ref{subsec:priority}). 
The CTL is provided for use through the Mikulski Archive for Space Telescopes (hereafter, MAST) and for interactive use via the Filtergraph data visualization system \citep{Burger:2013} at {\tt \url{http://filtergraph.vanderbilt.edu/tess_ctl}}.
A summary of the quantities included in the CTL on the Filtergraph portal is provided in Appendix~\ref{sec:appendix_filtergraph}.

\subsection{Assembly of the CTL and Determination of Additional Stellar Parameters\label{subsec:assembly}}

\begin{figure}[!ht]
    \centering
    \includegraphics[width=\linewidth]{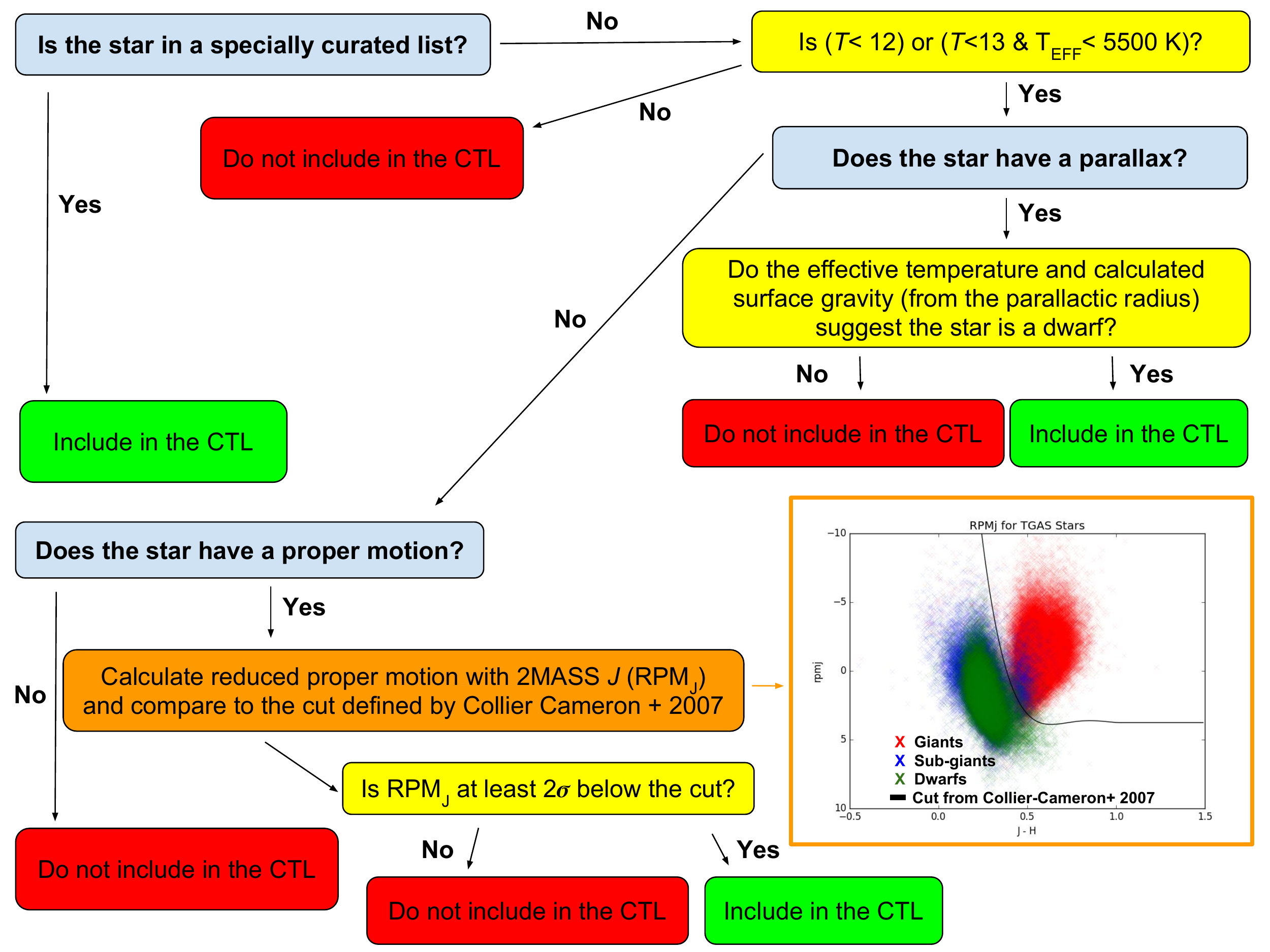}
    \caption{Overview of procedure for selecting stars for the CTL. The inset uses {\it Gaia\/} DR1 stars to show the effectiveness of separating red giants (red points) in the RPM$_J$ diagram using the cut defined by \citet{Collier-Cameron:2007}. To the left/below the dividing line are dwarfs and subgiants; the RPM$_J$ cut removes $\gtrsim$95\% of red giants but removes only $\sim$10\% of subgiants (see Section~\ref{sec:giant_removal}). 
    }
    \label{fig:giant_removal}
\end{figure}

As illustrated in Figure~\ref{fig:giant_removal},
from the 470 million point sources in the TIC we initially select stars for the CTL if they are: (1) identified as unlikely to be giants according to a reduced proper motion (RPM) criterion (see \S~\ref{sec:giant_removal}); and (2) satisfy the \teff/magnitude conditions ($T<12$ and $\teff \neq$ {\tt null}) or ($T<13$ and $\teff < 5500$~K). We implement the \teff/magnitude criteria to reduce the CTL to a manageable size, emphasizing the cool bright dwarfs that are likely to be the highest priority targets. 

Next, we calculate \logg\ for all stars with parallax measurements using the stellar radius and mass estimation algorithms described in Sec.~\ref{sec:mass_radius}, and either include or exclude them from the CTL depending on whether the estimated \logg\ is $\ge$4.1 or $<$4.1, respectively \citep{Paegert:2015}.
We do not include stars in the CTL if we are unable to determine their \teff\ spectroscopically or from dereddened colors (see Sec.~\ref{sec:dereddening}), or if we are unable to estimate their radius (Sec.~\ref{sec:mass_radius}) or the flux contamination from nearby stars (Sec.~\ref{sec:flux_contam}) since these are essential to setting target priorities (see Sec.~\ref{subsec:priority}). 
Finally, all stars in the specially curated Cool Dwarf, Hot Subdwarf, Bright Star, Known Exoplanet Host, and Guest Investigator target lists (Appendix~\ref{sec:spec_cat}) are included in the CTL, version~7 at present, comprises 3.8~million stars. {\it We emphasize that at the present time---and until the {\it Gaia\/} DR2 parallaxes are incorporated---the CTL includes both dwarfs and subgiants because of the RPM method is unable to reliably screen out subgiants (see Section~\ref{sec:giant_removal})}. The CTL of course also includes other non-dwarf stars inherited from the above-mentioned specially curated lists.  

Strictly speaking, the CTL as delivered to NASA is simply a list of candidate target stars with associated relative targeting priorities. We are providing an enhanced version of the CTL with all relevant observed and derived stellar quantities described here, through the Filtergraph Portal system as a tool for the community to interact with this unique data set. Appendix~\ref{sec:appendix_filtergraph} describes each quantity in the CTL that can be found on the Filtergraph Portal system.

\subsection{Reduced Proper Motion\label{sec:giant_removal}}

A key step for TESS targeting is the elimination of likely red giant stars, whose very large radii would make detection of transits by Earth-size planets very difficult.
Thus we begin with the $J$-band RPM diagnostic (RPM$_J \equiv J + 5 \log\mu$, where $\mu$ is the total proper motion in arcsec~yr$^{-1}$), adopting a slightly modified boundary from that proposed by \citet{Collier-Cameron:2007} to differentiate among dwarfs ($ \logg \ge 4.1$), subgiants ($3.0\le \logg < 3.5$ and $\teff \ge 5000$; or $3.5 \le \logg < 4.1$), and giants ($3.0\le \logg < 3.5$ and $\teff < 5000$; or $\logg \le 3.0$). 
We define the boundary by the following polynomial relation:
$${\rm RPM}_{J,cut} \equiv -58 + 313.42 (J-H) - 583.6 (J-H)^2 + 473.18 (J-H)^3 - 141.25 (J-H)^4$$
for $J-H \le 1$, and ${\rm RPM}_{J,cut} \equiv 3.75$ for $J-H > 1$,
such that stars at smaller (more negative) RPM$_J$ values than the relation are considered to be giants, and those at larger (more positive) RPM$_J$ values are considered to be dwarfs/subgiants. 

We tested the efficacy of the RPM$_J$ method using stars with valid proper motions and with spectroscopically measured \logg\ in the TIC (Sec.~\ref{sec:teff}). We checked the degree to which the RPM$_J$ cut correctly classified the stars according to their spectroscopic \logg. The results are shown as inset in Fig.~\ref{fig:giant_removal} for $\sim$160,000 stars with proper motions from {\it Gaia\/} DR1. 
A detailed breakdown of this comparison as a function of \teff\ is provided in Table~\ref{tab:rpm_comp}.
Overall, we find that 2\% of dwarfs are misidentified by the RPM$_J$ method as giants, 4\% of giants are misidentified as dwarfs, and 88\% of subgiants are misidentified as dwarfs; thus the contamination of apparent giants by actual dwarfs is 1\%, the contamination of apparent dwarfs by actual giants is 3\%, and the contamination of apparent dwarfs by actual subgiants is 53\%. 
These results are consistent with the expectation that the RPM method is robust at separating out giants but is essentially unable to distinguish subgiants from dwarfs. 

\begin{center}
\begin{longtable}[c]{|c|c|c|}
 \hline
\teff Range [K] & Subgiants/Giants Contamination in RPM$_{J}$ Dwarfs & Dwarf Contamination in RPM$_{J}$ Giants\\
 \hline
    3840--5000 &      0.34514 &      0.03020\\
    5000--6000 &      0.38192 &      0.00004\\
    6000--7000 &      0.43435 &      0.00028\\
    7000--8000 &      0.46203 &      0.00046\\
    8000--9000 &      0.44736 &      0.00069\\
    9000--10000 &     0.47707 &      0.00135\\
 \hline
  \caption{Contamination in RPM$_{J}$ classifications as a function of \teff.}
 \label{tab:rpm_comp}
\end{longtable}
\vspace{-0.25in}
\end{center}

\begin{figure}[!ht]
    \centering
    \includegraphics[width=0.75\linewidth,clip,trim=0 60 0 20]{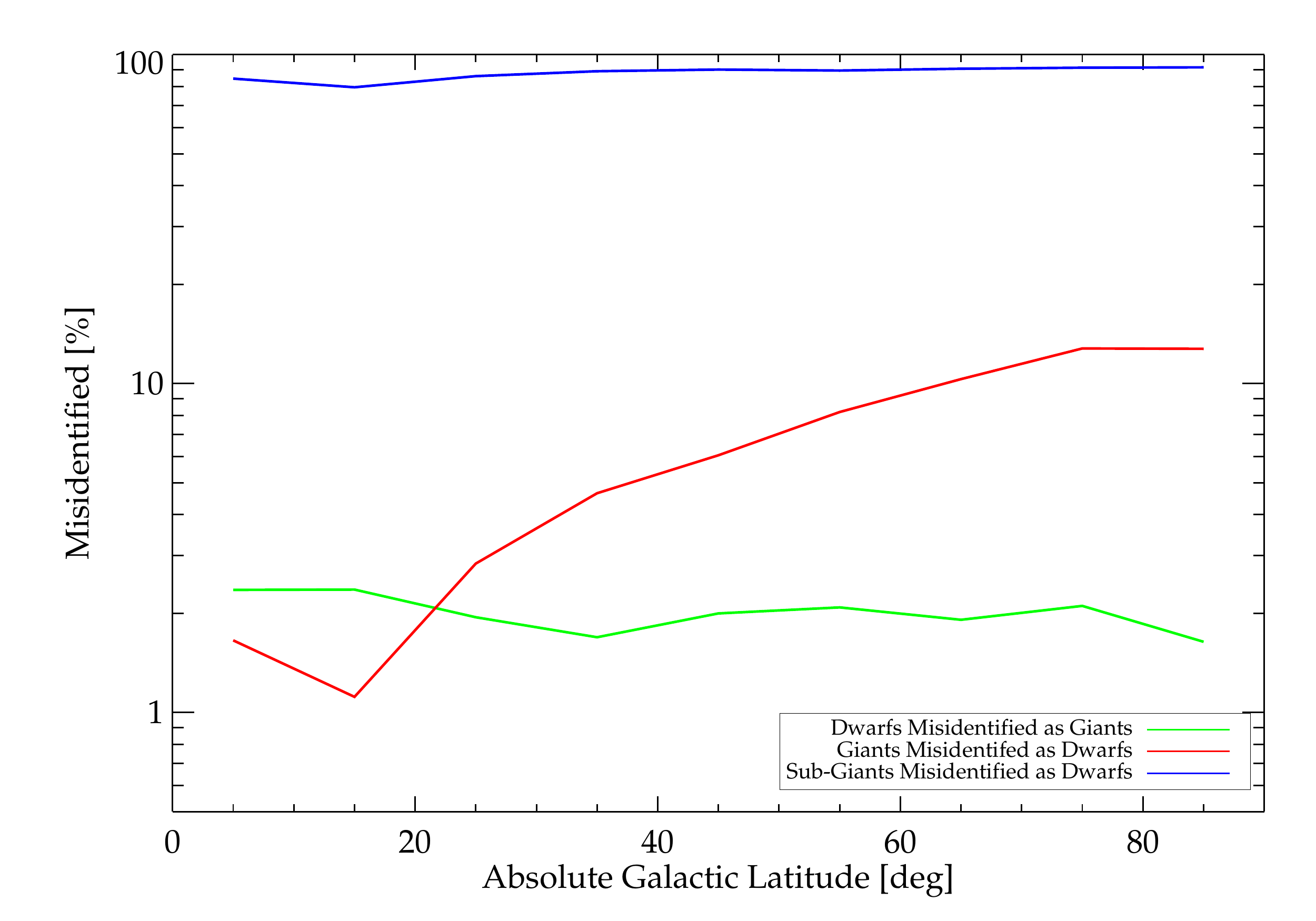}
    \includegraphics[width=0.75\linewidth,clip,trim=0 15 0 25]{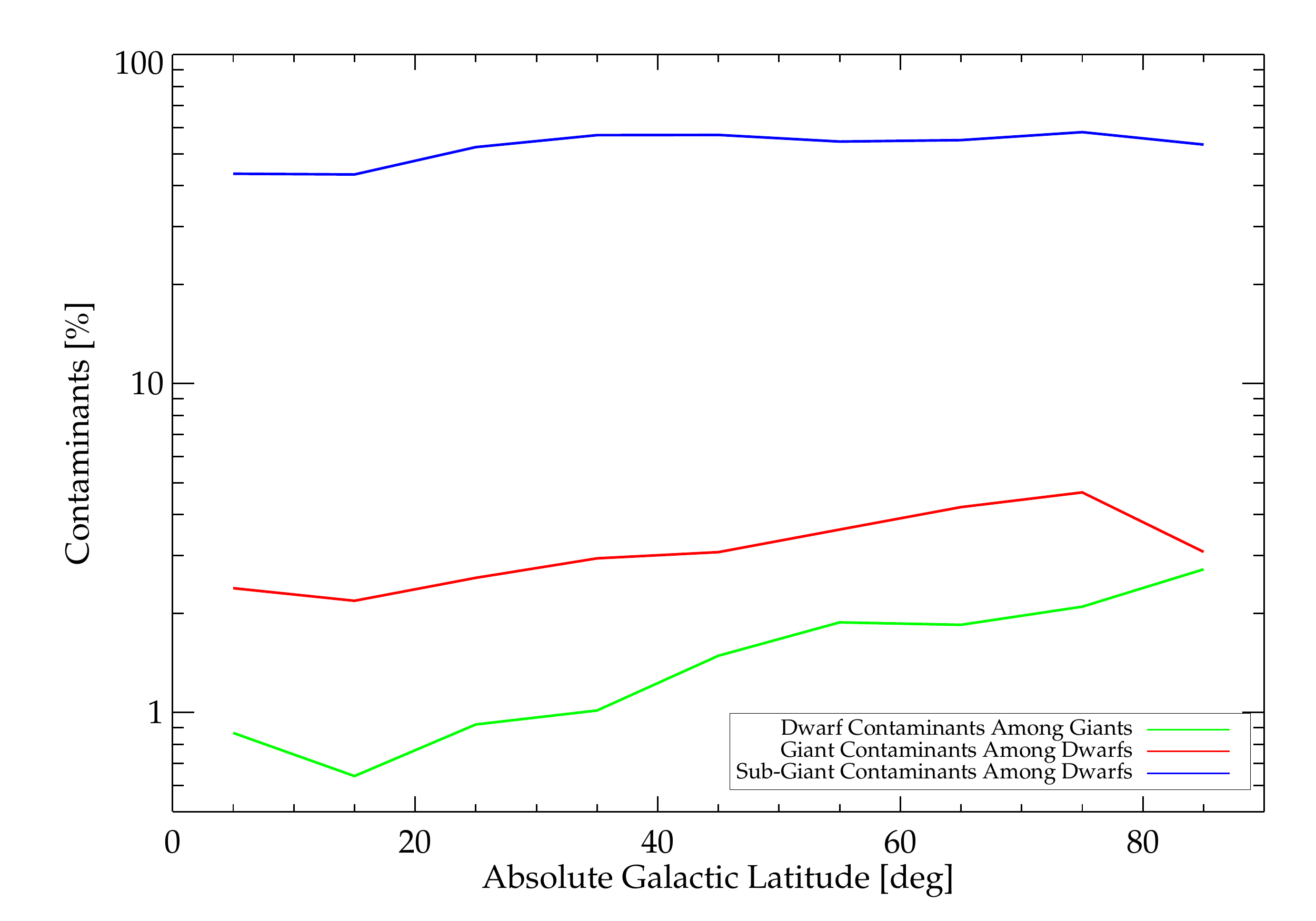}
    \caption{RPM$_J$ misidentification and contamination rates among {\it Gaia\/} DR1 stars with spectroscopic \logg. 
    {\it (Top:)} The fractions of dwarfs, subgiants, giants misidentified as other types, as a function of Galactic latitude. The apparent increase of giants misidentified as dwarfs may be due to the small number of giants toward the Galactic poles. {\it (Bottom:)} The fractions of dwarfs, subgiants, giants contaminating the others. 
    }
    \label{fig:rpmj_false}
\end{figure}

We also show these misidentification and contamination fractions as a function of Galactic latitude in Figure~\ref{fig:rpmj_false}. There is a mild trend (note the vertical axes are logarithmic) for the misidentification of giants as dwarfs to be larger at high Galactic latitudes, reaching $\sim$10\% at the Galactic poles. The origin of this trend is likely due to the fact that the RPM, as defined, is equal to absolute magnitude only if all stars have the same transverse velocity, however this assumption starts to break down away from the galactic plane. Again, the fraction of giants misidentified as dwarfs {\it overall} is 4\%.

\begin{figure}[!ht]
    \centering
    \includegraphics[width=\linewidth,trim=30 20 0 5,clip]{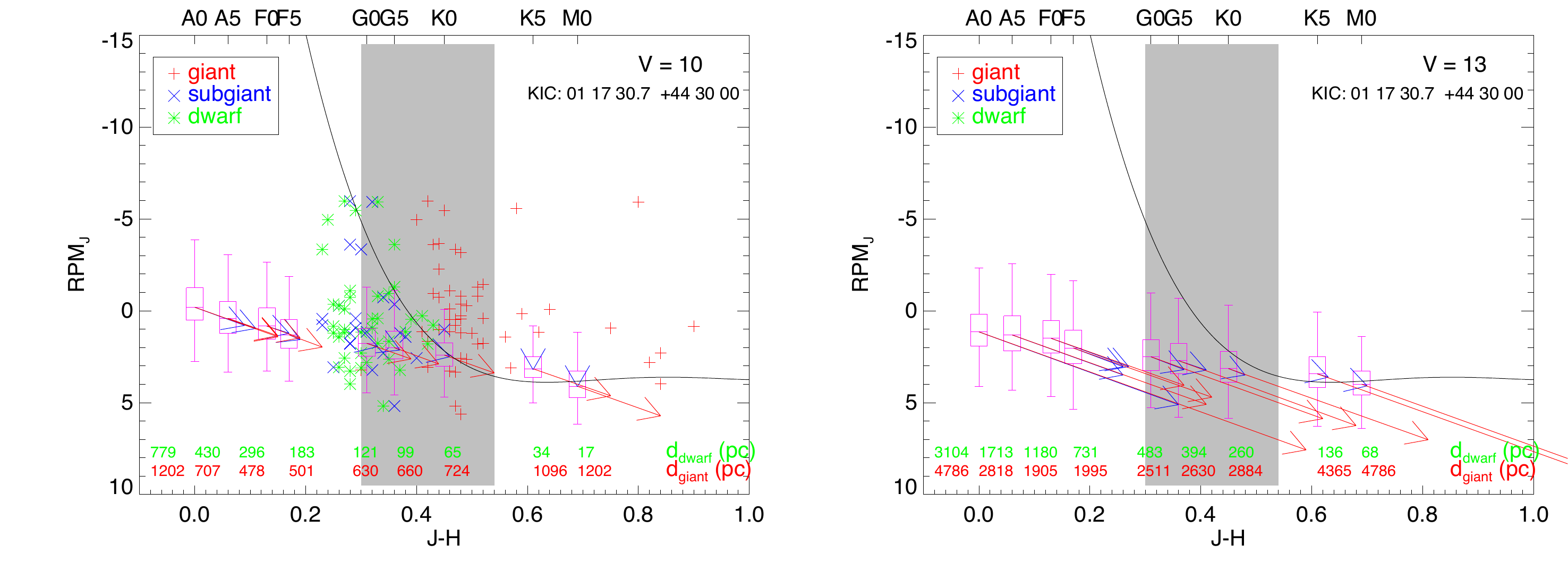}
    \caption{Effects of reddening on stars in the RPM$_J$ diagram. Reproduced from \citet{Paegert:2015}. Reddening vectors are shown (in this case for the direction of the {\it Kepler} field) for representative stars. Reddening can shift some hot dwarfs into the red giant region (thus making them false-positive giants) but it is very unlikely for a red giant to be shifted into the dwarf region.
    }
    \label{fig:rpmj_reddening}
\end{figure}

In any event, the RPM$_J$ method is evidently able to remove red giants from the candidate dwarf sample with very high fidelity, but at the same time leaves a very high contamination of the candidate dwarf sample by subgiants of approximately 50\%. Thus, subgiants remain as major contaminants among the putative dwarf sample.  \citep[See, e.g.,][for a discussion of the reasons for this in the context of the bright {\it Kepler} sample.]{Bastien:2014} In subsequent versions of the TIC we plan to incorporate the {\it Gaia\/} DR2 parallaxes so as to more completely screen out subgiants according to their radii \citep[see, e.g.,][]{Stassun:2017}, for those stars that have parallaxes available. Finally, we checked that reddening in the RPM$_J$ diagram does not adversely affect our ability to screen out likely red giants (Figure~\ref{fig:rpmj_reddening}).

\begin{figure}[!ht]
    \centering
    \includegraphics[width=\linewidth,clip,trim=80 50 85 70]{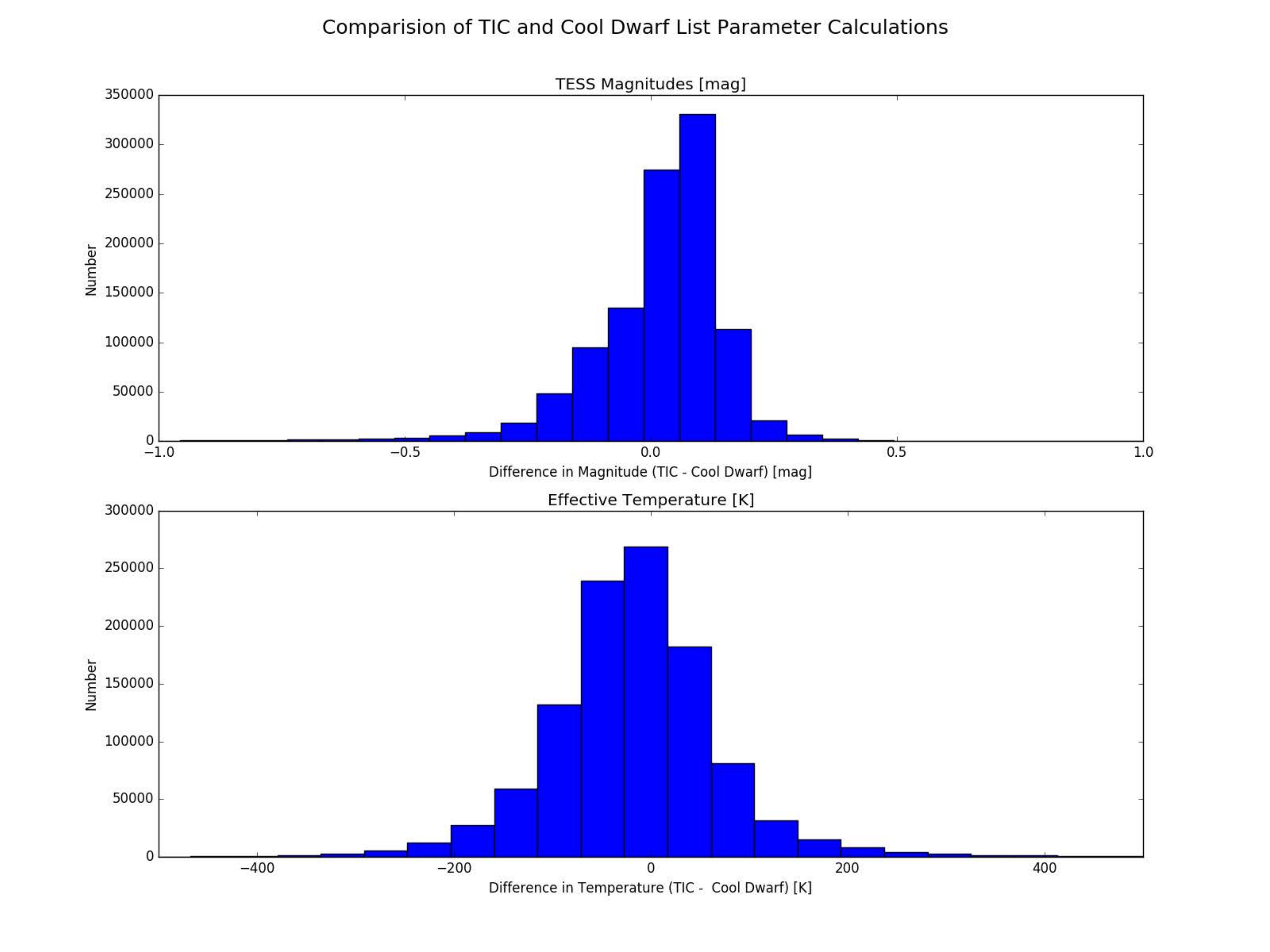}
    \caption{Comparison of calculated TESS magnitudes and \teff\ for stars in the SUPERBLINK catalog known to be cool dwarfs ($\teff < 4200$~K). We recover 99\% of the stars as dwarf stars. Their spread in the predicted TESS magnitudes is $\sim0.12$~mags, while the spread in the predicted \teff\ is $\sim63$~K.}
    \label{fig:superblink}
\end{figure}

As we did above with \teff, we verified using the specially curated cool-dwarf list that our RPM$_J$ method is reliable for the high-value M dwarfs. The TIC recovers more than 99\% of these curated cool dwarfs, suggesting a high level of completeness for this subset of stars in the TIC.
We also checked our calculated \tmag\ and \teff\ values for these stars against those supplied in the cool-dwarf list (Figure~\ref{fig:superblink}), finding an r.m.s.\ difference in \tmag\ of $\sim$0.12~mag and in \teff\ of $\sim$63~K.

\subsection{Algorithms for calculated stellar parameters\label{sec:ctl_algorithms}}

\subsubsection{Dereddening\label{sec:dereddening}}
Because we estimate stellar \teff\ principally from empirical color relations, and especially because the favored relations involve the $V-K_S$ color (Section~\ref{sec:teff}), which is highly susceptible to reddening effects, it is necessary to deredden the colors used for \teff\ estimation, as we now describe. 

\subsubsubsection{Basic approach}

\begin{figure}[!ht]
    \centering
    \includegraphics[width=0.525\linewidth]{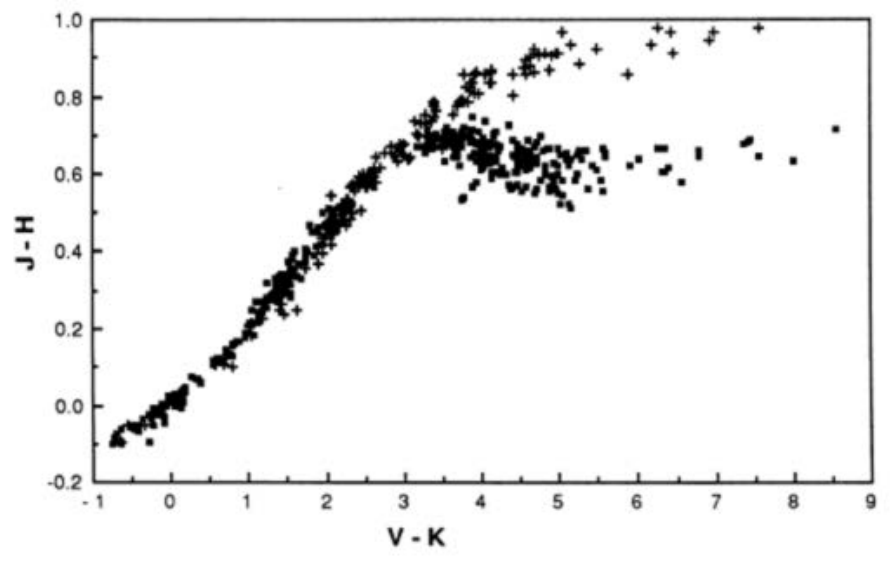}
    \caption{Reproduction of Figure 4 from \citet{Bessell:1988} showing the bifurcation of dwarfs (lower) and giants (upper) in the $J-H$ vs.\ $V-K$ color-color plane.
    }
    \label{fig:bessell}
\end{figure}

Figure \ref{fig:bessell} \citep[reproduced from][]{Bessell:1988} shows that stars in the $V-K_S$ vs.\ $J-H$ color-color plane bifurcate between dwarfs and giants at $J-H \approx 0.7$.
Any star with zero reddening should fall close to one of the two curves. Stars with reddened colors therefore will appear displaced from these curves along a reddening vector toward the upper right. Therefore, to {\it deredden\/} a star, we can move an observed star's colors backward along the reddening vector until it falls on one of the two curves.
Note that we update the relation from \citet{Bessell:1988}, which used $V$ with $JHK$ to a relation that uses $V$ with 2MASS\,$JHK_S$.

\subsubsubsection{Reddening vector}

We adopt a ratio of total-to-selective extinction of $R_V = 3.1$, and then calculate the corresponding extinctions in the near-IR colors ($A_J$, $A_H$, $A_{K_S}$) from \citet{Cardelli:1989} to determine a unit reddening vector from the color excesses $E(V-K_S)$ and $E(J-H)$. These values allow us to define the direction of the dereddening vector, where the length of the vector is given by the usual reddening $E(B-V)$.

\subsubsubsection{Dereddening procedure \label{subsec:dered}}

\begin{figure}[!ht]
    \centering
    \includegraphics[width=0.90\linewidth,trim=50 20 50 15,clip]{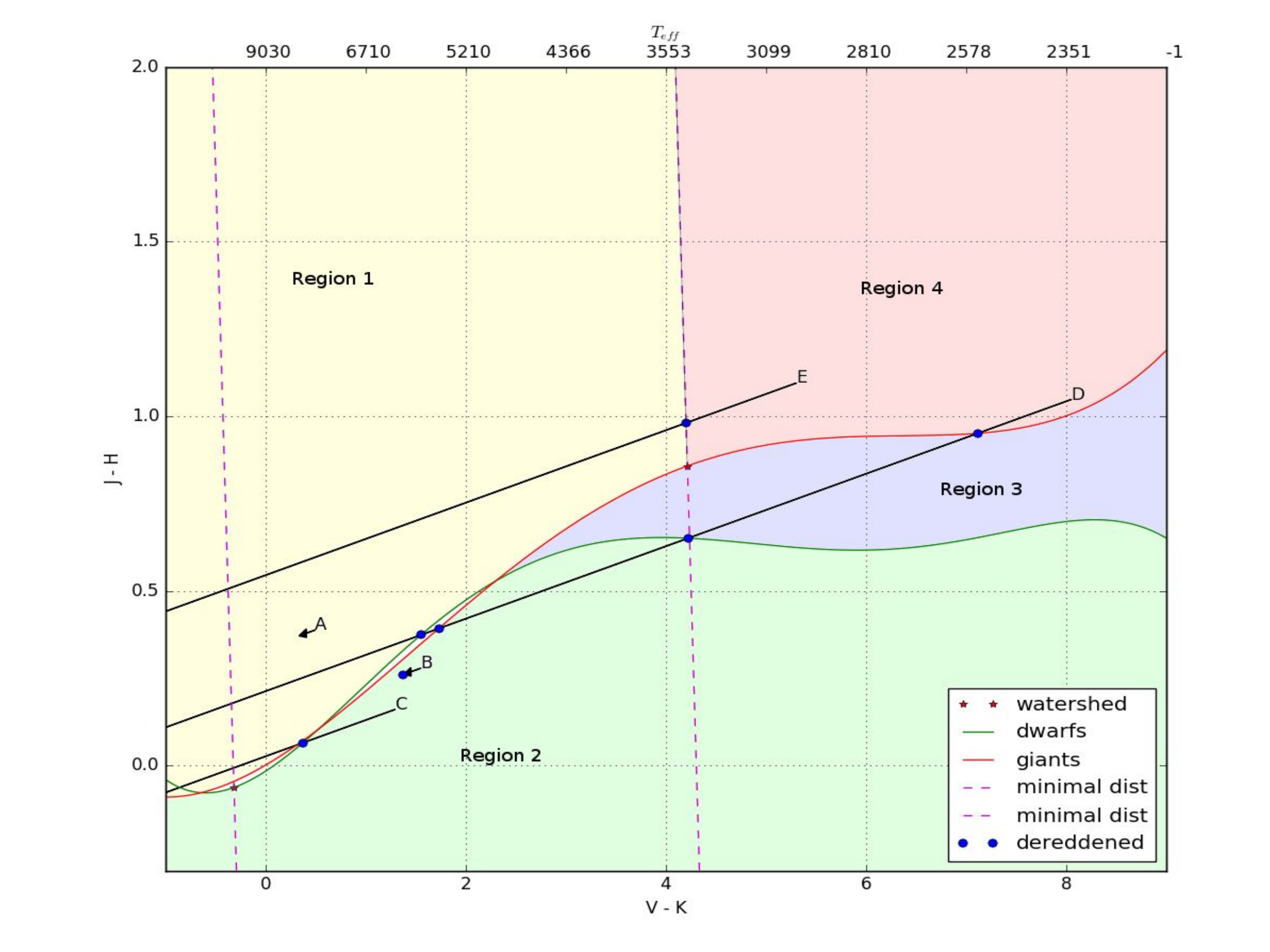}
    \caption{Illustration of dereddening for stars in different parts of the $J-H$ vs.\ $V-K_S$ color-color diagram.  
    }
    \label{fig:dereddening_procedure}
\end{figure}

First we fit the dwarf and giant sequences from \citet{Bessell:1988} with polynomial functions (Figure \ref{fig:dereddening_procedure}, green and red curves, respectively). 
For dwarf stars:
$$J-H=-0.048007+ 0.20983\:X + 0.067020\:X^2 - 0.036222\:X^3 + 0.0049886\:X^4 - 0.00021864\:X^5$$
where $X = V-K_S$.
For giant stars:
$$J-H=-0.033479 + 0.18300\:X + 0.040622\:X^2 - 0.011824\:X^3 + 0.00071399\:X^4$$
where $X = V-K_S$.

Dereddening involves shifting the colors of a given star along the dereddening vector to one of the polynomial curves, or to a certain maximum if dereddening does not intersect one of the curves. Outside the Galactic plane ($|b| > 16^\circ$) we take $E(B-V)$ from the \citet{Schlegel} dust maps as the maximum, while within the Galactic Plane we arbitrarily adopt a maximum allowed $E(B-V) = 1.5$. 

We consider stars in the four regions of the color-color plane in Fig.~\ref{fig:dereddening_procedure}. In general, we shift the star along the dereddening vector until the star either: (i) intersects the dwarf sequence (green curve), or else (ii) we take the tip or tail of the vector, whichever lies closest to one of the (dwarf or giant) sequences. Specifically, we treat their dereddening as follows, using examples of stars in each region. 
\begin{itemize}
\item Star A: No dereddening applied, as this would only move the star farther away from the dwarf sequence. These stars are flagged as `nondered' in the CTL, and the adopted $E(B-V)$ set to 0. 
\item Stars B and C: We shift the star along the dereddening vector until the star either: 
(i) intersects the dwarf sequence (green curve), or
(ii) reaches the maximum dereddening (see above), or
(iii) reaches closest approach to the dwarf sequence (magenta dashed line). 
Star B is typical for stars outside the Galactic plane that therefore have small reddening values. Star C is an example of a star in the plane for which we therefore adopt a maximum $E(B-V) = 1.5$ (see above); we adopt the point at which it intersects the dwarf sequence (marked by a dot). These stars are flagged as `dered' in the CTL, and the adopted $E(B-V)$ is recorded.
\item Stars D and E: We shift the stars along the dereddening vector until the star either:
(i) intersects either the dwarf or giant sequence (green or red curve), or
(ii) reaches the maximum possible dereddening, or
(iii) reaches the closest approach to one of the sequences (magenta dashed lines).
The vector for star D crosses the dwarf/giant sequences four times; in such cases we store all values (indicated by blue dots) but adopt the smallest reddening value. Star E is an example where the vector crosses neither the dwarf nor giant sequence; in these cases we instead adopt the value of the closest distance (magenta dashed line) between the dereddening vector and the giants sequence as indicated by the blue dot. These stars are flagged as `dered' in the CTL, and the adopted $E(B-V)$ is recorded.
\item For stars earlier than M-type ($V-K_S \leq 2.2241$) and within the typical $J-H$ error (0.05 mag) of the dwarf relation (green curve in Figure~\ref{fig:dereddening_procedure}), we assume zero reddening. These stars have been flagged with `dered0' in the CTL, and the adopted $E(B-V)$ set to 0.
\end{itemize}

Once dereddened as described here, the colors are used to determine an updated \teff\ following the procedures described in Section \ref{sec:teff}. 
In cases where the reddening cannot be determined (i.e., dereddening would move the star to a region of color space beyond the range of applicability of our color-\teff\ relations), no dereddening is attempted and the star is excluded from the CTL Effective temperatures included in a specially curated target list supersede temperatures derived by this method.

\begin{figure}[!ht]
    \centering
    \includegraphics[width=\linewidth,trim=0 0 0 20,clip]{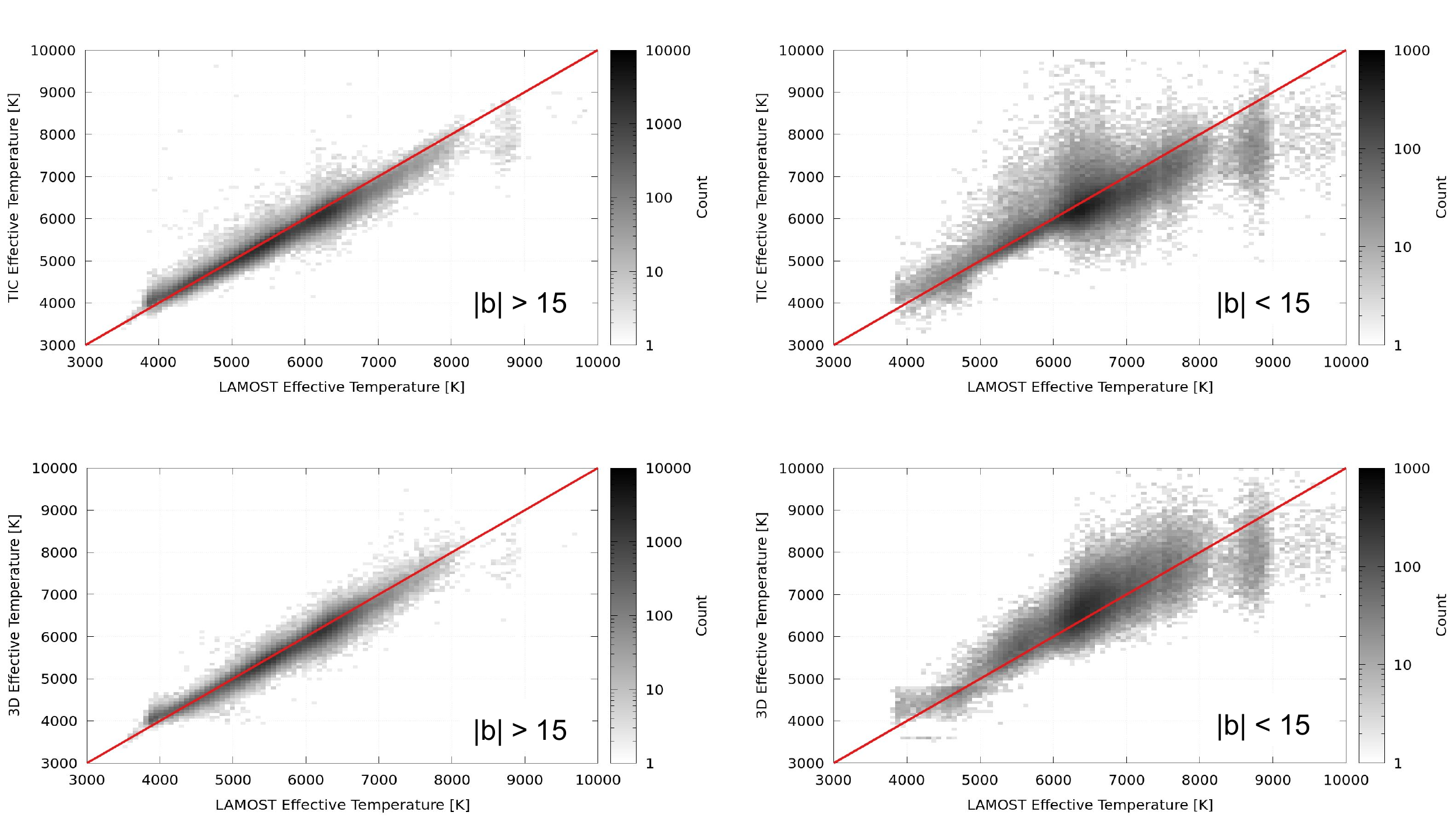}
    \caption{Comparison between our photometrically dereddened \teff\ and LAMOST spectroscopic \teff\. {\it Top:} The dereddened \teff\ calculated as described in Section~\ref{subsec:dered}. {\it Bottom:} The dereddened \teff\ calculated using the 3D dust maps of \citet{Bovy:2016}. We find the two methods to be comparable and thus adopt the method of moving stars in the $J-H$ vs.\ $V-K_S$ plane (Sec.~\ref{subsec:dered}) for computational expedience. 
    }
    \label{fig:dered_lamost_check}
\end{figure}

We have compared our photometrically estimated \teff\ values to the spectroscopically determined values from the LAMOST survey, which provides \teff\ estimates for AFGK stars ($\teff \gtrsim 3850$~K) as a general check on our \teff\ estimates and as a specific check on the dereddening procedure. We did this comparison for two different photometric dereddening techniques: (1) the scheme described above, and (2) using a 3D Galactic dust model from \citet{Bovy:2016} through which we iteratively estimate the distance on the assumption that the star is a main-sequence star. The latter procedure was examined because the \citet{Schlegel} maps do not provide reliable maximum line-of-sight extinctions within 15$^\circ$ of the Galactic plane, and also do not contain information about the relative amounts of extinction along the line of sight (i.e., they are not 3D). Figure~\ref{fig:dered_lamost_check} shows the results of these comparisons for 
the LAMOST spectroscopic 
sample.  

We find that the method of moving stars in the $J-H$ vs $V-K_S$ plane shows results similar to the 3D dereddening approach outside of the plane (differences relative to LAMOST of $-64\pm115$~K, compared to $-104\pm119$~K) and inside the plane ($178\pm370$~K compared to $-157\pm337$~K). This difference is not large, and because the computational effort required for the 3D approach is enormous for the many millions of stars in the CTL, we adopt the method of dereddening the stars in the $J-H$ vs.\ $V-K_S$ plane.

\begin{figure}[!ht]
    \centering
    \includegraphics[width=0.8\linewidth,trim=0 10 0 105,clip]{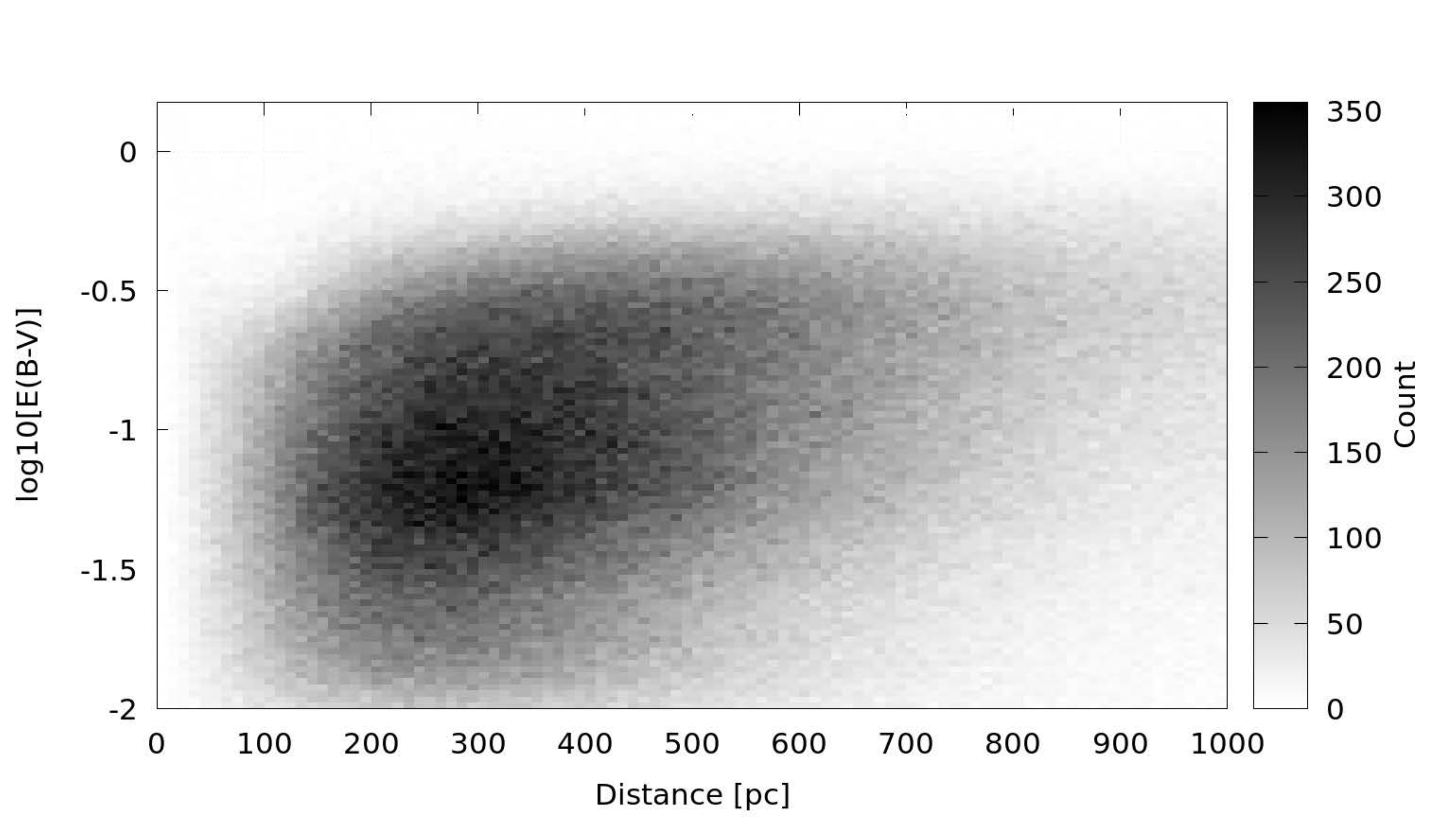}
    \caption{Comparison of the extinction values calculated for stars in \textit{Gaia} DR-1 as a function of distance in pc.
    }
    \label{fig:reddening_distance}
\end{figure}

Finally, to provide a sense for the typical range of extinctions for the distances probed by the CTL, Figure~\ref{fig:reddening_distance} shows our derived extinctions versus distance from {\it Gaia\/} for representative CTL stars.

\subsubsection{Stellar Mass and Radius\label{sec:mass_radius}}

In order to prioritize the stars based on the ability of TESS to observe transits by Earth-size planets (see Section \ref{subsec:priority}), it is essential to estimate the stellar radii and masses. When available, we simply adopt the radii and masses from the specially curated catalogs (Appendix~\ref{sec:spec_cat}),
such as those provided in the cool-dwarf list. These values are always accepted as the best representation of the true parameters and are given preference over any other values in the TIC. 
When such specially curated information is not available we use the procedures described below to calculate each parameter, in order of preference. 
Note that this approach does introduce heterogeneity in the stellar mass and radius scales, especially as a function of \teff, however we have opted for this approach in order to utilize the best available information wherever possible.

\subsubsubsection{Stars with parallax\label{subsec:rad_plx}}

When a parallax is available, we calculate the radius from the Stefan-Boltzmann equation, as follows. We first calculate the $V$-band bolometric correction, BC$_V$, using a polynomial formulation by \citet{Flower:1996}, which is purely empirical and has been found to work reasonably well for solar-type and hotter stars. 
(The choice to use the $V$ band was driven by the availability of bolometric corrections over the widest possible range of \teff.)
The coefficients for three different \teff\ ranges are shown in Table~\ref{tab:flower} \citep[see][]{Torres:2010}, where 
${\rm BC}_V = a + b\;\log \teff + c\;(\log \teff)^2 + d\;(\log \teff)^3 + e\;(\log \teff)^4 + f\;(\log \teff)^5$.
We add the following offsets to the above polynomial relations to make the three \teff\ ranges meet smoothly: 
For $\log \teff < 3.7$, add $-0.022$ mag to the polynomial result; 
for $3.7 \le \log \teff < 3.9$, no offset; and 
for $\log \teff \ge 3.9$, add $-0.003$ mag to the polynomial result. 

\begin{center}
\begin{longtable}[c]{|c|c|c|c|}
 \hline
 & Cool Regime & Middle Regime & Hot Regime \\
 & $\log \teff$ $< 3.7$ & $3.7 \le \log \teff < 3.9$ & $\log \teff \ge 3.9$\\
 \hline
$a$ & $-$0.190537291496456d+05   &  $-$0.370510203809015d+05   &  $-$0.118115450538963d+06 \\
$b$ &  +0.155144866764412d+05  &   +0.385672629965804d+05  &   +0.137145973583929d+06 \\
$c$ & $-$0.421278819301717d+04  &  $-$0.150651486316025d+05   &  $-$0.636233812100225d+05 \\
$d$ & +0.381476328422343d+03  &   +0.261724637119416d+04  &   +0.147412923562646d+05 \\
$e$ &   \nodata     &   $-$0.170623810323864d+03  &   $-$0.170587278406872d+04 \\
$f$ &  \nodata      &   \nodata   &   +0.788731721804990d+02 \\
 \hline
  \caption{BC$_V$ Relation Coefficients Adopted from \citet{Flower:1996}.}
 \label{tab:flower}
\end{longtable}
\vspace{-0.3in}
\end{center}


With BC$_V$ in hand, we next calculate the bolometric luminosity, \lbol, as follows:
\begin{enumerate}
\item Correct the apparent $V$ magnitude for extinction ($A_V$): $V_0 = V - A_V$;
\item Calculate absolute $V$-band magnitude with: $ M_V = V_0 - 10 + 5 \log\pi$, where $\pi$ is the parallax in milli-arcseconds; 
\item Compute the absolute bolometric magnitude with $M_{bol} = M_V + {\rm BC}_V$;
\item Compute the bolometric luminosity in solar units
with $\log L/L_{\odot} = -0.4 (M_{\rm bol} - M_{\rm bol, \odot})$, where $M_{\rm bol, \odot} \equiv 4.782$.\footnote{This is the value appropriate for the scale of the \citet{Flower:1996} bolometric corrections, such that the measured apparent visual magnitude of the Sun is reproduced exactly \citep{Torres:2010}. Note that the apparent visual magnitude of the Sun implicitly adopted here ($V_{\odot} = -26.71$) in order to derive $M_{\rm bol,\odot} = 4.782$ is the currently accepted value, and is different from the value of $V_{\odot} = -26.76$ adopted by \citet{Torres:2010}.  Note also that the value of $M_{\rm bol, \odot}$ above is not the same as the one recently defined by the IAU ($M_{\rm bol, \odot}$ = 4.75; 2016 Resolution B2), because the \citet{Flower:1996} scale is not the same as the scale recently defined by the IAU. For TESS we must use $M_{\rm bol, \odot} = 4.782$, or there would be a systematic error when adopting the bolometric corrections from \citet{Flower:1996}.}
\end{enumerate}
\noindent{The final formula for \lbol\ is therefore:}
$\log L/L_{\odot} = -0.4 (V - A_V - 10 + 5 \log\pi + {\rm BC}_V - 4.782)$. 

Next, we calculate the radius from \teff\ and the Stefan-Boltzmann law using ${\teff}_{,\odot} \equiv 5772$~K, as recommended by the IAU (2015 Resolution B3).
The above methodology is valid for $\teff \ge 4100$~K. The stellar radii for cooler stars will be obtained using other methods, as described in \S~\ref{subsubsubsec:spec_rad}.

To compute the error in the radius we propagate the observational uncertainties for all of the above quantities. For the {\it Gaia\/} DR1 parallax the we have adopted as the total error a quadrature sum of the nominal uncertainty and 0.3~mas systematic error \citep{Gaia:2016}, to be conservative. For BC$_V$ we have chosen to add 0.10 mag in quadrature to the BC$_V$ uncertainty that comes from the \teff\ error via the \citet{Flower:1996} polynomials. The error in $M_{\rm bol,\odot}$ is taken to be the estimated error in $V_{\odot}$, which is 0.02\;mag.

Using the radius and stellar mass calculated as discussed below,  we derive \logg\ .  A \logg\ derived from a known and measured parallax supersedes a spectroscopic \logg\ and is reported in the CTL.

\subsubsubsection{Stars without parallax but with spectroscopic parameters\label{subsubsubsec:spec_rad}}

When spectroscopic \teff\ and \logg\ are available, we calculate mass and radius from empirical relations \citep{Torresetal:2010}, with a reported scatter of 6.4\% in mass and 3.2\% in radius.

\subsubsubsection{Stars with neither spectroscopic parameters nor a parallax available\label{subsubsubsec:allen_rad}}

When none of the above sources of stellar radii are available, we estimate the stellar radii as well as the masses using a single relation for each quantity based on \teff. These relations were derived from high-precision empirical measurements of masses and radii of eclipsing binaries \citep{Torresetal:2010}, which in the case of the radii show relatively little scatter for hot and cool stars, but a considerably larger scatter for intermediate-temperature stars where subgiants are more common. However, in this regime the sample of eclipsing binaries was found to be sparse, and was therefore supplemented with simulations based on the TRILEGAL code \citep{Girardi:2005} to generate a population of stars that is complete in both mass and radius at a given \teff. 
The TRILEGAL sample was created using 12 representative sight lines at a Galactic longitude of 90$^{\circ}$ and Galactic latitudes between 30$^\circ$ and 85$^{\circ}$. Each sight line covered 10~deg$^2$ and excluded binaries, which led to $\sim390,000$ objects brighter than $\tmag = 20$. We included both dwarfs and subgiants, since these are the largest sample populations in the TIC. 

\begin{figure}[!ht]
    \centering
    \includegraphics[width=0.495\linewidth,clip,trim=40 190 40 75]{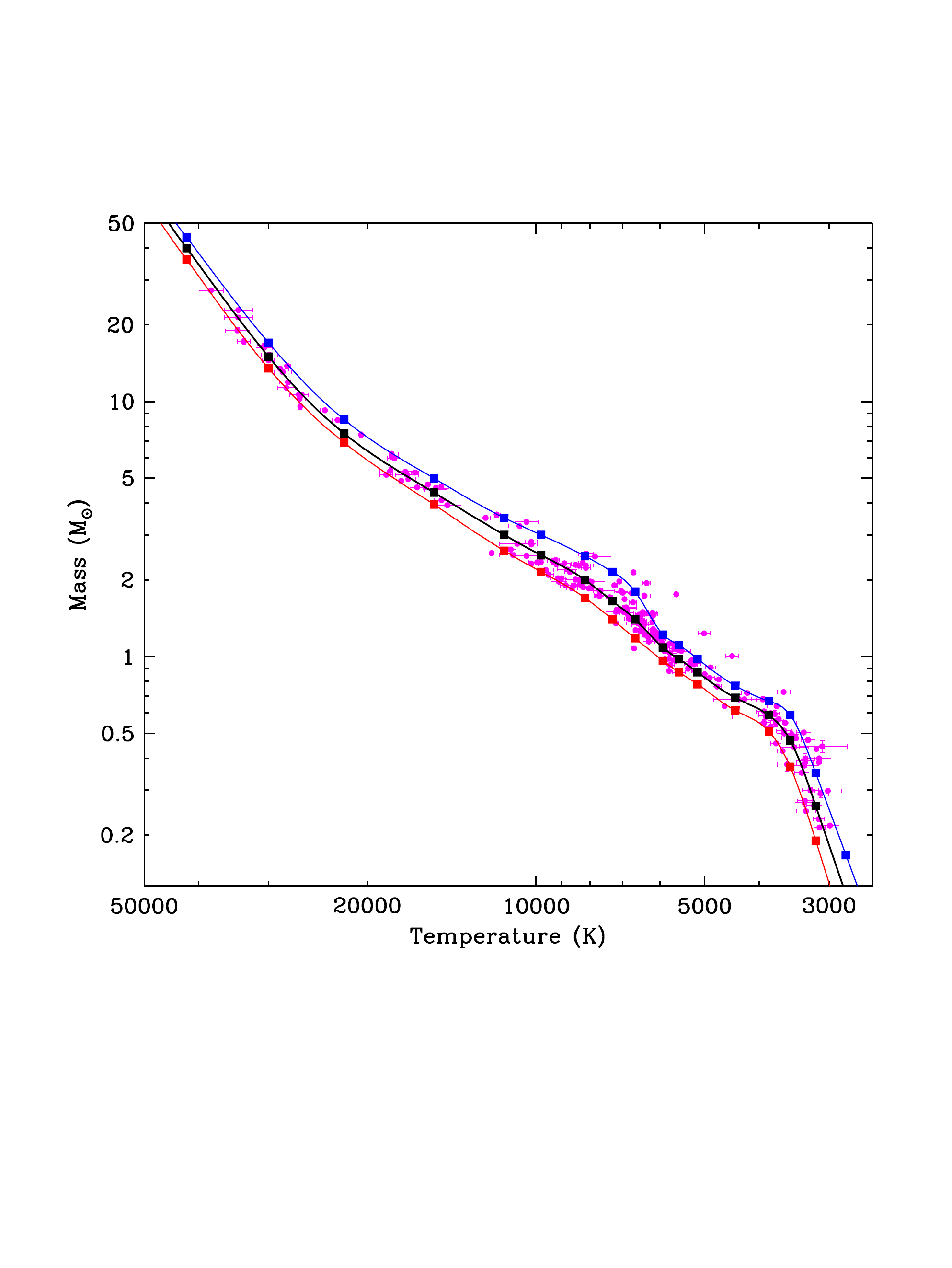}
    \includegraphics[width=0.495\linewidth,clip,trim=40 190 40 75]{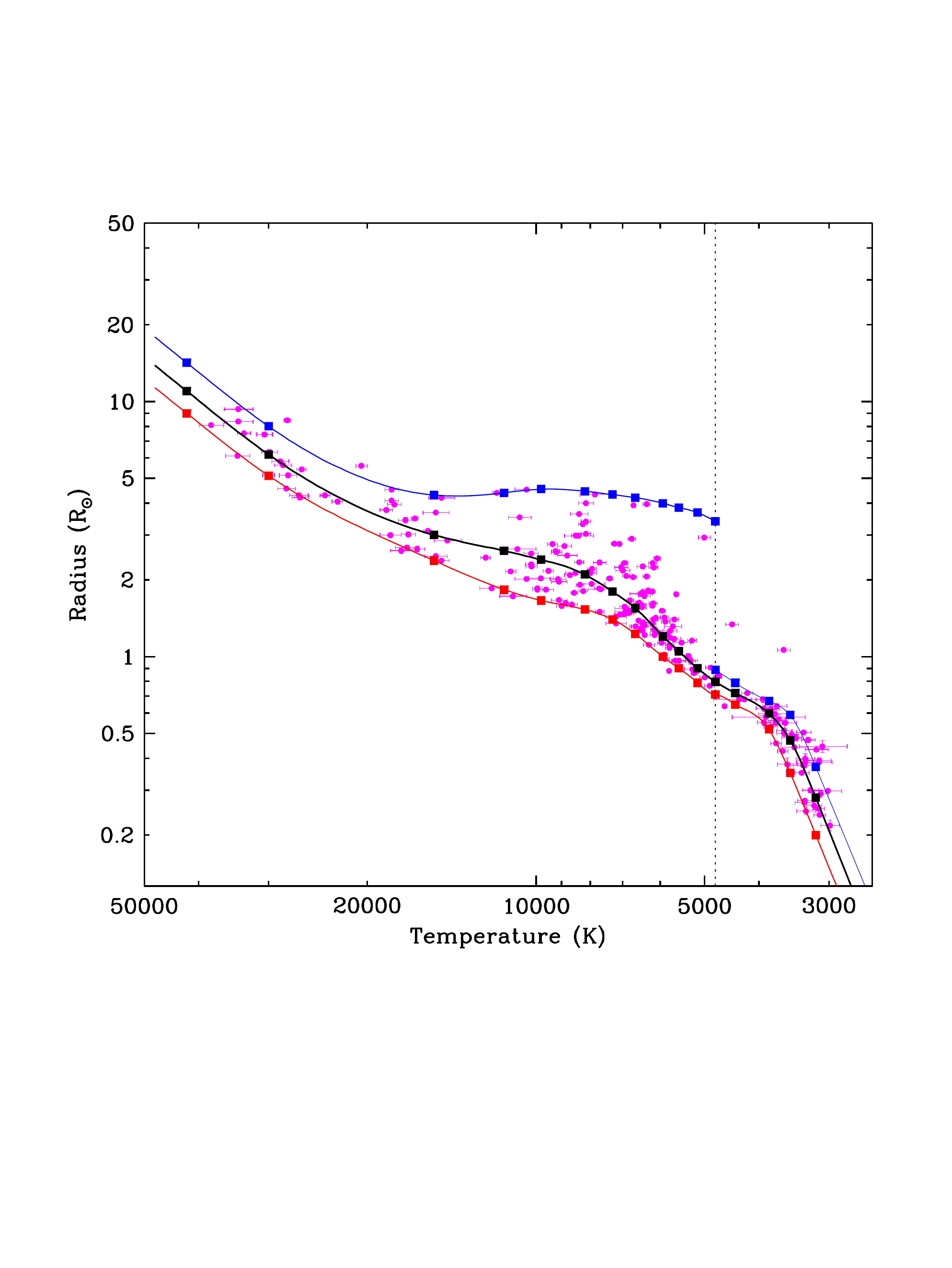}
    \caption{Derived mass-\teff\ (left) and radius-\teff\ (right) relations for the TIC based on eclipsing binary measurements from \citet[][magenta symbols]{Torresetal:2010}. The nodal points of the spline curves are given in Tables~\ref{tbl:nodes_radii} and \ref{tbl:nodes_mass}. In both figures the blue points are the upper mass/radius limits, the black points are the mean values, and the red points are the lower mass/radius limits. The vertical dotted line in the radius-\teff\ diagram marks the break in the upper envelope at $\sim$4800\;K caused by larger subgiant stars hotward of that limit. 
    }
    \label{fig:ref-mass}
\end{figure}

We drew spline curves through the middle of the distribution of points in the mass-temperature and radius-temperature diagrams, and also along the upper and lower envelopes (considering both the eclipsing binaries and the simulated stars from TRILEGAL) so as to provide a means to quantify the mass and radius uncertainties from their spread. The nodal points of these spline functions are provided in Tables~\ref{tbl:nodes_radii}--\ref{tbl:nodes_mass}, and the final relations are shown in Figure~\ref{fig:ref-mass}. 

In the range of \teff\ occupied by significant numbers of subgiants, the asymmetric spread toward larger radii is very large, but then decreases sharply below about 4800~K (because for the age of the Milky Way, all stars at such cool \teff\ are either main-sequence dwarfs of very low mass or else evolved red giants of higher mass). It was found convenient to split the upper envelope of the radius-temperature relation into two regimes at 4800~K, which introduces a discontinuity that matches what is seen in Figure~\ref{fig:ref-mass}. 

As discussed in Section~\ref{sec:giant_removal}, we find that our RPM$_J$-based procedure for removal of red giants does not effectively remove subgiants, and therefore the largest radius errors are for G-type stars, since there are large numbers of G subgiants. On the other hand, since there are no cool subgiants, typical radius errors for cool stars are small, reflecting only the small spread on the main sequence.  Radius errors are also small for hotter stars (A and F types), as mentioned before, because massive subgiants evolve extremely quickly through the Hertzsprung gap and there are few blue giants in the local neighborhood.
As currently defined, the TIC (and therefore the CTL) permits only a single value for radius error or mass error; we are currently unable to supply asymmetric errors. Therefore, for the radius error we adopt the average of the upper and lower asymmetric errors from Table~\ref{tbl:nodes_radii}, capped at 100\% of the radius, and for the mass error we report the average of the upper and lower errors from Table~\ref{tbl:nodes_mass}, which is always smaller than 100\% of the mass.

\begin{footnotesize} 
\begin{center}
\begin{longtable}[c]{|c|c|c|c|c|}
 \hline
 Type & \teff[K] & Mean Radius [\rsun] & Lower Radius Limit[\rsun] & Upper Radius Limit[\rsun]  \\
 \hline
O5 & 42000  & 11   & 9.0  &  14.2\\
B0 & 30000  & 6.2  & 5.12 &  8.0\\
B5 & 15200  & 3    & 2.38 &  4.3\\
B8 & 11400  & 2.6  & 1.83 &  4.39\\
A0 &  9790  & 2.4  & 1.66 &  4.54\\
A5 &  8180  & 2.1  & 1.53 &  4.45\\
F0 &  7300  & 1.8  & 1.40 &  4.32\\
F5 &  6650  & 1.55 & 1.23 &  4.2\\
G0 &  5940  & 1.2  & 1.00 &  4.0\\
G5 &  5560  & 1.05 & 0.90 &  3.84\\
K0 &  5150  & 0.9  & 0.79 &  3.67\\
K2 &   ..   &  ..  & ..   & 0.889/3.4\\
K5 &  4410  & 0.72 & 0.65 & 0.79\\
M0 & 3840   & 0.60 & 0.52 & 0.67\\
M2 &  3520  & 0.47 & 0.35 & 0.59 \\
M5 &  3170  & 0.28 & 0.20 & 0.37 \\
 \hline
\caption{Nodal points of the \teff-radius spline relations. Note the K2 spectral type has a discontinuity in the upper radius limit; see Fig.~\ref{fig:ref-mass}.\label{tbl:nodes_radii}}
\end{longtable}
\vspace{-0.5in}
\end{center}
\end{footnotesize}

\begin{footnotesize} 
\begin{center}
\begin{longtable}[c]{|c|c|c|c|c|c|}
 \hline
 Type & \teff[K] & Mean Mass [\msun] & Lower Mass Limit [\msun] & Upper Mass Limit [\msun]\\
 \hline
O5 & 42000 &   40.0  &  36.0  &  44.0\\ 
B0 & 30000 &   15.0  &  13.5  &  17.0 \\
.. & 22000 &    7.5  &   6.9  &   8.5 \\
B5 & 15200 &    4.4  &   3.95 &   5.0 \\
B8 & 11400 &    3.0  &   2.6  &   3.5 \\
A0 &  9790 &    2.5  &   2.15 &   3.0 \\
A5 &  8180 &    2.0  &   1.7  &   2.48\\
F0 &  7300 &    1.65 &   1.40 &   2.15\\
F5 &  6650 &    1.4  &   1.18 &   1.80\\
G0 &  5940 &    1.085 &   0.965 &  1.22\\
G5 &  5560 &    0.98 &   0.87  &  1.11\\
K0 &  5150 &    0.87 &   0.78  &  0.98\\
K5 &  4410 &    0.69 &   0.615 &  0.77\\
M0 &  3840 &    0.59 &   0.51  &  0.67\\
M2 &  3520 &    0.47 &   0.37  &  0.59\\
M5 &  3170 &    0.26 &   0.19  &  0.35\\
\hline
 \caption{Nodal points of the  \teff-mass spline relations. \label{tbl:nodes_mass}}
\end{longtable}
\vspace{-0.5in}
\end{center}
\end{footnotesize}

\subsubsection{Flux contamination\label{sec:flux_contam}}

If a TESS target is blended with a foreground or background star, the flux from the nearby source will fall into the TESS aperture of the target star, decreasing the ability to detect transits of the target. We use the catalogs described in Section~\ref{sec:overview} to identify all flux-contributing sources near each TESS target.

We identify all potential point-source contaminants for each TESS target, down to the limiting magnitudes of APASS and 2MASS($\tmag \sim 17$--19), and we calculate the fraction of contaminating flux in the aperture for the TESS target. That calculation relies on three quantities: (1) the distance out to which a star might possibly contaminate a target, (2) the shape and size of the TESS point-spread-function (PSF), and (3) the size of the TESS aperture.  

In order to calculate expected contamination ratios for TESS targets before the launch of the mission in a computationally practical way, we made a set of assumptions for each of the quantities involved.  For the maximum angular distance out to which a source  might contribute contaminating flux, we adopted a distance of 10 TESS pixels.  Although especially bright stars at larger distances will have wings of the PSF extending to distances significantly larger than that, their density on the sky is small.  

For the shape and size of the TESS PSF, we used a preliminary empirical PSF determined by the TESS mission.  Although the small focal ratio of the TESS optics means that there will be significant non-circularity of the PSF and focal plane distortions in the TESS images, the exact location of the targets on the TESS cameras was not known at the time of this writing.  Therefore, we selected the empirical TESS PSF determined for the center of the TESS field of view, which generally represents the most compact PSF, and therefore represents a lower limit of the rate of flux contamination for a given star.
We have considered but ultimately declined to add the calculation of the flux contamination for other choices of assumed PSF size, for two reasons. First, the flux contamination calculation is by far the most computationally expensive operation in the creation of the catalog and it is not feasible to do this calculation for multiple choices of assumed PSF size. Second, the precise PSF properties are in fact not yet sufficiently well determined from in-flight tests to warrant a more detailed treatment at this time.

\begin{figure}[!ht]
    \centering
    \includegraphics[width=0.6\linewidth]{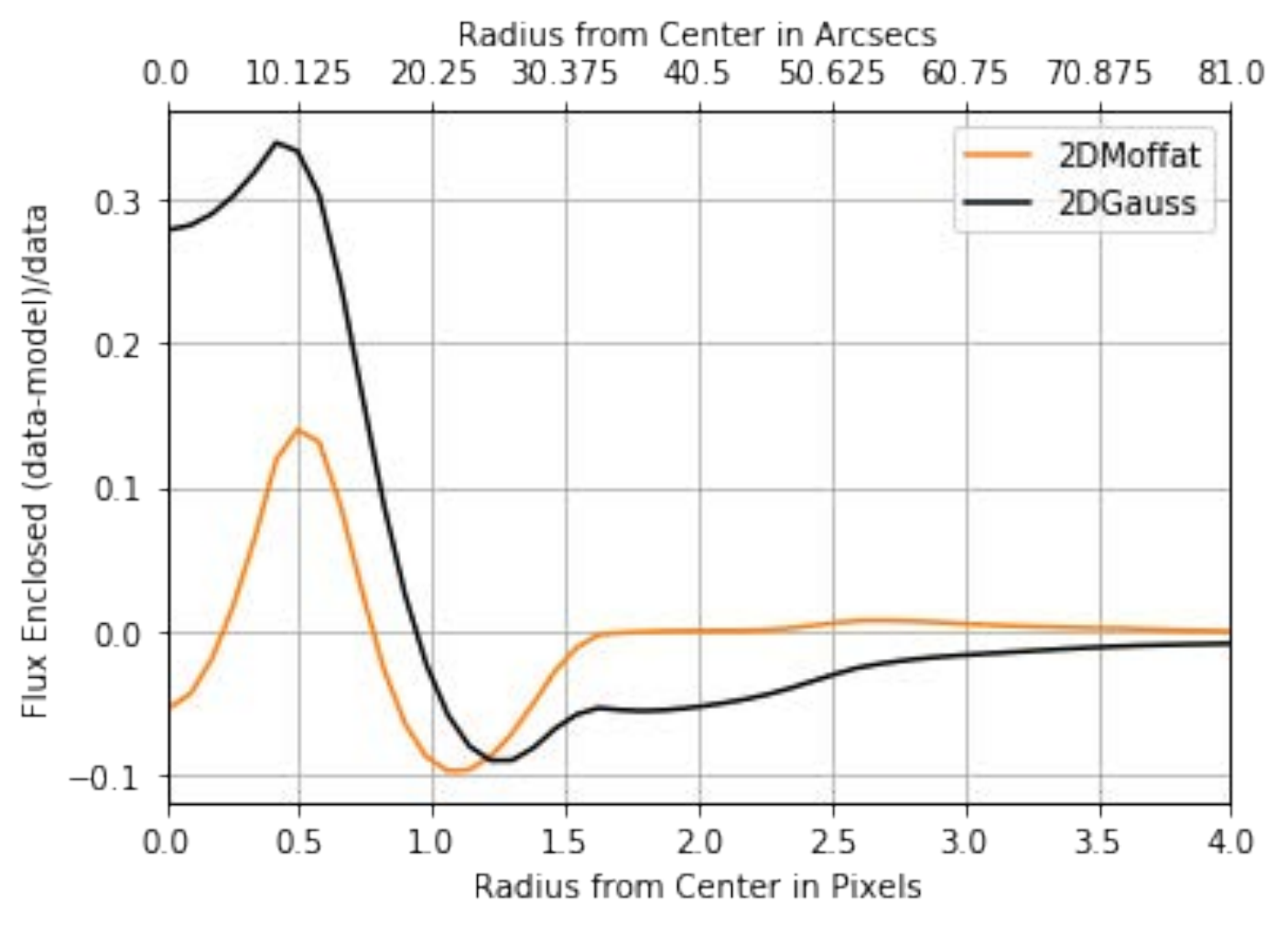}
    \caption{The difference in integrated fluxes between the preliminary TESS PSF and the 2D Gaussian and 2D Moffat models. The function used to calculate the PSF is provided as a tool by the TESS Guest Investigator program.
    }
    \label{fig:flux_contam_flow}
\end{figure}

We fit both a Moffat model and a 2D Gaussian model to the empirical PSF.  While there is virtually no difference between the empirical PSF and both the Moffat and Gaussian models in integrated flux, the relative errors reveal that the Gaussian model underestimates the total flux by 5\%. However, because it does this both for the target and the nearby contaminating stars, the effect partially cancels out. The main difficulty with a 2D Moffat profile is that there is no analytical solution for the integral over the function, and we would have to integrate numerically which requires considerable computing resources. Therefore we select a circular 2D Gaussian model (see Figure~\ref{fig:flux_contam_flow}). 

Just as the TESS PSF is not fully determined at this time, the size of the aperture used for a given TESS target is not known precisely.  As TESS prepares to observe each sector, the mission will define the set of pixels to be acquired for each 2-minute target.  After those pixels are downloaded, the SPOC will determine the optimal pixels to be used from the downloaded pixel set for the mission-derived light curves.  The size of the pixel set depends on the size and shape of the PSF, the brightness of the target star, the placement of the target stars in the TESS field, and the location of nearby objects.  

For the analysis here, the size of the aperture is adaptive based on the TESS magnitude of the star. We determine the radius of a circle around a given target for a given \tmag, and then derive the side-length of an enclosing square and the side length of an area preserving square. Both are averaged and this average is the size of the aperture.
We calculate a given star's PSF using the formula below, requiring the PSF to be no smaller than 1 pixel and limiting the brightest stars to have a PSF size equal to that of a $\tmag = 4$ star:
$$N_{\rm pix} = c_3\;\tmag^3+c_2\;\tmag^2+c_1\;\tmag+c_0$$
\noindent{where $c_3=-0.2592$, $c_2=7.7410$, $c_1=-77.7918$, and $c_0=274.2898$.}

\begin{figure}[!ht]
    \centering
    \includegraphics[width=\linewidth,clip,trim=0 0 0 80]{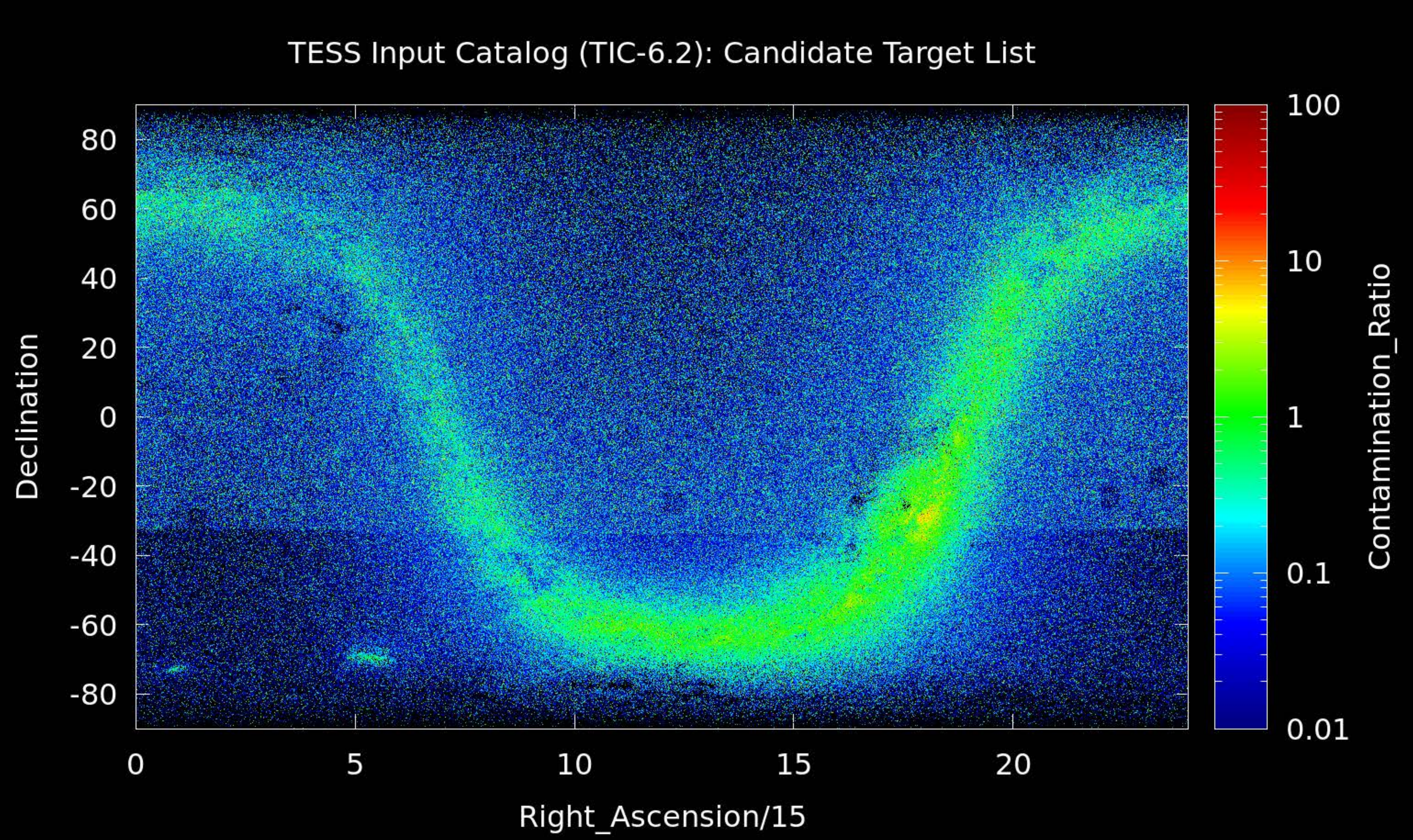}
    \caption{Estimated flux contamination ratio for stars in the CTL with contamination between 0.01 and 100. The contamination ratio clearly increases toward the Galactic plane and Magellanic Clouds. Features, such as the line at $\delta\sim35^{\circ}$ and small underpopulated squares, are adopted features from proper motion catalogs. 
    }
    \label{fig:flux_contam_galactic}
\end{figure}

We calculate the contamination ratio as the ratio of flux from nearby objects that falls in the aperture of the target star, divided by the target star flux in the aperture.  
The nominal parameters we adopt for these calculations are as follows: 
\begin{itemize}
    \item Pixel size: 20.25 arcsec
    \item Contaminant search radius: 10 pixels
    \item Standard deviation for the 2D Gaussian model: ${\rm FWHM}/(2 \sqrt{2 \ln 2})$. 
\end{itemize}
\noindent The flux contamination for $\sim3.8$~million stars in the CTL is shown in Figure \ref{fig:flux_contam_galactic}.  

\subsection{Target prioritization\label{subsec:priority}}

One of the key purposes of the CTL is to use it to prioritize targets for the TESS 2-min cadence postage stamps.  A full discussion of the considerations behind that process and the methodology used for it appears in Pepper et al. (in prep).  Here we summarize the content of that paper, and examine the implications of the target prioritization process on the distribution of stars in the CTL.

The priorities within the CTL are built to maximize the detection of small transiting planets.  The signal-to-noise ratio of a transit signal is:
\begin{equation}
\frac{S}{N} = \frac{\delta}{\sigma}\sqrt{N_{\rm data}}
\end{equation}
where $\delta$ is the fractional transit depth,
$\sigma$ is the photometric noise per data point,
and $N_{\rm data}$ is the number of data points during transit, which is $\sim R/(\pi a) N_{\rm tot}$ for a circular orbit, where $R$ is the stellar radius, $a$ is the semimajor axis, and $N_{\rm tot}$ is the total number of data points.

The per-point noise $\sigma$ depends on the flux $F$ of the star in the TESS bandpass, with the fractional uncertainty scaling as $\sigma \propto F^{-1/2}$.  In reality there will also be noise due to background or contaminating light in the target star aperture, detector noise, and the star's intrinsic variability.  The location of the target star within the focal plane of the TESS cameras will affect the TESS point-spread function (PSF) and in turn the size of the aperture and the amount of flux contamination.  For simplicity we assume that $\sigma$ is a function of $T$, the apparent magnitude in the TESS bandpass, with contributions from flux contamination, as well as detector noise.  We denote
this function as $\sigma_T$.

The number of data points in transit scales according to number of observing sectors $N_S$ in which the star is located.  We can then define the prioritization metric as a quantity proportional to the signal-to-noise ratio of a transit signal,
\begin{equation}
\frac{\sqrt{N_S}}{\sigma_T\, R^{3/2}}.
\label{eqn:pri}
\end{equation}
We use this metric to rank all the CTL stars in order of achievable $S/N$ of the transit signal.  This metric accounts for the properties of the star and the ability to detect a planet orbiting that star of a fixed size, with no prior information about the physical or orbital properties of the planet itself.  The metric requires us to know $T$ and $R$, along with the ecliptic latitude, which determines the likely value of $N_S$.

For details of the calculation of $\sigma_T$, see Pepper et al.\ (in prep).  We compute the priority metric from Equation \ref{eqn:pri} for all stars in the CTL.  We account for sector overlaps and thus $N_S$ by taking the approximate number of sector observations that a target star would have based on the target ecliptic latitude, which naturally boosts the priority of stars near the ecliptic poles. In addition, we deliberately de-prioritize objects within the Galactic plane ($|b|<15$) by a factor of 0.1 (chosen arbitrarily but with the intent of making very unlikely that such stars will be selected over other targets).  This is done because in that region the crowding of sources means we have less certainty about the cross-matching of stars between catalogs and the determination of the stellar proper motions.  Also, since we have less knowledge of the total reddening and extinction toward a given target near the plane, our determination of the object's $V$-magnitude and color is less reliable, which affects the ability to calculate the stellar temperatures and other physical properties of the targets.  We note the de-prioritization in the Galactic plane does not apply to stars in the cool dwarf list, for which we have greater confidence in the identification of the stars and their physical parameters.

The current prioritization scheme can be qualitatively summarized as prioritizing small and bright stars, with preference for high ecliptic latitude stars, and penalties for stars near the Galactic plane or those with significant flux contamination.  There are a number of reasons that a given star that might otherwise be of interest for a transit search may have a low or zero priority.  Possible reasons might be:
\begin{itemize}
    \item Stars with absolute ecliptic latitudes less than 6$^\circ$ will not be observed during the prime mission due to a gap in camera coverage between the Southern and Northern observations. Therefore, their $N_S$ values are 0 and the resulting priority is 0.
    \item Stars located within 15 degrees of the Galactic plane experience the de-prioritization effect described above.
    \item Stars near a much brighter star will have a large degree of flux contamination, decreasing the priority of the potential target star.  The relatively large size of the TESS pixels ($\sim20$ arcsec) means that bright stars several arcmin away can contribute significant flux to a given star's aperture.
    \item We attempt to exclude evolved stars from the CTL since their radii are generally both large and uncertain.  That process is described in \S~\ref{subsec:rad_plx}.
    \item While bright stars and small stars are both favored with this prioritization metric, both conditions contribute to the priority.  Therefore, some stars will not be highly prioritized with this metric no matter how bright due to large radii or our inability to reliably calculate the radii, such as hot O and B stars or giants.  Some stars will simply be too dim, such as many late type M dwarfs.
    \item Stars with \logg\ $>$ 4.8 and for which the \teff\ source is not a specially curated list, have their priority values set to 0 to avoid stars with problematic parameters (see notes on column 88 in Appendix~\ref{sec:relnotes} for more details).
\end{itemize}

Stars in the specially curated bright star list always have their priority set to 1, so as to ensure targeting of stars for which TESS will obtain the most exquisite light curves.




\section{Summary of stellar properties, known limitations, and future work\label{sec:discussion}}


\subsection{Representative properties of stars in the TIC}

Table~\ref{tbl:summary} summarizes the numbers of stars in the TIC and CTL for various representative subsets. 


\begin{center}
\vspace{-0.2in}
\begin{longtable}[c]{|c|c|c|c|}
 \hline
 Quantity & Number of Stars & Sub-population & Number of Stars \\
 \hline
 \tmag\ magnitude & 470,995,593  & $\tmag < 10$ & 966,297\\
 \teff & 331,414,942 & CTL stars with $\teff < 4500$~K & 991,868 \\
 Radius & 27,302,067  & CTL stars with $R < 0.5$\;\rsun & 787,924 \\
 Mass & 27,302,066 & CTL stars with  $M < 0.5$\;\msun & 741,483 \\
 Spectroscopic \teff\ and/or \logg\ & 572,363  & Spectroscopic \teff\ $<6000$ and $\logg > 4.1$ & 923,671\\ 
 Proper motion & 316,583,013 & Proper Motion $>1000$~mas & 655 \\
 Parallax & 2,045,947 &  &  \\
 \hline
 \caption{Summary of basic stellar properties in the TIC and CTL. 
 \label{tbl:summary}}
\end{longtable}
\vspace{-0.45in}
\end{center}


\subsection{Structure within the TIC and CTL\label{subsec:structure}}

Because of the manner in which the TIC and CTL are assembled---where identification of the most promising targets takes priority over catalog completeness or statistical uniformity---there are structures to be found within many of its observed and calculated parameters. Here we briefly identify some of the known structures for some of the most important catalog parameters, and attempt to explain their origin.

\subsubsection{Factors affecting target priorities}

We have made a set of decisions to assemble comprehensive information about potential TESS targets, and also to construct the CTL with a goal of maximizing the detection of small planets.  That process has led to a CTL that has a great deal of heterogeneity in the sources of star information, and in some places it displays discontinuities in the distribution of stellar parameters.  

\begin{figure}[!ht]
    \centering
    \includegraphics[width=0.95\linewidth,clip,trim=10 5 -10 60]{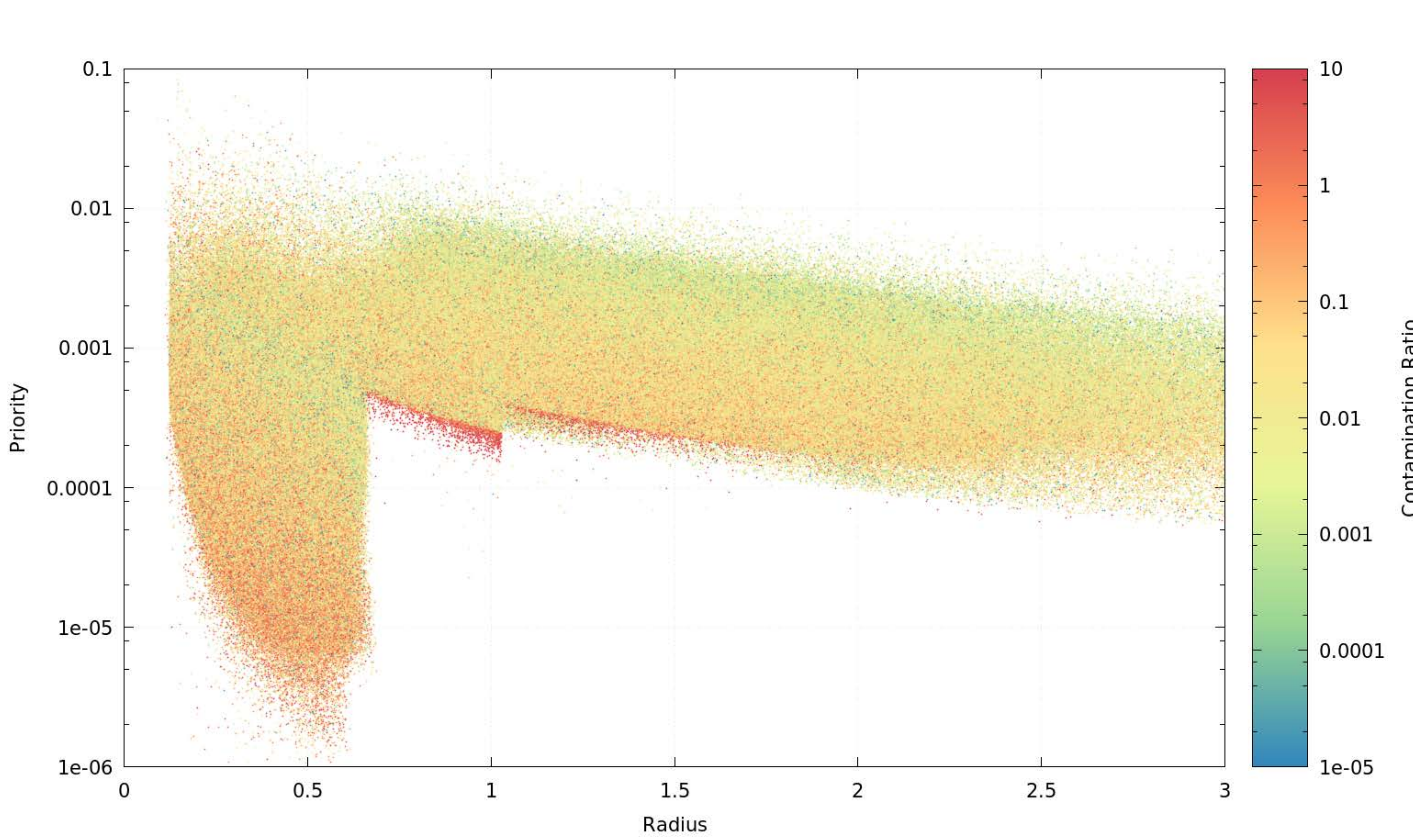}
    \includegraphics[width=0.95\linewidth,clip,trim=10 5 -10 45]{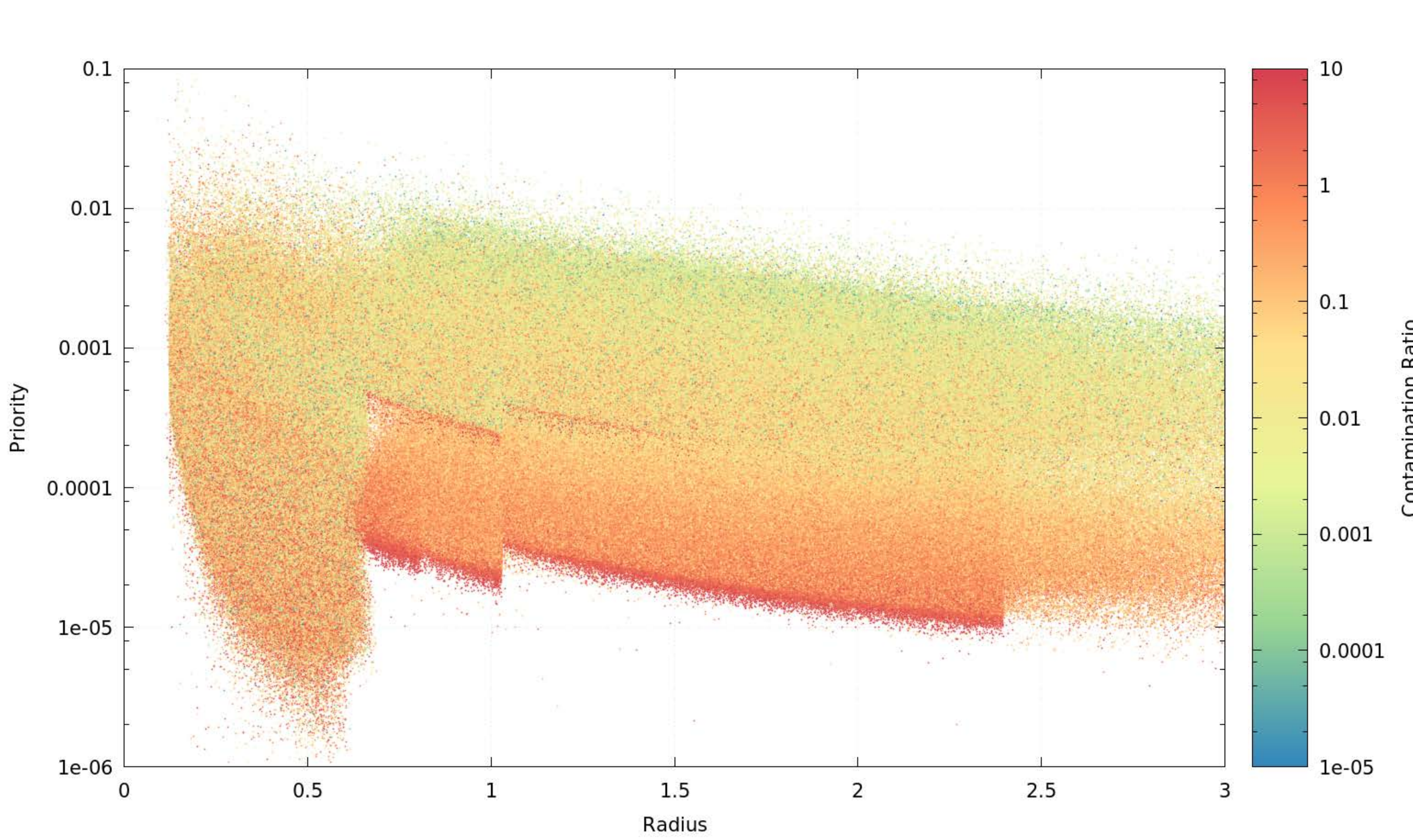}
    \caption{CTL priority as a function of stellar radius, with color indicating average flux contamination ratio. In general, the smaller the radius the higher the priority. {\it (Top:)} stars within 15 degrees of the Galactic Plane are excluded. {\it (Bottom:)} All stars included. The patterns seen in both panels are described in the text.
    \label{fig:prioritization} }
\end{figure}

A good example of such structure is shown in Figure~\ref{fig:prioritization}.  Here we plot the priority of CTL stars versus stellar radius, portrayed in a heatmap with color indicating average contamination ratio in each 2-dimensional bin.  Overall, the highest priority targets typically have small radii, and relatively low contamination ratios.  In the top panel, only stars outside the Galactic plane ($|b|>15$ degrees) are displayed, and in the bottom panel, no such cut is imposed.  Sharp vertical edges in the distributions are visible around $R = 0.65 R_{\odot}$ and $R = 1.03 R_{\odot}$.  Those features are due to the elimination of stars with \tmag\ $> 12$ and \teff\ $> 5500$~K, and \tmag\ $> 13$, except for those in special lists as described in \S~\ref{subsec:assembly}.  The bottom panel includes the stars near the Galactic plane.  Since those stars had their priorities de-boosted, they show up below the existing population, with similar patterns.

\subsubsection{The Effect of Effective Temperature\label{subsubsubsec:bestteff}}
\begin{figure}[!ht]
    \centering
    \includegraphics[width=\linewidth,clip,trim=0 10 0 100]{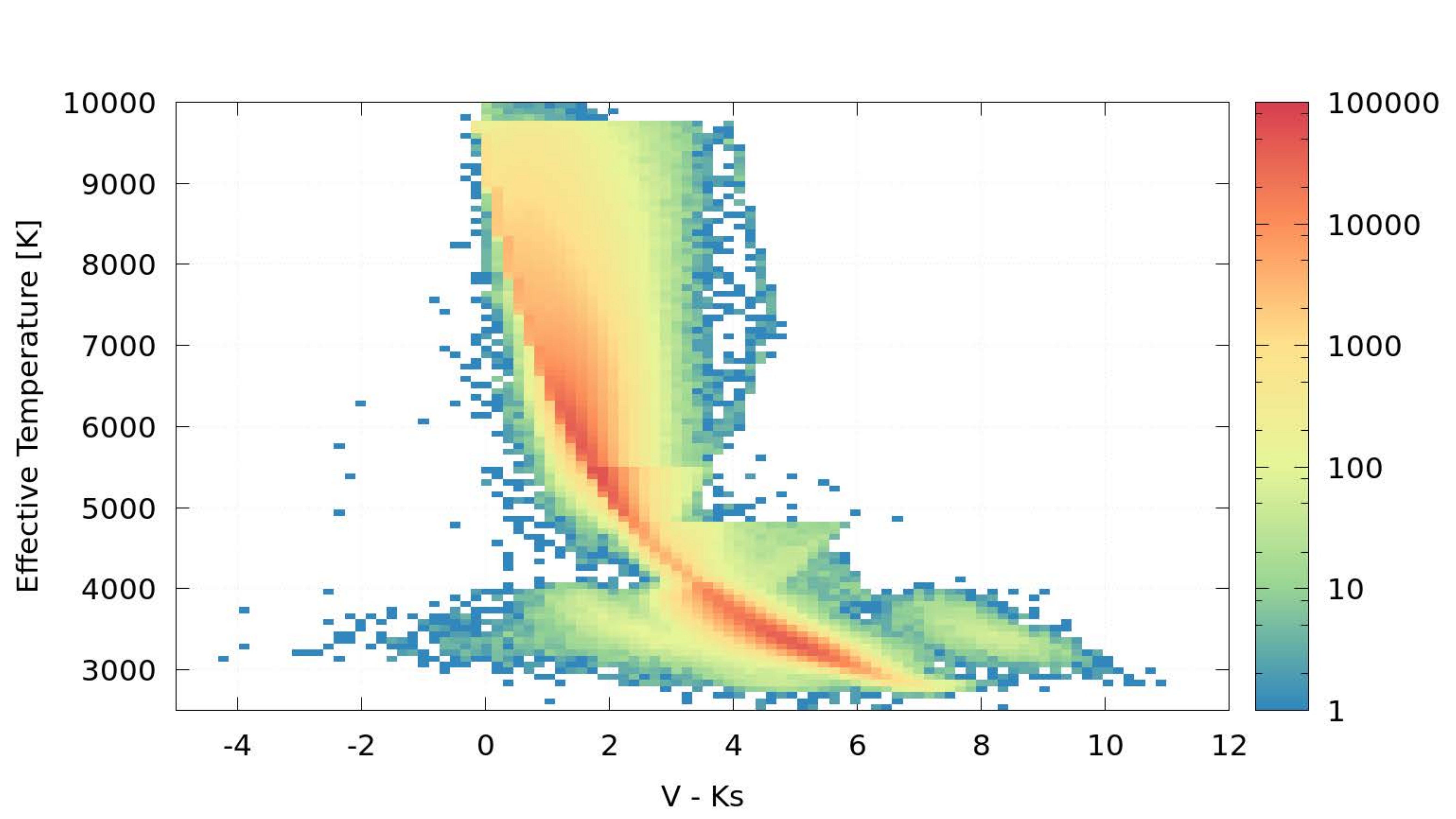}
\caption{Heatmap of \teff\ (selected as described in Sec.~\ref{subsubsubsec:bestteff}) and each star's $V-K_S$ color for stars in the CTL. The points follow a basic trend (the $V-K_S$ relation described previously) but there are noticeable features in the distribution which are described by: the specially curated cool dwarf list (stars with $\teff<3840$~K); stars below the relation (spectroscopic values); stars above the relation (de-reddened objects); stars far to the left of the relation (stars without $V$ magnitudes which defaulted to the $J-K_S$ relation); jagged edges at 5500~K and 4500~K (CTL cutoffs in at $T=12$ and $T=13$ respectively; and the jagged edge near $\sim9800$~K limit placed on the de-reddening relations.}
    \label{fig:bestteff}
\end{figure}

As described in Section~\ref{sec:teff}, stars with valid $V$ and $K_S$ magnitudes from 2MASS should have a temperature calculated from the $V-K_S$ color. While the final reported \teff\ follows the preference order of (1) specially curated lists (cool dwarf and hot subdwarf), (2) spectroscopic \teff, (3) de-reddened \teff, and (4) non-dereddened \teff, we expect \teff\ to generally follow the trend of our initial $V-K_S$ relation. Indeed, Figure~\ref{fig:bestteff} shows general agreement between the $V-K_S$ color and the ``best" selected \teff\ using the order of precedence above.  

However, as also shown in Figure~\ref{fig:bestteff}, there is a significant amount of structure present which cannot be explained by the $V-K_S$ relation. This structure is the result of the combination of numerous selection criteria explained in previous sections and the forced validity ranges of parameter calculations. Each feature is well understood, as follows:
\begin{itemize}
    \item Stars with $V-K_S<-0.1$ and $\teff>3840$~K. These stars do not have valid $V$ magnitudes or were too blue for the validity range of the $V-K_S$ relation, and were required to fall back to the $J-K_S$-to-\teff\ relation. Given that their $V-K_S$ color is so blue and yet their \teff\ is cool, this suggests either $V$ or $K_S$ are incorrect, due either to a mismatch or a problem with the original photometry. At present we do not have a mechanism to flag such cases---the quality flags from the original photometric catalogs not otherwise suggesting problems---and so we simply caution that such cases do occasionally make their way into the TIC despite our best efforts.
    \item The three populations with $\teff<3840$~K. These values come from the cool dwarf list and the variety of color relations used to determine \teff. See \citet{Muirhead:2018} for more details.
    \item Stars just to the left and below the $V-K_S$ relation. These temperatures come from spectroscopic sources and will \textit{generally} follow the relation but not exactly land on it.
    \item Stars to the right and above the $V-K_S$ relation. These temperatures come from the de-reddening routine -- a star's effective temperature increases when de-reddened. 
    \item The jagged edges at 5,500 and 4,800~K. These result from the $T<13$ and $T<12$ limits imposed on the CTL.
    \item The jagged edge at $\sim9,800$~K. This edge is the result of the limits on the de-reddening routine.
    \item Stars with $\teff>10,000$~K. These temperatures mainly come from spectroscopic values.
    
\end{itemize}

\begin{figure}[!ht]
    \centering
    \includegraphics[width=\linewidth,clip,trim=10 0 0 60]{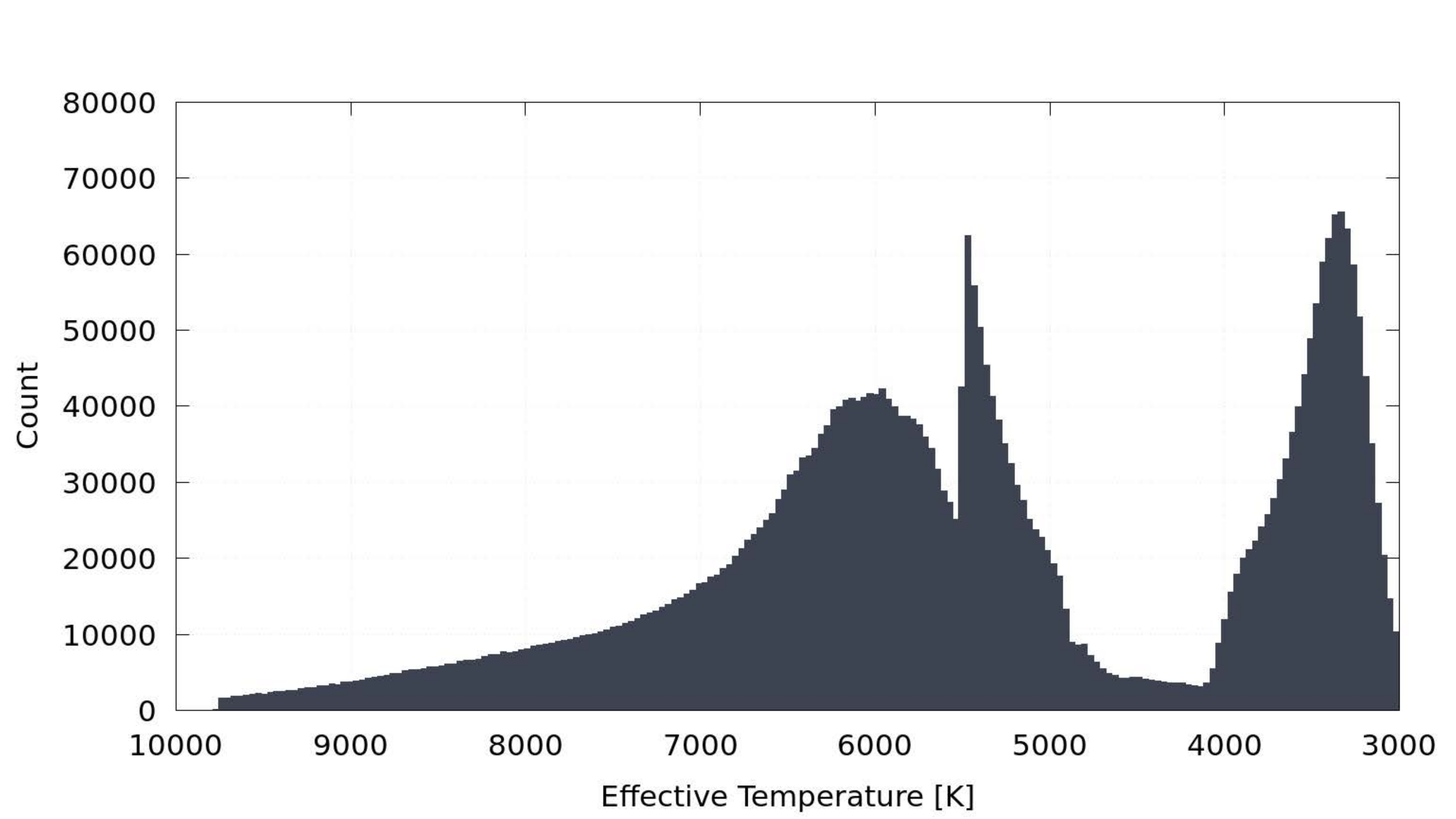}
    \caption{The distribution of the stars in the CTL according to \teff. Various peaks and valleys are shown in the distribution of values due to effects described in \S~\ref{subsubsubsec:bestteff}.}
    \label{fig:full_teff}
\end{figure}

If we make a simple histogram of the \teff~values in the CTL, we also notice a significant deviation from a smooth distribution, as shown in Figure~\ref{fig:full_teff}. There are 4 particularly noticeable features of the distribution: 1) the distribution of targets with $\teff<4000$; 2) the lack of targets with $5000<\teff<4000$~K, i.e. ``missing" K dwarfs; 3) a peak and sharp drop near $\teff=5,500$~K; and 4) a peak and smooth decline in the number of targets with $\teff>5,500$~K. 

\begin{figure}[!ht]
    \centering
    \includegraphics[width=\linewidth,clip,trim=5 100 0 50]{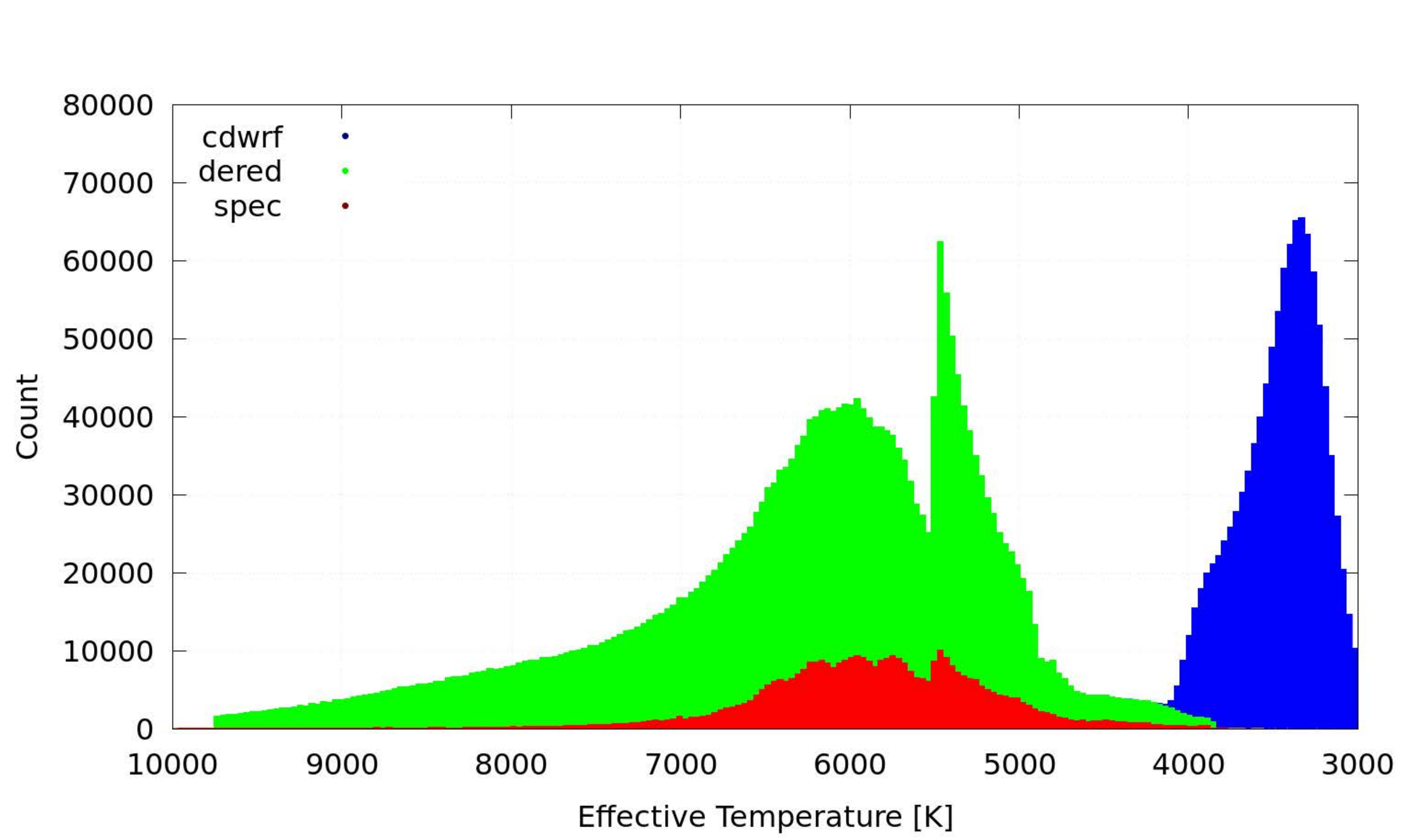}
    \includegraphics[width=\linewidth,clip,trim=5 10 0 50]{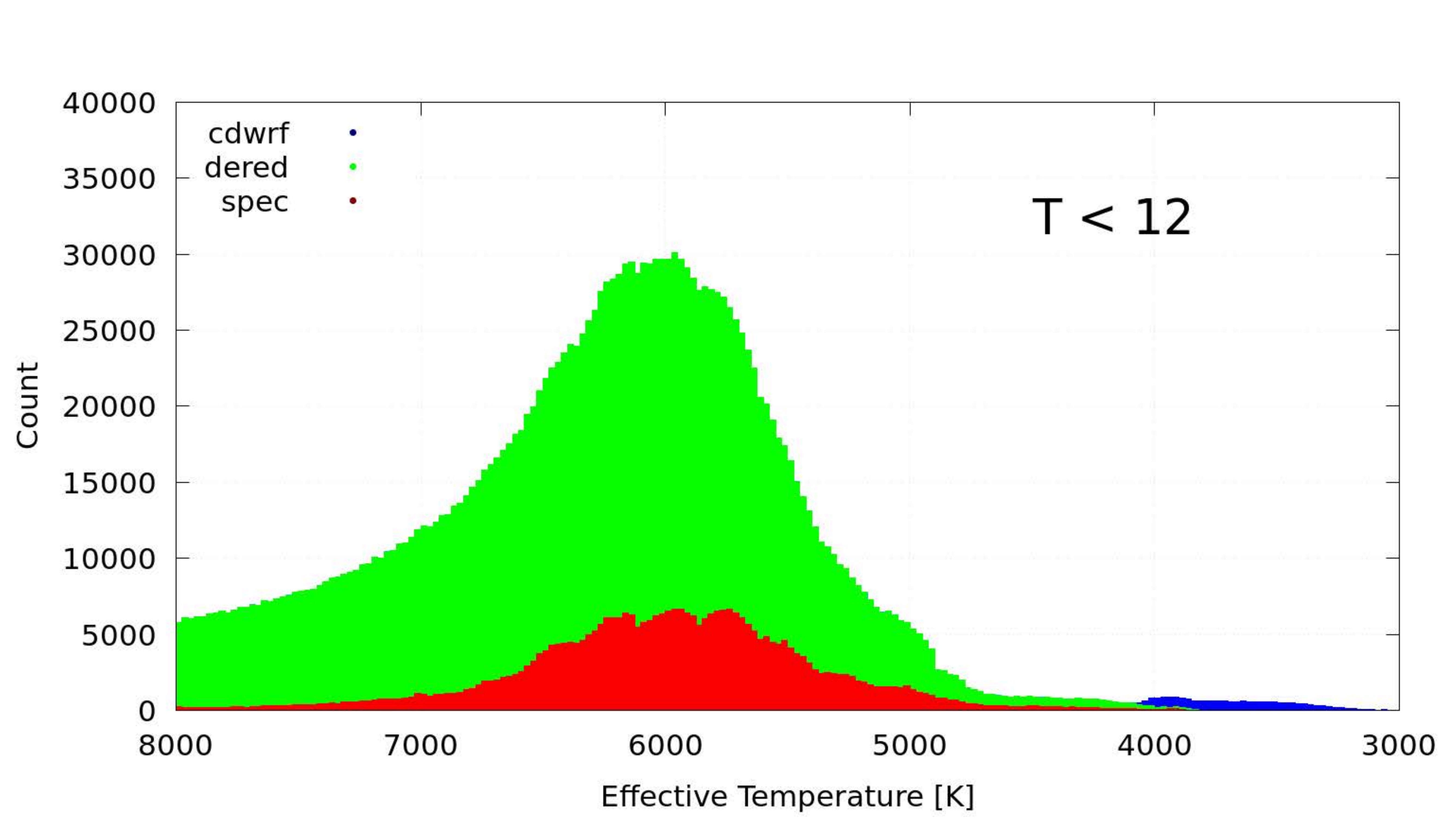}
    \caption{(\textit{Top}:) The distribution of \teff\ split into the three main sources: de-reddened photometric (green), spectroscopic (red), and specially curated cool-dwarf list (blue). The cool dwarf list clearly is the source of the large number of sources with $\teff<4000$. (\textit{Bottom}:) The distribution of \teff\ split into the three main sources, but only for stars with $T<12$. This eliminates the spiked feature at $5,500$~K which is from the magnitude and \teff\ limit placed by the selection function of the CTL.
    \label{fig:teff_src}}
\end{figure}

Features 1, 3, and 4 can be explained by the selection function of the CTL (feature 2 is discussed in Sec.~\ref{subsubsec:kdwarf}). The large distribution of targets with $\teff<4,000$~K (1) comes from the specially curated cool dwarf list, as shown in the top part of Figures~\ref{fig:teff_src}. The sharp drop off of targets near $\teff=5,500$~K is from the combination of magnitude limits of $T<12$ for all stars with $T>5,500$ and $T<13$ for $\teff<3480$~K. In fact, if we \textit{only} display stars with $T<12$, this feature disappears, as shown in the bottom part of Figure~\ref{fig:teff_src}.

\subsubsection{``Missing" K dwarfs \label{subsubsec:kdwarf}}
\begin{figure}[!ht]
    \centering
    \includegraphics[width=0.9\linewidth,clip,trim=0 90 0 50]{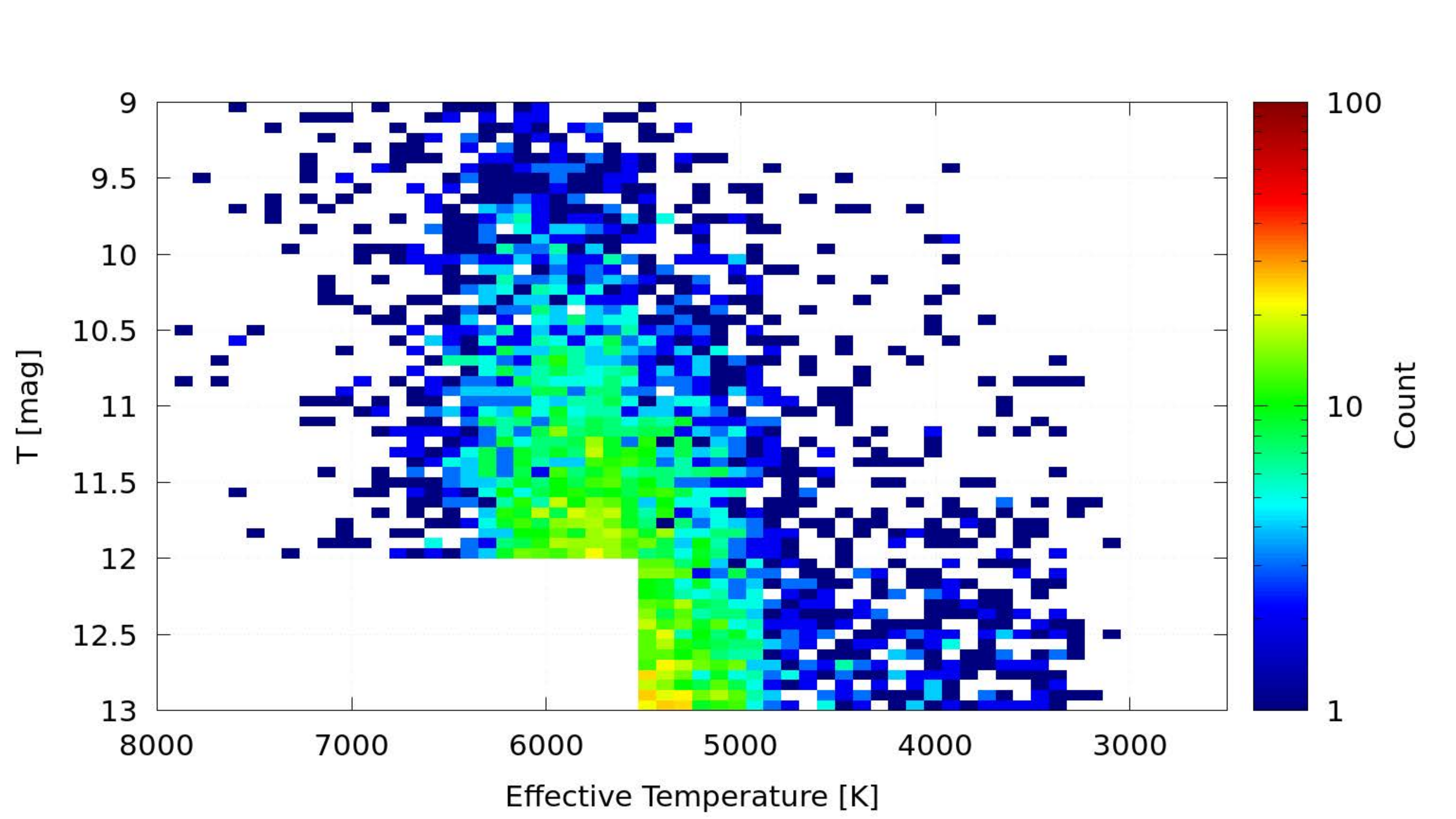}
    \includegraphics[width=0.9\linewidth,clip,trim=0 10 0 50]{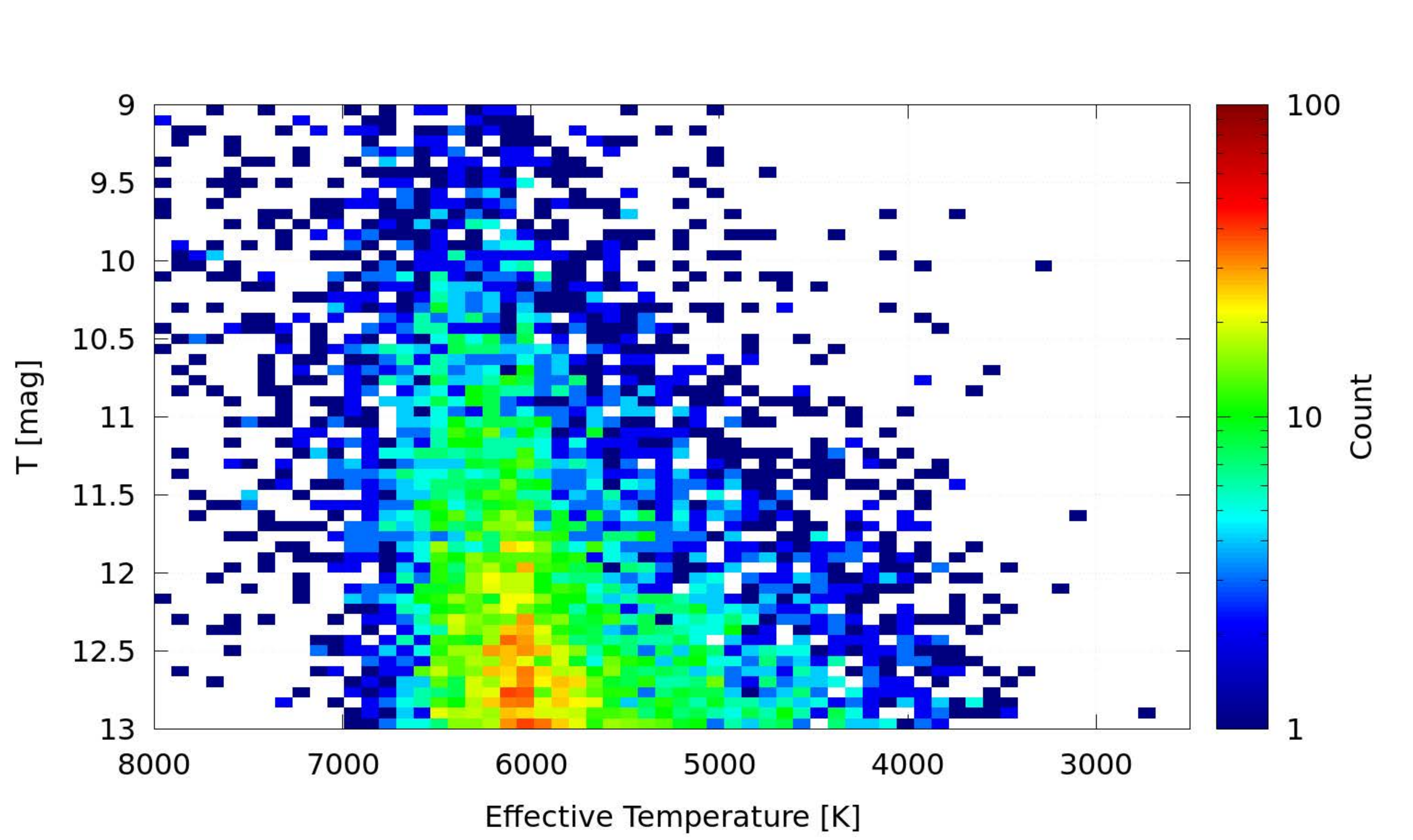}
    \caption{Heatmap of star counts in the CTL according to the \tmag\ magnitude and \teff, for real stars in the CTL (top) and for a comparable simulated population of stars from the TRILEGAL population synthesis model (bottom). The chunk of stars missing with $T>12$ and $\teff>5500$ were removed as described in Sec.~\ref{subsec:assembly}. 
    \label{fig:missing_kdwarfs_comp} }
\end{figure}

As noted above, the distribution of stellar temperatures in the CTL shows a ``gap" among the K dwarfs (Figure~\ref{fig:full_teff}). This may seem surprising, as surely K dwarfs are more abundant than G dwarfs. However, we have verified using a TRILEGAL population synthesis simulation (Figure~\ref{fig:missing_kdwarfs_comp}) that this is mostly an expected consequence of the fact that we prioritize according to a combination of stellar brightness (brighter stars receive higher priority) and stellar radius (smaller stars receive higher priority). The G dwarfs benefit from the boost according to brightness, whereas M dwarfs benefit from the boost according to size, where we have intentionally inserted a specially curated sample of (generally faint) M dwarfs into the CTL. K dwarfs suffer in priority due to being relatively faint but not as small as M dwarfs.

The TRILEGAL simulation does, nonetheless, predict somewhat more K dwarfs than we observe in the real CTL.
Another important factor may be that the RPM cuts are more prone to fail for K dwarfs (since the RPM relation ``flattens" out at colors corresponding to K stars, small errors in RPM can push a star above or below the cut). A similar effect was reported in the stellar properties for the EPIC \citep[cf.\ Sec.~5.4 in][]{Huber:2016}.

It is also especially worth noting the nature of the \teff\ distribution in Figure~\ref{fig:full_teff}  between the stars in the CVZ and elsewhere.  The area outside the CVZs includes a set of early-type stars in the F and G regime, a smaller population of K dwarfs, and a large number of cool dwarfs in the late-K and early- to mid-M ranges.  However, in the CVZs, there are simply not many late-type stars above the TESS detection limits, so the priority boost in those areas results in larger numbers of relatively fainter G stars.  The sharp cutoff at 5500~K is an artifact of the cuts applied in selecting stars for the CTL. 

Note that these features, which will complicate statistical analyses of the ensemble planet properties discovered by TESS, could in principle be ameliorated by different choices in the targeting strategy. However, as noted in the Introduction, the primary purpose of TESS as defined by the mission is to ensure discovery of a certain number of small planets; statistical purity of the sample is a secondary consideration. Thus, certain features of the target sample, such as those discussed here, are likely to remain in future updates to the TIC and CTL as well.

\subsubsection{Stellar radius and mass}
\begin{figure}[!ht]
    \centering
    \includegraphics[width=\linewidth,clip,trim=0 80 0 50]{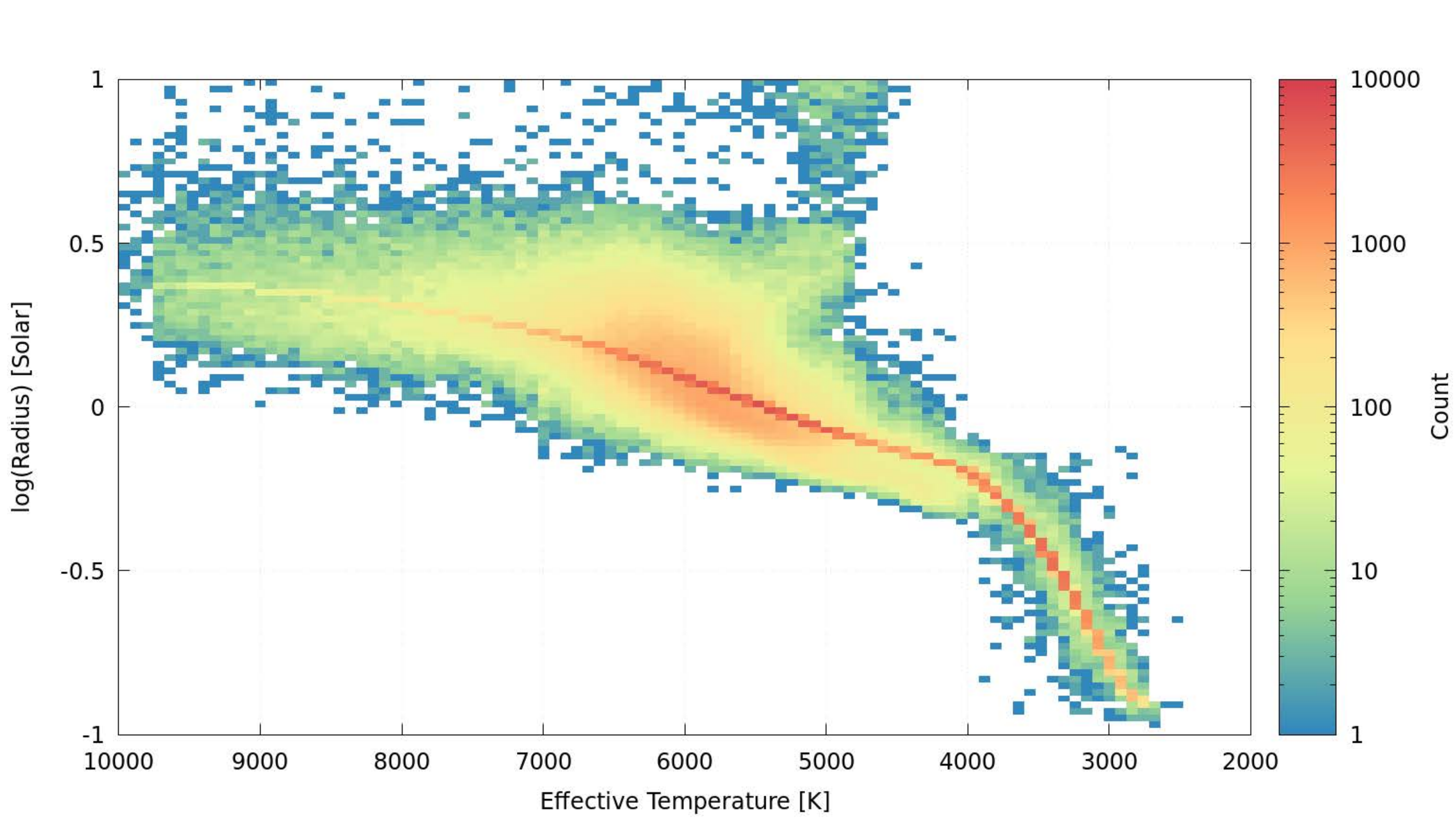}
    \includegraphics[width=\linewidth,clip,trim=0 5 0 50]{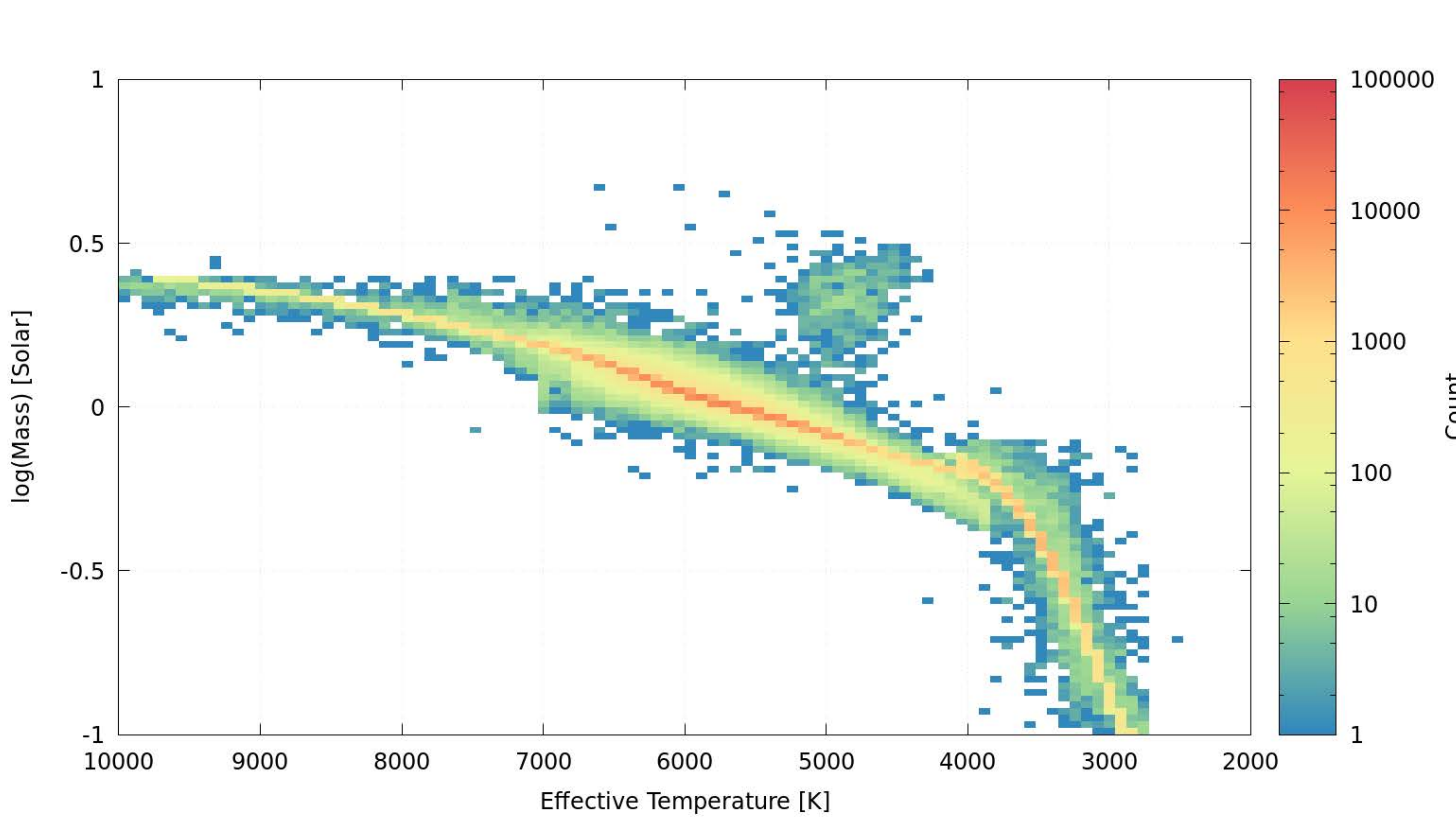}
\caption{Heatmaps of the calculated radius (\textit{top}) and mass (\textit{bottom}) for the top $\sim400,000$ stars in the CTL. The second stellar loci for values with $\teff\ < 4000$~K, comes from the cool-dwarf list. The over-density of points (stellar loci) in both figures comes from the unified spline relations. While the  The ``cloud" of points in the radius plot is due to a combination of radii from the cool-dwarf list, the use of parallaxes, and the spectroscopic relation of \cite{Torresetal:2010}. The distribution of points in the mass plot results from a combination of cool-dwarf list values and values from the spectroscopic relation.
}
    \label{fig:CTL_mass_radius}
\end{figure}

Figure~\ref{fig:CTL_mass_radius} shows the calculated mass and radius for the top $\sim400,000$ stars in the CTL. While the majority of CTL stars had their radii and masses calculated from the unified spline relations described above, there are a number of objects that had their parameters calculated from spectroscopy, from parallaxes, taken from the independent cool-dwarf list, or a combination of these methods. Therefore Fig.~\ref{fig:CTL_mass_radius} shows a large number of stars with masses and radii from the spline relation but also shows a cloud of points around the relation.
While there is a significant spread of radii and masses for a particular \teff, it is encouraging that in general the distribution of values is more or less centered on the ridge coming from our spline relation. Thus, a TIC/CTL user should expect to find stars with similar \teff\ but differing radii or masses.

\subsubsection{Sky distribution of top priority objects}

\begin{figure}[!ht]
    \centering
    \includegraphics[width=\linewidth,clip,trim=0 0 30 80]{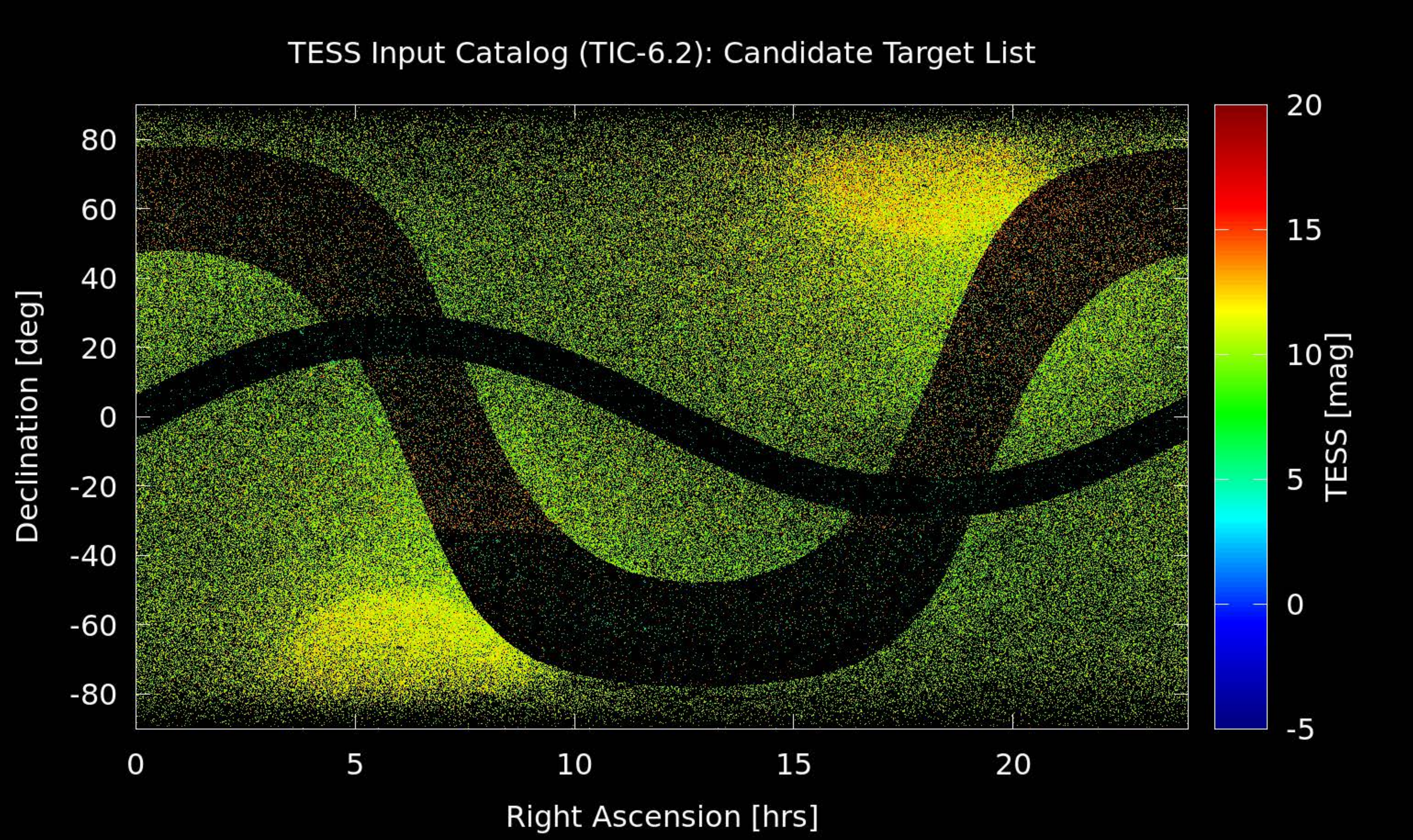}
\caption{The distribution of top CTL targets in right ascension and declination, colored by \tmag\ magnitude. Clear patterns arise due to de-boosting in the Galactic Plane ($|b|<15^{\circ}$), the boosting towards the ecliptic poles ($|\beta| > 78^{\circ}$), and the coverage of the proper motion catalogs (the lines near $-35^{\circ}$).}
    \label{fig:top_prio}
\end{figure}

One of the primary purposes of the TIC and CTL is to provide a list of the top 200,000 to 400,000 targets to be observed in the 2-min cadence. If we investigate the distribution of these targets in the celestial sphere significant features begin to arise as shown in Figure~\ref{fig:top_prio}. 

The most notable feature is the dearth of stars in the Galactic Plane. Because confusion is large in the plane, we de-boosted stars that do not have spectroscopic \teff\ or that are not included in any specially curated list.
Secondly, there is a clear demarcation at about $-35^{\circ}$ in declination. This is due to the proper motions of the majority of TIC stars coming from the UCAC4/5 survey, which is based on the PPMXL catalog, and as described above this catalog is deficient at these southern declinations. 

Third, the stars in the ecliptic plane ($|\beta|<6$) are bright stars with priorities set to 1.
Finally, the two elliptical regions at the northern and southern ecliptic poles ($|\beta|>78^{\circ}$) are the CVZs.  Because the priority increases with ecliptic latitude, they are quite overdense compared too the rest of the sky. 

\subsection{Known limitations and plans for improvement}

\subsubsection{Stellar Activity and RV Followup}

So far the prioritization of candidate transit targets has been based on estimates of the detectability of transits produced by small planets, and assumptions on the geometric parameters (e.g., stellar radius) and measurement noise (e.g., stellar brightness and flux contamination by neighbors). However, in principle it should be possible to also prioritize based on the likely photometric and/or radial-velocity quiescence of the star. Photometric noise from, e.g., magnetic activity could complicate the detection of small transit signals, whereas radial-velocity noise (e.g., `jitter') could complicate the confirmation of planets and the measurement of stellar masses. 
With regards to photometric noise, there are a number of long-term photometric monitoring campaigns of bright stars across the sky that could be used to obtain a measure (or at least an upper limit) on the amplitude of photometric variability. Surveys that could be utilized for this purpose include SuperWASP and KELT, as well as ASAS \citep{Pepper:2007,Pepper:2012,Smith:2014,Pojmanski:1997}.

With regards to radial-velocity jitter, most RV surveys cover relatively few stars. However, in principle it would be possible to use a proxy such as stellar rotation, either spectroscopic ($v\sin i$) or photometric (rotation period). For example, \citet{Vanderburg:2016} gives empirical relations linking radial velocity jitter to stellar rotation period as a function of stellar \teff. 

We have determined stellar rotation periods and variability amplitudes for nearly the entire KELT data set \citep{Oelkers:2018}. Because the KELT and TESS pixel scale, field of view, and depth are comparable \citep[see][]{Pepper:2007,Pepper:2012}, we anticipate that the KELT photometric variability measures may prove useful in further winnowing the CTL of stars that might otherwise be too photometrically or Doppler noisy.

\subsubsection{Binarity}

It is known that most stars are in gravitationally bound pairs or stellar systems of higher multiplicity \citep[e.g.,][]{Duchene:2013}.  All-sky or wide-field catalogs of stars are often able to separately identify the individual stellar components of especially wide binary pairs, in which both stars are bright enough for detection, but generally are not able to resolve the components below a certain angular separation, or detect fainter companions.  Furthermore, even when all components in a multiple system are individually identified, it is typically not possible to reliably match the bound stars without good proper motion or other dynamical information.  Therefore, we expect that a large fraction of the stars in the TIC to represent blended, unresolved multiple stars.

Binarity and higher order multiplicity affects the reliability and integrity of the TIC, since an unresolved companion will cause the photometric magnitudes ascribed to a star to be incorrect, along with any derived stellar properties such as \teff, mass, and radius.  Furthermore, the presence of stellar companions specifically affects the suitability of a star as a TESS mission target, even if the stellar properties of the target star are correctly described.  That can happen by introducing signals that complicate or confuse the transit search, or by creating conditions that would not permit the presence of any planet that TESS is capable of detecting.  In the first category, if a stellar companion is in orbit about the target with a period shorter than the TESS dwell time on the target and happens to be eclipsing, the system can mimic the photometric signal of a transiting planet.  That can happen if the companion is in a grazing orbit, or if the companion is a small enough star to create an eclipse comparable to the transit of a planet.  Or, if there is a planet transiting the target on a stable interior orbit, the photometric signal of the eclipsing companion can interfere with the ability to detect the transit signal.  Also, the presence of a luminous stellar companion unresolved from the target will dilute the photometric transit signal.

In the second category, the presence of a stellar companion indicates that there could not exist a planet in an orbit around the target star that TESS could detect, meaning that the target could potentially be excluded from the 2-min target list if the presence of the companion were known.  Generally, the presence of a stellar companion in an orbit of a given size will dynamically exclude planetary companions in a range of orbits \citep[e.g.,][]{Kraus:2016}.  Also, a stellar companion with a short orbital period may indicate the absence of any circumbinary planets.

For those reasons, it would be advantageous to know about the existence of all stellar multiplicity for each potential target star.  That information is generally not available but will be critical for establishing the correct planetary radii \citep{Ciardi:2015,Bouma:2018}.  We are endeavoring to obtain whatever information is available, through lists of known multiple systems from stellar spectroscopy that can identify single- and double-lined spectroscopic binaries, photometric surveys that can identify eclipsing systems, proper motion catalogs that can identify co-moving companions, and eventually time-series astrometric catalogs to identify astrometric binaries.  However, we expect all such efforts to be quite incomplete due to various observational biases. 

While beyond the scope of this paper, in principle one could attempt to estimate the fraction of binaries in the TIC and CTL, using empirical estimates of binarity as functions of spectral type and other factors. The study by \citet{Duquennoy:1991} was concerned with solar-type stars; though we do have many solar-type dwarfs in the CTL, TESS emphasizes M-dwarfs as being more valuable targets, and those have a lower intrinsic binary frequency. \citet{Duchene:2013} provides a more updated and comprehensive view of the frequency of binaries. In lieu of performing such an analysis, we suggest that the CTL is likely missing essentially all physical binary companions, so the percentage of unknown binaries in the catalog should be essentially the same as the frequency of binaries of different spectral types as given by \citet{Duchene:2013}.

Therefore, such information about stellar multiplicity will be primarily useful for the evaluation of TESS transit candidates after observations are taken, rather than usable for target selection during the prime mission.
Indeed, our selection of targets independent of multiplicity considerations could well prove to be a major benefit for studying the effect of multiplicity on planet occurrence.

\subsubsection{{\it Gaia\/} DR2}\label{sec:gaiadr2}
As mentioned in Sec.~\ref{sec:tic},
it had been hoped from the start of planning for TESS target selection that the highly anticipated {\it Gaia\/} DR2 parallaxes might become available sufficiently in advance of TESS launch to be incorporated into the TIC. Unfortunately, the DR2 parallaxes did not become public until after launch, which necessitated creation of a TIC and CTL for use in the TESS Year\;1 operations without the benefit of DR2 parallaxes. Nonetheless, we are now working to incorporate parallaxes and other information from {\it Gaia\/} DR2.  We expect that the {\it Gaia\/} DR2 data set will be incorporated into the next version of the TIC before Year\;2 of the mission when TESS observes the northern ecliptic sky.

There is at least one specific area in which the incorporation of the {\it Gaia\/} DR2 parallaxes could dramatically enhance target selection: elimination of subgiant contaminants. As discussed by \citet{Bastien:2014} in the context of the bright {\it Kepler\/} targets, roughly half of the putative dwarf stars are in fact modestly evolved subgiants. Recent assessments of the full {\it Kepler\/} planet sample find subgiant contamination of $\sim$25\% \citep{Berger:2018}. In any event, the slightly lower \logg\ of subgiants can be too subtle for discernment by photometric and even some spectroscopic methods. Moreover, as discussed above (Section~\ref{sec:giant_removal}), the reduced-proper-motion method that we employ to screen out giants is also ineffective at removing most subgiants. 
As laid out in Section~\ref{sec:mass_radius}, with an accurate parallax for most if not all TESS targets, we can accurately determine the stellar radius and thereby screen out subgiants with high fidelity. In the meantime, a large proportion of subgiant contaminants are to be expected among the putative dwarf stars in the CTL. We should note that such a step should only be undertaken after careful consideration, because there is also significant value specifically in detecting planets transiting subgiants, which are prime targets for asteroseismic studies \citep{Campante:2016}.

Finally, with {\it Gaia\/} magnitudes available for virtually all stars in the TIC, it should be possible in principle to re-determine accurate \tmag\ magnitudes for all TIC stars.

\subsubsection{Full-frame images}

In addition to the standard 2-min cadence of measurements for the primary 200,000--400,000 bright transiting planet candidates, TESS will provide full-frame images (FFIs) with a cadence of 30~min. These FFIs can in principle be used to also identify planetary transits, and specialized pipelines for extracting precise light curves from the FFIs are available or in preparation \citep[e.g.,][]{Lund:2017,Oelkersa:2018}.
Thus it may be beneficial to consider which types of transit hosts can be effectively studied at 30-min cadence so as to reserve as many 2-min slots for those types of systems that most require the higher cadence. Here we consider two types of situations where such a consideration could lead to further optimization of the CTL.

\subsubsubsection{Transits of M dwarfs and White Dwarfs by Earth-sized planets}

There are systems where there could be transits with durations less than an hour or so, i.e., short enough that the SNR for detection starts to be significantly compromised if they are only observed with a 30-min cadence.  It is fairly likely that transits could be non-equatorial and have durations as short as 30--50\% that of an equatorial transit.  Especially considering the possibilities of non-equatorial transits, and that orbits could have semimajor axes as tight as $\sim3\;R_\star$, some systems could even have transits that are much shorter than 30 minutes. 

One approach would be to propose that the highest priority for inclusion in the CTL be those systems where it is reasonably possible that there could be transits shorter than one hour.  This would include all white dwarfs and all low-mass main-sequence stars.  The faint end of the magnitude range would tentatively be $\tmag \sim 14$, and the upper mass cutoff would be $\sim$0.5\;\msun\ for main-sequence stars.  These notional parameter cutoffs are justified as follows.

We have performed crude calculations regarding the transits of 0.1--1.0\;\msun\ main-sequence stars by small planets, i.e., planets with radii that are much smaller than those of their host stars.  For these stars we take the mass-radius relation to be
$M_\star/\msun \approx R_\star/\rsun$.

The equatorial transit duration is then:
$t_{\rm eq-dur} = P_{\rm orb}\;2 R_\star/(2 \pi a)$, where $P_{\rm orb}$ is the orbital period, $R_\star$ is the radius of the host star, and $a$ is the semimajor axis of the planetary orbit.  The close-in orbits are the ones that might make very short transits.  We take the ``relatively likely" extreme close-in case to be $a = 3 R_\star$. Then $t_{\rm eq-dur} = P_{\rm orb}/10$, roughly speaking. Using Kepler's third law with $a = 3 R_\star$ and the above mass-radius relation, we obtain
$P_{\rm orb} \approx 0.6 (R_\star/R_\odot)^{1/2}$ days. For a star with $M = 0.2\,\msun$ this gives $P_{\rm orb} = 6.4$ hrs. Indeed, at least one planet has been found around a low-mass star with an orbital period in the range of 4--5~hrs.

Continuing, we obtain
$t_{\rm eq-dur} \approx P_{\rm orb}/10 =  0.06 (R_\star/\rsun)^{1/2}$ days = $1.4 (R_\star/\rsun)^{1/2}$ hrs.
This gives a 1-hr transit duration for a planet orbiting a star with $M=0.5\,\msun$ at 3\,$R_\star$. It gives a duration of $\sim$0.5~hr for a planet orbiting a 0.1~M$_\odot$ star at 3~R$_\star$. These are the durations of equatorial transits; non-equatorial transits will be a bit shorter.  A transit impact parameter of 0.86, which is admittedly extreme, gives a transit that is half the duration of an equatorial transit. 

A star with $M=0.1\,\msun$ may be very unusual in TESS observations.  Stars with $M \sim 0.2$--0.3\;\msun\ should be more frequent targets.
This suggests that $0.5\,\msun$ may serve as a reasonable upper mass limit for this category.  There should also be a faint magnitude limit based on detectability of these short periods in an FFT or Box Least Squares \citep[BLS;][]{Kovacs:2002} search of a 27-day or longer observation.  We have not yet attempted that calculation, which would depend on the planet radius as well as the other assumptions made above.  We have also neglected the planet radius in computing the transit duration.  Large planets on such orbits would be detectable even around fairly faint stars, but the transits would be a bit longer than calculated.  Small planets would of course be harder to detect.

In principle, stellar mass and radius estimates are also needed. However, in practice, one might assume all targets are main-sequence (except for white dwarfs) and the above mass-radius relation may be sufficient.

\subsubsubsection{Transits of hot subdwarfs and white dwarfs by Earth-sized planets}

\begin{figure}[!ht]
    \centering
    \includegraphics[width=0.49\linewidth]{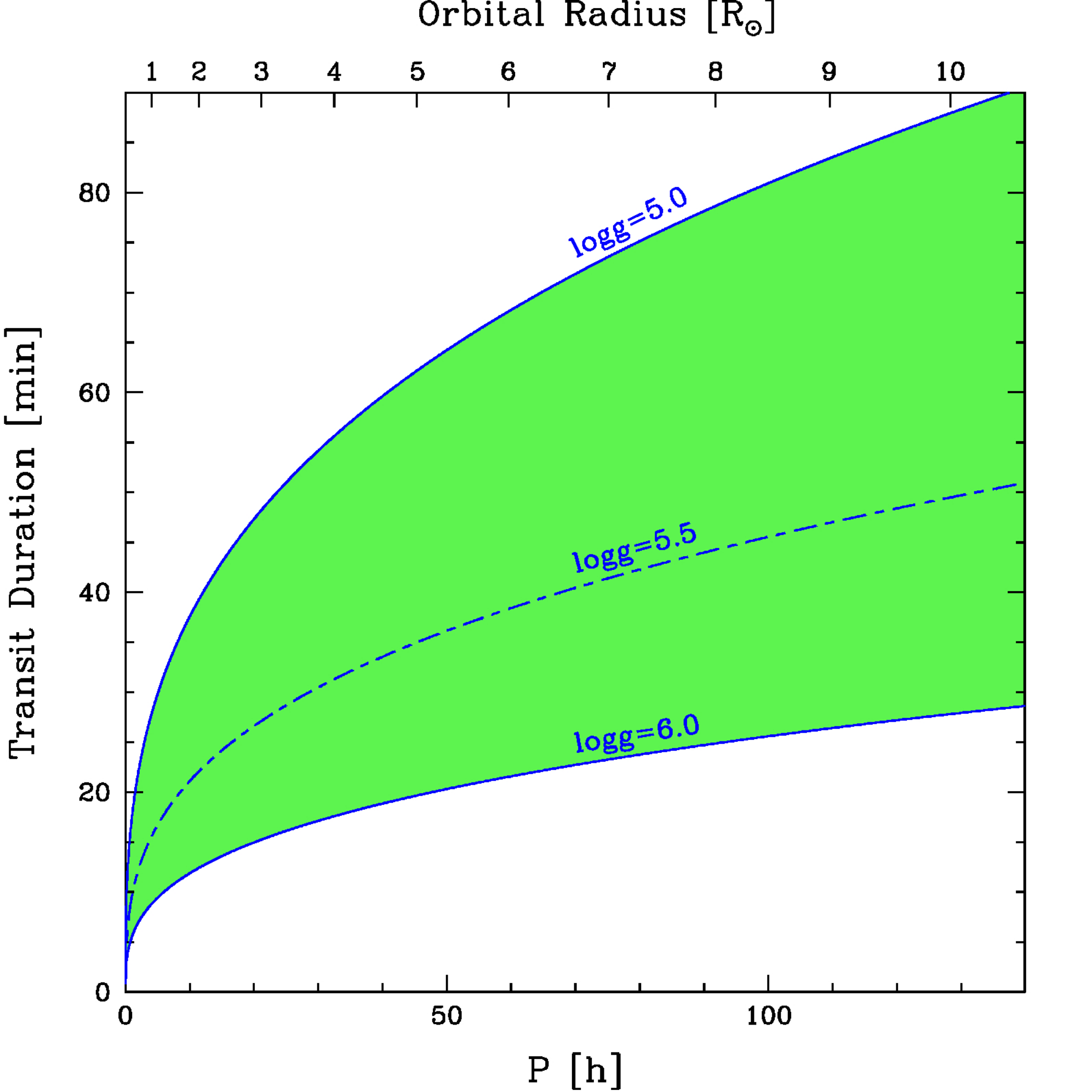}
    \includegraphics[width=0.49\linewidth]{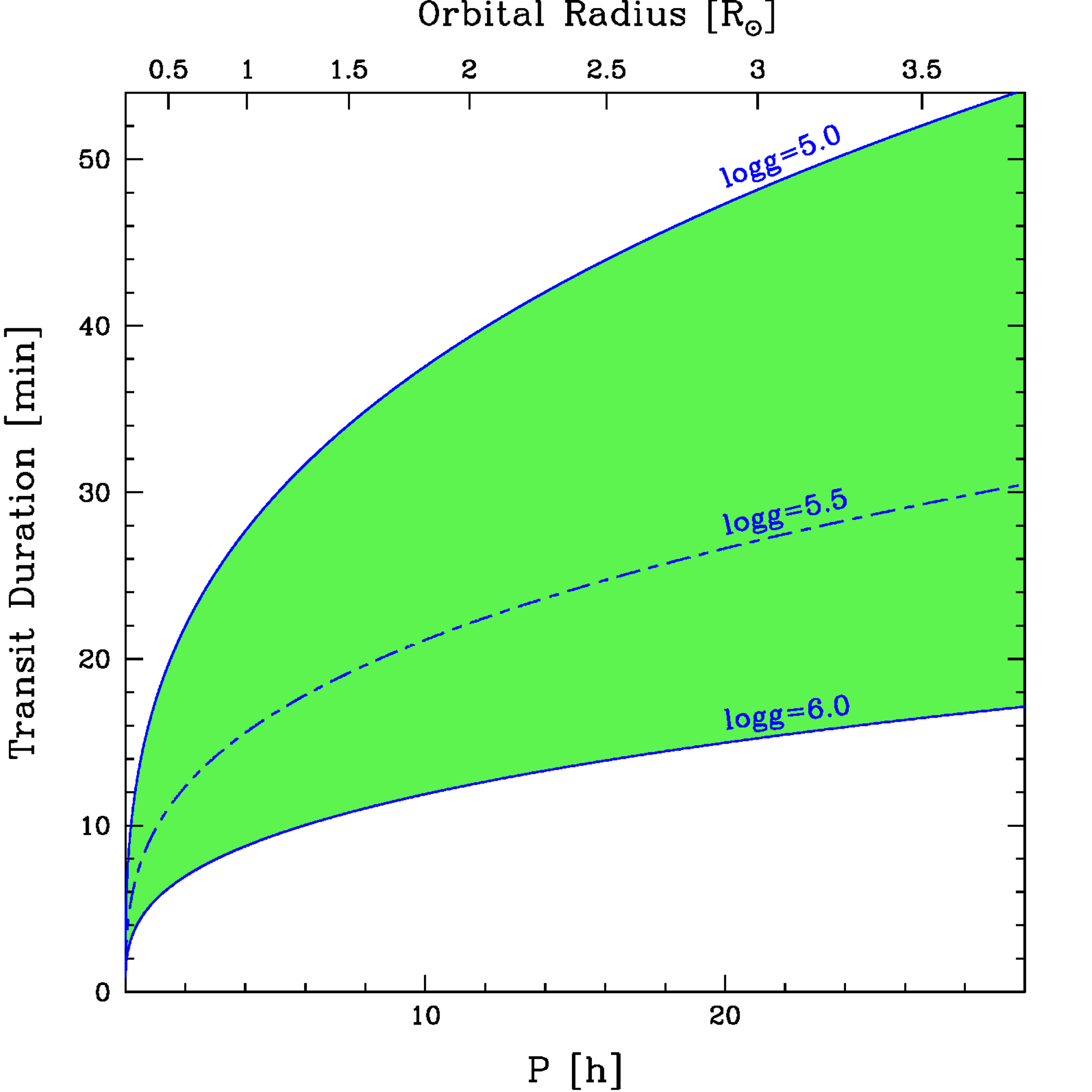}
    \caption{(Left) Duration of the transit of a 0.47-\msun\ hot subdwarf by a planet with a radius of $1\;\rearth$, as a function of orbital period and \logg\ of the host star. (Right) Detail at short orbital periods.}
    \label{fig:subdwarfs}
\end{figure}

\begin{figure}[!ht]
    \centering
    \includegraphics[width=0.49\linewidth]{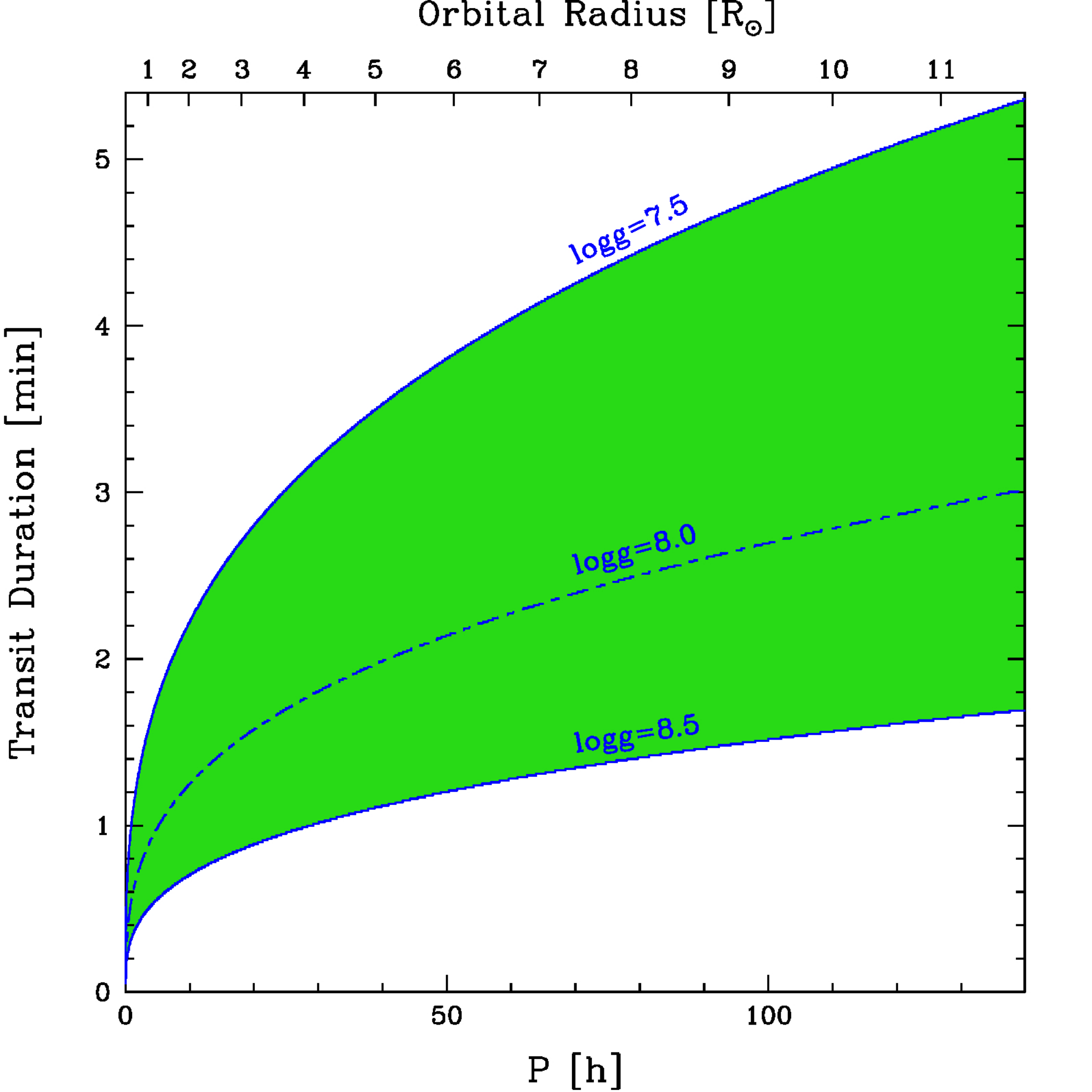}
    \includegraphics[width=0.49\linewidth]{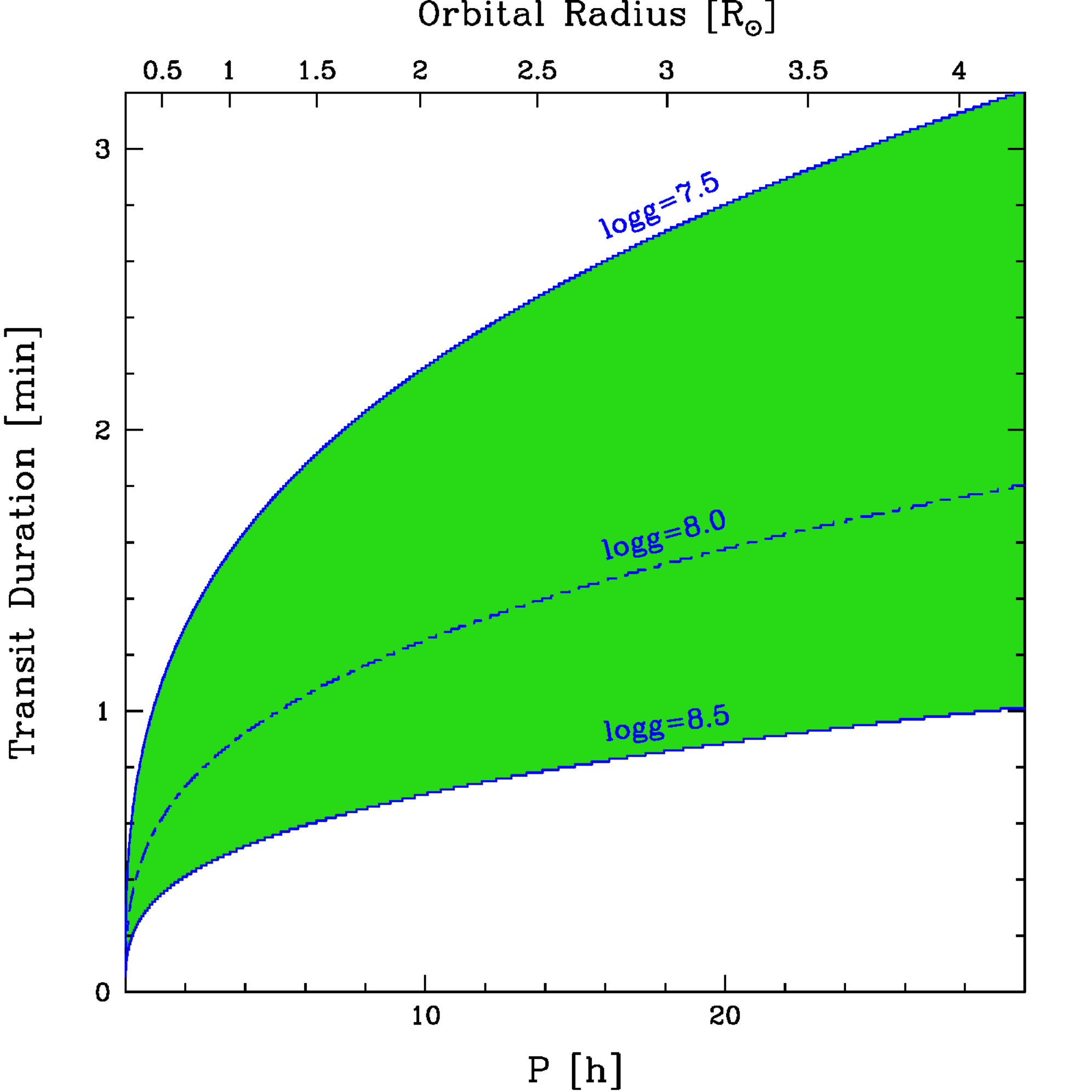}
    \caption{(Left) Duration of the transit of a 0.64-\msun\ white dwarf by a planet with a radius of $1\;\rearth$, as a function of orbital period and \logg\ of the host star. (Right) Detail at short orbital periods.}
    \label{fig:wds}
\end{figure}

Hot subdwarfs and white dwarfs represent other potentially interesting classes of transiting planet host stars that, because of the small stellar radius, may require the 2-min cadence observations in at least a subset of cases. The transit duration of a 0.47-\msun\ subdwarf star and of a 0.64-\msun\ white dwarf by a $1\:\rearth$ planet is shown in Figure~\ref{fig:subdwarfs} and Figure~\ref{fig:wds}, respectively, as a function of orbital period and the \logg\ of the host star. 

\subsubsection{Effects of reddening on the \tmag\ magnitude}
\label{sec:redT}

In a future version of the TIC, we will need to account for the effects of reddening and extinction on \tmag. 
The color relations we are using are derived from model atmospheres that assume the fluxes are not affected by extinction, which is not true in practice.
We plan to overcome this approximation by correcting our $V+JHK_S$ magnitudes for extinction and then, using the de-reddended fluxes, we will derive an extinction-free TESS magnitude with the previously described relations and coefficients. We will then apply the appropriate extinction for the TESS band to recover an ``observed" TESS magnitude. Using the \citet{Cardelli:1989} extinction law with a weighted mean wavelength for the passband and ignoring the spectrum of the star, we estimate $A_T=0.656\,E(V-K_S)$, $A_J = 0.325\,E(V-K_S)$, and $A_G = 0.901\,E(V-K_S)$, where $G$ is the {\it Gaia\/} magnitude we expect to incorporate into a future version of the TIC. The conversion from $E(V-K_S)$ to $E(B-V)$ can be made with $E(B-V) = 0.372\,E(V-K_S)$, and similarly $E(G-J) = 0.576\,E(V-K_S)$. For stars that do not have a reliable $V$ magnitude no extinction correction will be computed. 

\begin{figure}[!ht]
    \centering
    \includegraphics[width=0.975\linewidth]{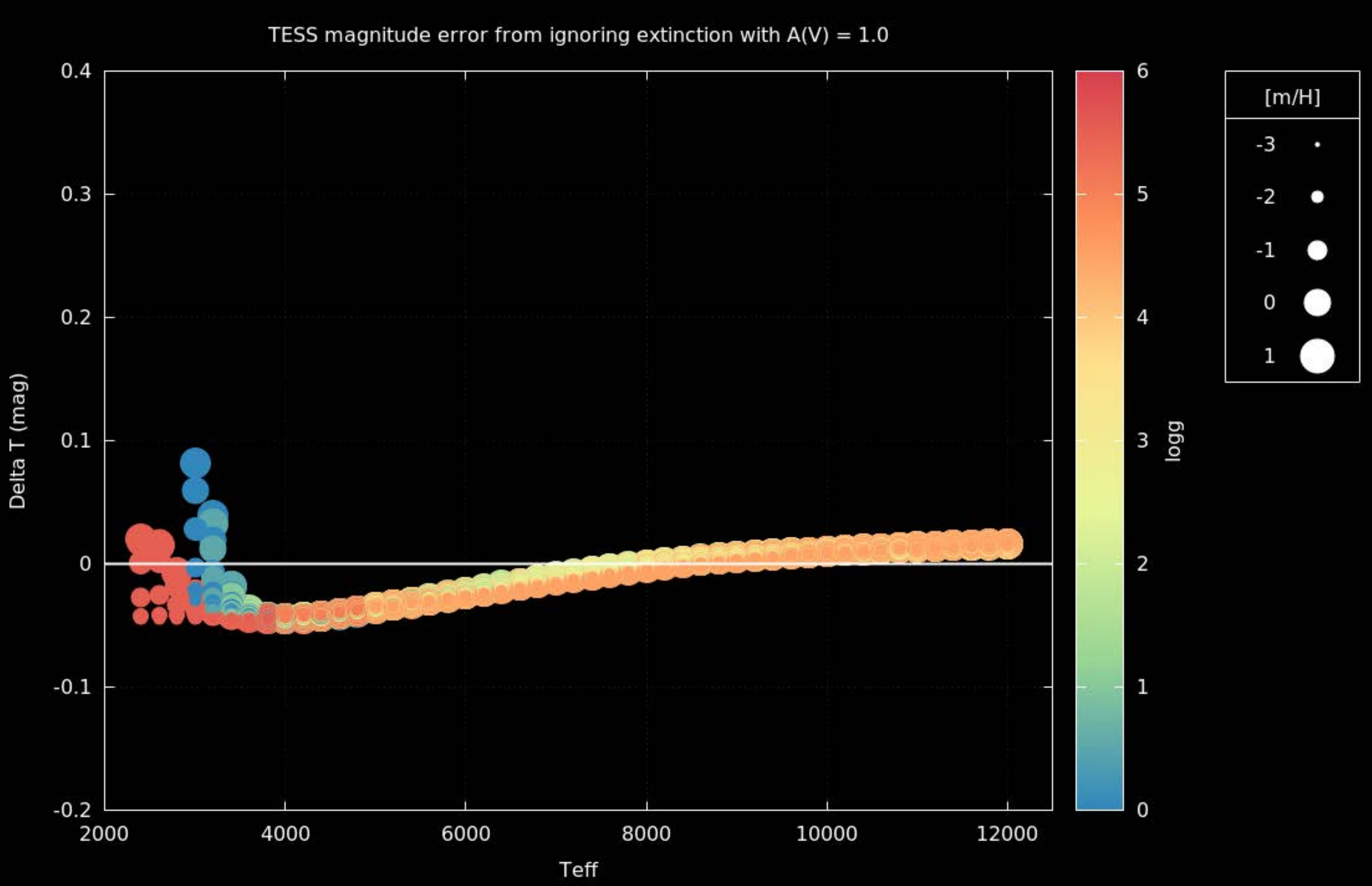}
    \includegraphics[width=0.975\linewidth]{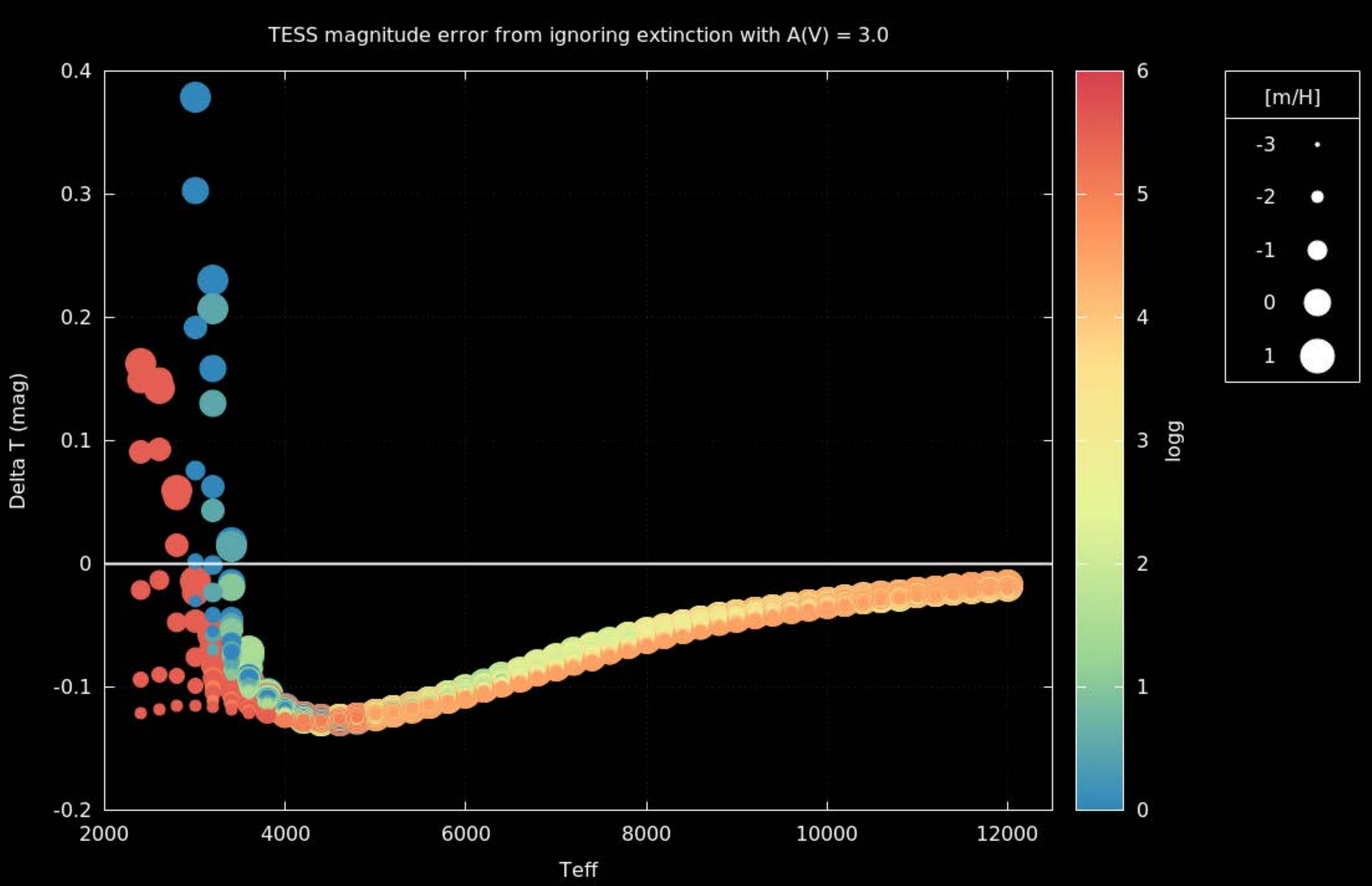}
\caption{(Top) Error in the calculated TESS magnitude if extinction is ignored and the true extinction is $A_V=1.0$. (Bottom) Same as above for $A_V=3.0$.}
    \label{fig:tmager1}    
    \label{fig:tmager3}
\end{figure}

Figure \ref{fig:tmager1} 
shows the expected differences between TESS magnitudes calculated without proper consideration of extinction and those with extinction accounted for, for extinction levels of $A_V =1$ and $A_V=3$. 
For most stars in the TIC experiencing modest extinction, the impact of not properly modeling the effect amounts to a systematic error of $\sim$0.1\;mag for the coolest stars. However, for areas of high extinction such as in the Galactic plane the effect can be as large as several tenths of a magnitude for $\teff \lesssim 4000$\;K.

\acknowledgments
It is a pleasure to thank the members of the TESS Target Selection Working Group for input into the development of the procedures described here, including especially contribution of the specially curated target lists. This work has been partially supported by the NASA TESS mission through a subaward to Vanderbilt University, and from NASA grant XRP17 2-0024. This research has made use of the NASA Exoplanet Archive, which is operated by the California Institute of Technology, under contract with the National Aeronautics and Space Administration under the Exoplanet Exploration Program. 

\software{TRILEGAL (Girardi et al. 2005), PHOENIX (Husser et al. 2016)}


\clearpage

\appendix
\section{TIC-7 Release Notes\label{sec:relnotes}}

\begin{center}
{\bf TESS Input Catalog Version 7 (TIC-7) Release Notes 2018-06-17}
\end{center}

This delivery contains the seventh version of the TESS Input Catalog (TIC) produced entirely by the Target Selection Working Group (TSWG), and was finalized and prepared for delivery to the TESS Science Office (TSO) on 2018 June 17.

The delivery has a number of minor issues (see below) which have not been fixed in this version due to time constraints during preparation. Specific details of the method of production and the contents of this TIC will be described in the full TIC-7 Documentation can currently be found on the arXiv at (https://arxiv.org/abs/1706.00495).

The design is the same as TIC-6, in that the columns and their format are the same, but there have been additional changes compared to previous versions of the TIC. TIC IDs have not been changed, and all future deliveries of the TIC will use the same IDs for specific objects. New objects added to the TIC will always receive new IDs. Objects may be removed from the TIC, if they are found to be spurious, but TIC IDs will always be unique and a new TIC object will never receive the ID of an old TIC object.

\noindent{\bf Changes compared to TIC-6}

This delivery contains major changes in computed quantities compared to TIC-6. It should be noted that the methods used to estimate a variety of stellar parameters are still under active development and can be affected by poor catalog photometry when there is no acceptable alternative photometry for a given star. The major changes compared to previous versions are:

\begin{enumerate}
\item Nearly all coordinates have been computed for epoch 2000.0, the exceptions are POSFlag (Column 16) hotsubdwarf (40 stars), and gicycle1 (1692 stars) for which an epoch for the coordinates was not provided to the Target Selection Working Group. For stars having been Gaia DR1 multiples in TIC-6 (POSFlag tmmgaia) we reverted to the 2MASS coordinates of the main component, and propagated the epoch given in 2MASS to 2000.0. Ecliptic and galactic coordinates (Column 25-28) have the same epoch as RA and Dec (Column 14 and 15).
\item Targets from the asteroseismology list, previously missing in the TIC, have been added to the TIC (278 stars). 
\item CTL6 stars with POSFlag (Column 16) hip and lepine have been rematched. 37 duplicates were identified. These stars were in TIC6 twice, once for hipparcos and once for 2MASS. Duplicates and artifacts have been deleted from the CTL and flagged as such in TIC. The disposition and duplicate\_id column (Column 86 and 87) point to the valid TIC-entry for every duplicate. The artifacts are stars from an earlier version of Superblink which do not occur in any later version of Superblink.
\item The 329 Hipparcos stars which are not in the CTL, have been propagated to epoch 2000.0, but have not been rematched. It is possible many of these stars are duplicates, but due to time constraints, they could not be matched by hand. These will be updated with the Gaia DR2-based TIC-8.
\item All CTL-stars now have a contamination ratio. Where a priority could not be computed due to missing information, the contamination ratio may serve as a guide for the suitability of target selection.
\item The 13 stars in TIC-6 which did not have TESS magnitudes, now have proper non-null values. 
\item Stars in the specially curated cool dwarf list were updated, and their stellar parameters have changed. Users are encouraged to double check the stellar parameters for these stars have been updated with the SPflag = `cdwrf' (Column 64).
\item Stars in the specially curated known planet host list now have their stellar parameters updated. Users are encouraged to investigate these parameters using the SPflag = `kplnt' (Column 64).
\item The legacy SPflag = `allen', have been replaced with `splin' to properly reflect the source of the stellar parameters. Similarly, SPflag = `spect' has been replaced with ‘spec.’
\item Stars in TIC-6 which had luminosity errors larger than the luminosity value, now show luminosity errors equal to the luminosity value. 

\end{enumerate}

\noindent{\bf Notes on the individual columns:}

Column No. 	\textbf{Name} 		Notes

\begin{enumerate}

\item  	\textbf{TICID} --	A unique identifier for every object in the TIC. The ID is unique and permanent. If an object is removed from the TIC in later versions, a new object will never inherit an old ID. 

\item       \textbf{Version} -- 	This column denotes the date YYYYMMDD, in which the TIC was finalized and prepared for delivery. 
\setcounter{enumi}{6}

\item	    \textbf{SDSS} -- 		The values given are the 64-bit "objID" values, not the IAU-format "SDSS J" identifiers.
\setcounter{enumi}{8}

\item 	    \textbf{GAIA} -- 		Gaia IDs in TIC-6 are included for stars which are found in the Gaia-provided Gaia-2MASS look-up table. For TIC stars with more than one associated Gaia magnitude, the ID of the brightest matching Gaia source is provided.

\item	    \textbf{APASS} --	    APASS stars do not have an identifier, only coordinates. We use the 
primary key of an internal TESS version of the APASS database table as a proxy identifier. 
\setcounter{enumi}{16}
\item	    \textbf{pmRA} --		The right ascension proper motions, in order of preference, are: (1) Gaia-TGAS, (2) Superblink, (3) Tycho-2, (4) Hipparcos. (5-7) Stars only found to have proper motions in UCAC4, UCAC-5 or HSOY were subject to a new set of requirements. Total UCAC-4 proper motion $> 1800$~mas/yr; total UCAC-5 proper motions $> 200$~mas/yr and $< 1800$~mas/yr; total HSOY proper motion $< 200$~mas/yr. If a star did not have a proper motion in these catalogs, it is not provided.

\item	    \textbf{pmRA$\_$e} --	    The right ascension proper motion errors are taken directly from the given proper motion catalog except in the case of SuperBlink, which in its delivered state, does not provide proper motion errors. In this case, we adopt an error of 2 mas/yr for stars with updated proper motions from Gaia and 8 mas/yr for stars without updated proper motions from Gaia.

\item	    \textbf{pmDec} --		See notes for column 17.

\item	    \textbf{pmDec$\_$e} --	    See notes for column 18.

\setcounter{enumi}{21}

\item	    \textbf{plx} --		    The parallax values, in order of preference, are: (1) Gaia-TGAS, and (2) Hipparcos. Some values are negative because of the way the parallaxes were measured in TGAS and Hipparcos.
\setcounter{enumi}{28}

\item	    \textbf{Bmag} --		Johnson B magnitude. When a Johnson B magnitude was not found in one of the optical catalogs, the TIC reports a Johnson B derived from the USNO-A2.0 magnitude given in the 2MASS catalog.
\setcounter{enumi}{30}

\item	    \textbf{Vmag} --		Johnson V magnitude. Observed V magnitudes are preferred when they are converted from Tycho $V_T,$ Hipparcos, or UCAC4. We now calculate a Johnson V magnitude from a G-Ks color for stars which do not have a reliable observed Johnson V magnitude.
\setcounter{enumi}{58}

\item	    \textbf{Gmag} --		Gaia magnitudes are now included for all stars with such values in Gaia 
DR-1.
\setcounter{enumi}{60}
\item 	    \textbf{Tmag} -- 		This column is never NULL. The Tmag values are typically based on relations that depend on J and V-Ks or J-Ks (see column 63 for method flag). TESS magnitudes for objects for which only poor catalog photometry was available were computed simply as offsets from a reference magnitude (see Documentation Section 2.2.1). 
\setcounter{enumi}{62}
\item 	    \textbf{TESSflag} -- 	These flags denote which relation or catalog provides the TIC TESS magnitude. Full descriptions can be found in Appendix C of the documentation:

\begin{enumerate}
        \item gaiak - magnitude calculated from G and 2MASS Ks
        \item gaiaj - magnitude calculated from G and 2MASS J 
        \item joffset2 - magnitude calculated from 2MASS J and an offset (+1.75 for $J-K_S > 1$)
        \item hipvmag - magnitude calculated Hipparcos V magnitude
        \item gaiaoffset - magnitude calculated from G and an offset
        \item hoffset - magnitude calculated from 2MASS H offset
        \item vjh - magnitude calculated from V and 2MASS J-H
        \item jhk - magnitude calculated from 2MASS J-Ks
        \item vjk - magnitude calculated from V and 2MASS J-Ks
        \item hotsubdwarf - magnitude adopted from hot subdwarf list
        \item vk - magnitude calculated from V and 2MASS Ks
        \item joffset - magnitude calculated from 2MASS J offset (+0.5 for $J-K_S < -0.1$)
        \item gaiav - magnitude calculated from G and V
        \item tmvk - magnitude calculated from V and 2MASS Ks (same as vk)
        \item from$\_$apass$\_$i - magnitude from cool dwarf list \citep{Muirhead:2018}
        \item from$\_$sdss$\_$ik - magnitude from cool dwarf list \citep{Muirhead:2018}
        \item gaiah - magnitude calculated from Gaia and 2MASS H
        \item jh - magnitude calculated from 2MASS J-H
        \item cdwarf - magnitude from cool dwarf list \citep{Muirhead:2018}
        \item bpjk - magnitude calculated from photographic B and 2MASS J-Ks
        \item voffset - magnitude calculated from V and offset
        \item koffset - magnitude calculated from 2MASS Ks and offset
        \item wmean$\_$vk$\_$jhk - magnitude from cool dwarf list \citep{Muirhead:2018}
        \item lepine - magnitude from Lepine catalog
        \item gicycle1 - magnitude from GI Cycle 1 proposal
        \item from$\_$sdss$\_$i - magnitude from cool dwarf list \citep{Muirhead:2018}
\end{enumerate}

See TIC-7 Documentation Section 2.2.1 for details of each method. While most of these relations (which are used for most Tmag values) are only appropriate for dwarf stars, some are applicable to giants. Extended objects were treated as if they were dwarfs. In general the dwarf relations are strictly valid between specific color ranges and tend to be less accurate for very blue stars ($J-K_S < -0.$) or very red stars ($J-K_S > 1$). 

\item 	    \textbf{SPFlag} -- 	    These flags denote the origin of stellar parameters:
\begin{enumerate}
        \item cdwrf - mass and radius adopted from the Cool Dwarf list \citep{Muirhead:2018}
        \item hotsdwrf - mass and radius taken from the Hot Subdwarf list
        \item tplx - parameters computed from measured TGAS parallax
        \item hplx - parameters computed from measured HIP parallax
        \item spec - parameters computed using Torres et al. 2010, A\&ARv, 18, 67
        \item splin - parameters computed from unified spline relations 
        \item kplnt - mass and radius adopted from the Known Planet Host list
\end{enumerate}
\item 	    \textbf{Teff} --	    The effective temperatures come from one of four sources, in the following order of preference: (1) the Cool Dwarf list or the Hot Subdwarf list; (2) spectroscopic catalogs (see Column 64); (3) dereddended V-Ks color; and (4) non-dereddened V-Ks color.  We no longer allow stars with effective temperature not corrected for reddening to enter the CTL, except for stars in the bright star list.

\item	    \textbf{e$\_$Teff} --		The SPOCS and GALAH catalogs do not provide uncertainties for 
effective temperatures; 25K and 41K were assigned, respectively, based on the reported statistical error from those catalogs.

\item 	    \textbf{Logg} --		Surface gravity is calculated using the nominal formula: $log_{10}(G*M * Msun/(R * Rsun)^2)$. Where Msun is the mass of the Sun, G is the gravitational constant, Rsun is the Radius of the Sun, M is the mass of the star (column 73) and R is the radius of the star (column 71). Some stars may have unphysical $log(g)$ values, such as $log(g) > 4.8$. If the star’s stellar characteristics were calculated from de-reddened effective temperature or from a spectroscopic temperature, their priorities have been set to 0 to not prioritize stars with low quality stellar characteristics but the log(g) value remains to keep the TIC internally consistent.

\item 	    \textbf{e$\_$Logg} --	    For stars which do not have spectroscopic $log(g)$ measured, we define the error in the surface gravity as $\sqrt((M_e/M)^2+(2*R_e/R)^2)$, where M is the mass of the star (column 73), $M_e$ is the mass error (column 74), R is the radius of the star (column 71) and R$\_$e is the radius error of the star (column 72). For stars with spectroscopic log(g) from a single observation, the error was copied. For stars with multiple observations in the same catalog, the error listed in the TIC is a combination of each single observation’s error added in quadrature. The SPOCS and GALAH catalogs do not provide uncertainties for surface gravities; 0.028 and 0.17 dex were assigned, respectively, based on the reported statistical error from those catalogs. 

\setcounter{enumi}{69}
\item 	    \textbf{e$\_$M/H} --		For stars with spectroscopic metallicity from a single observation, the error was copied from the relevant catalog. For stars with multiple observations in the same catalog, the error listed in the TIC is a combination of each single observation’s error added in quadrature. The SPOCS and GALAH catalogs do not provide uncertainties for metallicities; 0.10 and 0.05 dex were assigned, respectively, based on the reported statistical error from these catalogs.

\item 	    \textbf{Radius} --		The stellar radii were estimated using a variety of techniques, in the
following order of preference: (1) radii provided by the specially curated Cool Dwarf list or the Hot Subdwarf list; (2) using the Gaia parallax and bolometric corrections; (3) spectroscopic relations from Torres et al. 2010, A\&ARv, 18, 67; and (4) a unified relation based on measured radii for eclipsing binaries as well as simulations using Galactic structure models. 
\setcounter{enumi}{72}
\item 	    \textbf{Mass} --		If an object’s mass is provided in the specially curated cool dwarf list or hot subdwarf list it is included in the TIC. Otherwise, the stellar masses were estimated using an unified relation based on measured masses for eclipsing binaries as well as simulations using Galactic structure models (see section 3.2.2 in the full documentation for details).
\setcounter{enumi}{74}
\item 	    \textbf{Rho} -- 		The density in solar units is calculated using the formula $M/R^3$, where M is the mass of the star (column 73) and R is the radius of the star 
(column 71).

\item 	   \textbf{Rho$\_$e}  --	 	The error in the density is calculated using the following formula: $3.0*Rho*(R_e/R)$, where Rho is the density (column 75), $R_e$ is the error in the radius (column 72) and R is the radius of the star (column 71).

\item      \textbf{LumClass} -- 	This is a boolean dwarf flag. If this is set, LumClass = DWARF, or otherwise GIANT.  SUBGIANT is not used at present. However, the DWARF flag for TIC-6 effectively means that the star is either a dwarf or a subgiant, based on reduced proper motion cuts.

\item 	    \textbf{Lum} -- 		The luminosity is calculated using the following formula and defined in solar units: $R^2/(Teff/5772)^4$, where R is the radius of the star (column 73) and Teff is the effective temperature (column 65).

\item 	    \textbf{Lum$\_$e}  --	    The error in the luminosity is calculated using the following formula: $2.0* L * (R_e / R)$. Where L is the luminosity (column 78), R$\_$e is the radius error (column 74) and R is the radius (column 73). If the luminosity error was found to be larger than the luminosity, it was set to be equal to the luminosity.

\setcounter{enumi}{81}
\item 	   \textbf{E(B-V)} --		Stars for which $E(B-V) > 1.5$ have their E(B-V) values set to a maximum of 1.5.
\setcounter{enumi}{84}
\item 	    \textbf{contratio} --	The contamination ratio is defined as the nominal flux from the contaminants divided by the flux from the source. Flux contamination is calculated for all stars in the CTL. Contaminants are searched for within 10 TESS pixels of the target and the contaminating flux is calculated within a radius that depends on the target’s TESS magnitude (Tmag, column 61). The PSF is modeled using a 2D-Gaussian based on preliminary PSF measurements from the SPOC. See section 3.2.3 of the full documentation for details.
\setcounter{enumi}{87}
\item 	    \textbf{priority} --	Priority of target for observation. This is a floating-point value ranging 
from 0 to 1, where 1 is highest priority. The priority is based on the relative ability of TESS to detect small planetary transits, and is calculated using the radius of the star, the contamination ratio, and the total expected photometric precision. Stars are given a boost factor to their priority which scales with a probabilistic model of the expected number of sectors any given star could fall in. Typically, the closer the star is to the Ecliptic North or South pole, the larger the boost factor. Stars close to the Galactic Plane (|b|<15) have been de-boosted by a factor of 0.1 since we generally have a poor understanding of their true reddening, unless they are in the specially curated Cool Dwarf list (see Muirhead et al. 2018) or Known planet host list. 
 
The formula for CTL7.1 is defined as: $\sqrt{Ns}/(R^{1.5}\times\sigma)$ where Ns is the expected number of  TESS sectors to observe the star; R is the radius of the star (column 71), and $\sigma$ is the expected photometric precision of the star based on the TESS magnitude (column 61) calculated using the methodology of \citet{Sullivan:2015}. The priority is normalized by the priority for a star with R = 0.1 solar, Ns = 12.654 sectors, $\epsilon$ = 0 contamination and $\sigma$ = 61.75 ppm. 
Some stars will have distinct priorities:
\begin{enumerate}
    \item Stars with log(g) values that are greater than 4.8 and temperature sources from ‘dered’ or ‘spec’ have their priority values set to 0 to avoid biases from Giant stars masquerading as dwarfs.
    \item Stars in the bright star list always have their priority set to 1.
    \item Stars with absolute ecliptic latitudes (column 28) less than $\sim6^{\circ}$ are not expected to be observed as part of the main mission due to a gap in camera coverage between the Southern and Northern observations. Therefore, their Ns values are 0 and thus the priority is 0.
    \item Stars within the known planet list without a radius had their priority values set to 0.
\end{enumerate}

\end{enumerate}

\noindent{\bf The following columns are not populated:}

Column No. 	\textbf{Name}

\begin{enumerate}

\setcounter{enumi}{82}
    \item   \textbf{e$\_$EBV}
\setcounter{enumi}{85}
    \item   \textbf{Disposition}
\end{enumerate}
\noindent{\bf Known issues and pitfalls:}

There are a number of minor issues that have been identified by the TSWG. We expect to address these issues in a future version of the TIC. The issues include:
\begin{enumerate}

\item All coordinates are for the epoch of observation (often 2MASS or SDSS for extended objects). Epochs are not currently supplied.
\item Because some stars have poor quality 2MASS photometry flags (such as ‘D’, ‘U’), offsets where applied to G, V, J, H, or Ks magnitudes to provide a more realistic TESS magnitude but may be different from the true value by a magnitude or more.  
\item ‘allen’ is an older flag that should be replaced by the spline flag.
\item Some stars will show an error in the stellar density which is larger than the density itself. In these cases, the error should be interpreted as equal to the density.
\item The error in the luminosity currently only reflects the effect of the radius error but should also include the effects of temperature.
\item Due to the preference of proper motion catalogs which are based on PPMXL, there is structure in the distribution of high priority candidates mainly above declinations larger than -30 deg. 
\item Stars which have ecliptic latitudes between -6 and 6 degree have priorities set to 0, unless they are in the bright star list. This “gap” in priority is meant to mimic the expected gap in camera coverage for the 2 year primary TESS mission.
\item Some bright stars may have nearby impostor stars with similar magnitudes that lie along diffraction spikes from 2MASS photometry. Users can identify these impostors by checking 2MASS quality flags for very poor photometry (such as ‘D’, ‘E’, ‘F’, ‘U’). These objects should be removed in future versions of the TIC.
\item Some stars in the cool dwarf list and the known planet host list have effective temperature which are null, but still have calculated stellar parameters. These were adopted as is from each list for consistency. 
\item Stars in the known planet list which did not have a radius, had their priority values set to 0.

\end{enumerate}

\noindent{\bf Planned Improvements in Future Versions:}

There are a number of planned improvements for future versions of the TIC. At present these improvements include:
\begin{enumerate}
\item Inclusion of all known exoplanets reported at the NASA archives with a full set of CTL parameters wherever this is possible and feasible. 
\end{enumerate}

\section{TIC columns, formats, and minimum/maximum permitted values\label{sec:appendix_ticranges}}

In Table~\ref{tab:minmax} we describe each column found in the TIC, its data type, basic column description, and the minimum and maximum allowed values. If the column is never permitted to be null (NN = `never null'), it is also indicated.

\textit{It is important for users of the TIC to understand that, as described in Sec.~\ref{sec:tic}, the TIC deliberately includes point sources (stars) and extended sources (e.g., galaxies); positional searches of the TIC will in general return some extended sources as well as stars. These can be separated by use of the {\tt objtype} flag (column 12 of Table~\ref{tab:minmax}).}

\begin{center}
\begin{longtable}[c]{|llllllll|}
\hline 
Column & Name & Type & Units & Description  & Min  & Max   & NN? \\
\hline 
1  & ID          & I10   & --      & TESS Input Catalog identifier        & 1      & 10\textasciicircum 16  & NN \\
2  & Version     & A8    & yyyymmdd & Version Identifier for this entry    & --     & --                     & -- \\
3  & HIP         & I6    & --      & Hipparcos Identifier                 & --     & --                     & -- \\
4  & TYC         & A12   & --      & Tycho2 Identifier                    & --     & --                     & -- \\
5  & UCAC        & A10   & --      & UCAC4 Identifier                     & --     & --                     & -- \\
6  & TWOMASS     & A16   & --      & 2MASS Identifier                     & --     & --                     & -- \\
7  & SDSS        & A20   & --      & SDSS DR9 Identifier                  & --     & --                     & -- \\
8  & ALLWISE     & A20   & --      & ALLWISE Identifier                   & --     & --                     & -- \\
9  & GAIA        & A20   & --      & GAIA Identifier                      & --     & --                     & -- \\
10 & APASS       & A30   & --      & APASS Identifier                     & --     & --                     & -- \\
11 & KIC         & I8    & --      & KIC Identifier                       & --     & --                     & -- \\
12 & Objtype     & A10   & --      & Object Type [star, extended, etc]    & --     & --                     & -- \\
13 & Typesrc     & A12   & --      & Source of the object type            & --     & --                     & -- \\
14 & RA          & D10.6 & deg      & Right Ascension JD2000               & 0      & 360                    & NN \\
15 & Dec         & D10.6 & deg      & Declination JD2000                   & -90    & 90                     & NN \\
16 & Posflag     & A12   & --      & Source of the position               & --     & --                     & -- \\
17 & pmRA        & D10.3 & mas/yr   & Proper Motion in Right Ascension     & -15000 & -15000                 & -- \\
18 & e\_pmRA     & D10.3 & mas/yr   & Uncertainty in Right Ascension       & 0      & 15000                  & -- \\
19 & pmDec       & D10.3 & mas/yr   & Proper Motion in Declination         & -15000 & 15000                  & -- \\
20 & e\_pmDec    & D10.3 & mas/yr   & Uncertainty in Declination           & 0      & 15000                  & -- \\
21 & PMFlag      & A12   & --      & Source of the Proper Motion          & --     & --                     & -- \\
22 & plx         & D10.3 & mas      & Parallax                             & -100   & 1000                   & -- \\
23 & e\_plx      & D10.3 & mas      & Error in the parallax                & -100   & 1000                   & -- \\
24 & PARFlag     & A12   & --      & Source of the parallax               & --     & --                     & -- \\
25 & GalLong     & D10.6 & deg      & Galactic Longitude                   & 0      & 360                    & NN \\
26 & GalLat      & D10.6 & deg      & Galactic Latitude                    & -90    & 90                     & NN \\
27 & EcLong      & D10.6 & deg      & Ecliptic Longitude                   & 0      & 360                    & NN \\
28 & EcLat       & D10.6 & deg      & Ecliptic Latitude                    & -90    & 90                     & NN \\
29 & Bmag        & E6.3  & mag      & Johnson B                            & -25    & 50                     & -- \\
30 & e\_Bmag     & E6.3  & mag      & Uncertainty in Johnson B             & 0      & 50                     & -- \\
31 & Vmag        & E6.3  & mag      & Johnson V                            & -25    & 50                     & -- \\
32 & e\_Vmag     & E6.3  & mag      & Uncertainty in Johnson V             & 0      & 50                     & -- \\
33 & umag        & E6.3  & mag      & Sloan u                              & -25    & 50                     & -- \\
34 & e\_umag     & E6.3  & mag      & Uncertainty in Sloan u               & 0      & 50                     & -- \\
35 & gmag        & E6.3  & mag      & Sloan g                              & -25    & 50                     & -- \\
36 & e\_gmag     & E6.3  & mag      & Uncertainty in Sloan g               & 0      & 50                     & -- \\
37 & rmag        & E6.3  & mag      & Sloan r                              & -25    & 50                     & -- \\
38 & e\_rmag     & E6.3  & mag      & Uncertainty in Sloan r               & 0      & 50                     & -- \\
39 & imag        & E6.3  & mag      & Sloan I                              & -25    & 50                     & -- \\
40 & e\_imag     & E6.3  & mag      & Uncertainty in Sloan I               & 0      & 50                     & -- \\
41 & zmag        & E6.3  & mag      & Sloan z                              & -25    & 50                     & -- \\
42 & e\_zmag     & E6.3  & mag      & Uncertainty in Sloan z               & 0      & 50                     & -- \\
43 & Jmag        & E6.3  & mag      & 2MASS J                              & -25    & 50                     & -- \\
44 & e\_Jmag     & E6.3  & mag      & Uncertainty in 2MASS J               & 0      & 50                     & -- \\
45 & Hmag        & E6.3  & mag      & 2MASS H                              & -25    & 50                     & -- \\
46 & e\_Hmag     & E6.3  & mag      & Uncertainty in 2MASS H               & 0      & 50                     & -- \\
47 & Kmag        & E6.3  & mag      & 2MASS K                              & -25    & 50                     & -- \\
48 & e\_Kmag     & E6.3  & mag      & Uncertainty in 2MASS K               & 0      & 50                     & -- \\
49 & TWOMflag    & A20   & --      & Quality Flags for 2MASS              & --     & --                     & -- \\
50 & prox        & E6.3  & arcsec   & 2MASS Nearest Neighbor               & 0      & 500                    & -- \\
51 & W1Mag       & E6.3  & mag      & WISE W1                              & -25    & 50                     & -- \\
52 & e\_W1Mag    & E6.3  & mag      & Uncertainty in WISE W1               & 0      & 50                     & -- \\
53 & W2Mag       & E6.3  & mag      & WISE W2                              & -25    & 50                     & -- \\
54 & e\_W2Mag    & E6.3  & mag      & Uncertainty in WISE W2               & 0      & 50                     & -- \\
55 & W3Mag       & E6.3  & mag      & WISE W3                              & -25    & 50                     & -- \\
56 & e\_W3Mag    & E6.3  & mag      & Uncertainty in WISE W3               & 0      & 50                     & -- \\
57 & W4mag       & E6.3  & mag      & WISE W4                              & -25    & 50                     & -- \\
58 & e\_W4Mag    & E6.3  & mag      & Uncertainty in WISE W4               & 0      & 50                     & -- \\
59 & Gmag        & E6.3  & mag      & GAIA G Mag                           & -25    & 50                     & -- \\
60 & e\_Gmag     & E6.3  & mag      & Uncertainty in GAIA G                & 0      & 50                     & -- \\
61 & Tmag        & E6.3  & mag      & TESS Magnitude                       & -4     & 50                     & NN \\
62 & e\_Tmag     & E6.3  & mag      & Uncertainty in TESS Magnitude        & 0      & 50                     & NN \\
63 & TESSFlag    & A5    & --      & TESS Magnitude Flag                  & --     & --                     & -- \\
64 & SPFlag      & A5    & --      & Stellar Properties Flag              & --     & --                     & -- \\
65 & Teff        & E6.0  & K        & Effective Temperature                & 300    & 100000                 & -- \\
66 & e\_Teff     & E6.0  & K        & Uncertainty in Effective Temperature & 0      & 100000                 & -- \\
67 & logg        & E6.3  & cgs      & log of the Surface Gravity           & -5     & 10                     & -- \\
68 & e\_logg     & E6.3  & cgs      & Uncertainty in Surface Gravity       & 0      & 2.5                    & -- \\
69 & M/H         & E6.3  & dex      & Metallicity                          & -7     & 2                      & -- \\
70 & e\_M/H      & E6.3  & dex      & Uncertainty in the Metallicity       & 0      & 2.5                    & -- \\
71 & Rad         & E8.3  & solar    & Radius                               & 0.001  & 10000                  & -- \\
72 & e\_Rad      & E8.3  & solar    & Uncertainty in the Radius            & 0      & 10000                  & -- \\
73 & Mass        & E8.3  & solar    & Mass                                 & 0.01   & 500                    & -- \\
74 & e\_Mass     & E8.3  & solar    & Uncertainty in the Mass              & 0      & 500                    & -- \\
75 & rho         & E10.3 & solar    & Stellar Density                      & 0      & 10\textasciicircum 8   & -- \\
76 & e\_rho      & E10.3 & solar    & Uncertainty in the Stellar Density   & 0      & 10\textasciicircum 8   & -- \\
77 & LumClass    & A10   & --      & Luminosity Class                     & --     & --                     & -- \\
78 & Lum         & E10.3 & solar    & Stellar Luminosity                   & 0      & 10\textasciicircum 7   & -- \\
79 & e\_Lum      & E10.3 & solar    & Uncertainty in Luminosity            & 0      & 10\textasciicircum 7   & -- \\
80 & d           & E8.1  & pc       & Distance                             & 1      & 5\textasciicircum 9    & -- \\
81 & e\_d        & E8.1  & pc       & Uncertainty in the distance          & 0      & 5\textasciicircum 9    & -- \\
82 & e(b-v)      & E6.3  & mag      & Color Excess                         & 0      & 99\textasciicircum 999 & -- \\
83 & e\_e(b-v)   & E6.3  & mag      & Uncertainty in Color Excess          & 0      & 99\textasciicircum 999 & -- \\
84 & numcont     & I6    & --      & Number of Contamination Sources      & 0      & 999999                 & -- \\
85 & contratio   & E8.6  & --      & Contamination Ratio                  & 0      &                        & -- \\
86 & disposition & A10   & --      & disposition type                     & --     & --                     & -- \\
87 & dup\_id     & I10   & --      & Points to the TIC ID                 & --     & --                     & --\\
88 & pri         & EXX   & --      & Targeting Priority                     & 0       & 1                     & -- \\
\hline 
\caption{Brief description of TIC contents and permitted ranges for all values.\label{tab:minmax}}
\end{longtable}
\vspace{-0.6in}
\end{center}

\section{Provenance Flags in the TIC\label{sec:flags}}

\begin{center}
\begin{longtable}[c]{|rlll|}
\hline 
Column & Name & Flags & Description  \\
\hline 
12            & Objtype     & star  & object is a star \\
    ...       &  ...   & extended  & object is a galaxy/extended source \\
    
13            & Typesrc     & hip  & stellar source is hipparcos \\
    ...       &  ...   & cooldwarfs  & stellar source is the cool dwarf list \\
    ...       &  ...   & 2mass  & stellar source is 2MASS \\
    ...       &  ...   & lepine  & stellar source is Lepines All-sky Catalog of Bright M Dwarfs (2011) \\
    ...       &  ...   & tmgaia & stellar source from Gaia with unique 2MASS match \\  
    ...       &  ...   & tmmgaia & stellar source from Gaia without unique 2MASS match \\   
    ...       &  ...   & hotsubdwarf  & stellar source is the hot subdwarf list \\
    ...       &  ...   & gicycle1  & stellar source is the GI cycle 1 program \\
    
16            & Posflag     & hip  & stellar source is hipparcos \\
    ...       &  ...   & cooldwarfs  & stellar source is the cool dwarf list \\
    ...       &  ...   & 2mass  & stellar source is 2MASS \\
    ...       &  ...   & lepine  & stellar source is Lepines All-sky Catalog of Bright M Dwarfs (2011) \\
    ...       &  ...   & tmgaia & stellar source from Gaia with unique 2MASS match \\  
    ...       &  ...   & tmmgaia & stellar source from Gaia without unique 2MASS match \\   
    ...       &  ...   & hotsubdwarf  & stellar source is the hot subdwarf list \\
    ...       &  ...   & gicycle1  & stellar source is the GI cycle 1 program \\
    ...       &  ...   & 2MASSEXT  & extended source from 2MASS extended source catalog \\
    
21            & PMFlag & ucac4  & proper motions from UCAC4 \\
    ...       &  ...   & tgas  & proper motions from Tycho2-\textit{Gaia} Astrometric Solution \\
    ...       &  ...   & sblink  & proper motions from SuperBlink \\
    ...       &  ...   & tycho2  & proper motions from Tycho2 \\
    ...       &  ...   & hip  & proper motions from Hipparcos \\
    ...       &  ...   & ucac5  & proper motions from UCAC5 \\
    ...       &  ...   & hsoy & proper motions from Hot Stuff for One Year \\
        
24            & PARFlag & tgas  & parallax from Tycho2-\textit{Gaia} Astrometric Solution  \\
    ...       &  ...    & hip   & parallax from Hipparcos \\    

63            & TESSFlag    & gaiak & magnitude calculated from $G$ and 2MASS $K_S$\\
    ...       &  ...    & gaiaj & magnitude calculated from $G$ and 2MASS $J$ \\
    ...       &  ...    & joffset2 & magnitude calculated from 2MASS $J$ and an offset (+1.75 for $J-K_S > 1$)\\
    ...       &  ...    & hipvmag & magnitude calculated Hipparcos $V$ magnitude\\
    ...       &  ...    & gaiaoffset & magnitude calculated from $G$ and an offset\\
    ...       &  ...    & hoffset & magnitude calculated from 2MASS $H$ offset\\
    ...       &  ...    & vjh & magnitude calculated from $V$ and 2MASS $J-H$\\
    ...       &  ...    & jhk & magnitude calculated from 2MASS $J-K_S$\\
    ...       &  ...    & vjk & magnitude calculated from $V$ and 2MASS $J-K_S$\\
    ...       &  ...    & hotsubdwarf & magnitude adopted from hot subdwarf list\\
    ...       &  ...    & vk & magnitude calculated from $V$ and 2MASS $K_S$\\
    ...       &  ...    & joffset & magnitude calculated from 2MASS $J$ offset (+0.5 for $J-K_S < -0.1$)\\
    ...       &  ...    & gaiav & magnitude calculated from $G$ and $V$\\
    ...       &  ...    & tmvk & magnitude calculated from $V$ and 2MASS $K_S$ (same as vk)\\
    ...       &  ...    & from$\_$apass$\_$i & magnitude from cool dwarf list \citep{Muirhead:2018}\\
    ...       &  ...    & from$\_$sdss$\_$ik & magnitude from cool dwarf list \citep{Muirhead:2018}\\
    ...       &  ...    & gaiah & magnitude calculated from Gaia and 2MASS $H$\\
    ...       &  ...    & jh & magnitude calculated from 2MASS $J-H$\\
    ...       &  ...    & cdwarf & magnitude from cool dwarf list \citep{Muirhead:2018}\\
    ...       &  ...    & bpjk & magnitude calculated from photographic $ B$ and 2MASS J-$K_S$\\
    ...       &  ...    & voffset & magnitude calculated from $V$ and offset\\
    ...       &  ...    & koffset & magnitude calculated from 2MASS $K_S$ and offset\\
    ...       &  ...    & wmean$\_$vk$\_$jhk & magnitude from cool dwarf list \citep{Muirhead:2018}\\
    ...       &  ...    & lepine & magnitude from Lepine catalog\\
    ...       &  ...    & gicycle1 & magnitude from GI Cycle 1 proposal\\
    ...       &  ...    & from$\_$sdss$\_$i & magnitude from cool dwarf list \citep{Muirhead:2018}\\

64            & SPFlag      &   cdwrf & mass and radius from cool-dwarf list (see Sec.~\ref{subsubsec:cool} \& \citet{Muirhead:2018} \\
    ...       &  ...        & hotsd  & mass and radius from the hot subdwarf list (see Sec.~\ref{subsubsec:hotsb}) \\
    ...       &  ...        & tplx  & characteristics computed from measured TGAS parallax \\
    ...       &  ...        & hplx  & characteristics computed from measured HIP parallax \\
    ...       &  ...        & spect &  characteristics computed using the spectroscopic Torres relations \\
    ...       &  ...        & spec &  characteristics computed using the spectroscopic Torres relations \\
    ...       &  ...        & allen & characteristics computed from spline relations based on eclipsing \\
    ...       &  ...        &       & binary properties and TRILEGAL simulations -- old flag, same as splin \\
    ...       &  ...        & splin & characteristics computed from spline relations based on eclipsing \\
    ...       &  ...        &       & binary properties and TRILEGAL simulations  \\
\hline 
\caption{Brief description of flags in the TIC and CTL.\label{tab:flags}}
\end{longtable}
\vspace{-0.6in}
\end{center}

\section{Internal Consistency in the TIC}\label{sec:internal_consistency}

As described above, the TIC is created by compiling numerous independent catalogs and using this information to calculate a variety of stellar parameters, in a variety of ways. This can lead to the final version of the TIC not being entirely self-consistent. Here we explain a variety of checks aimed at making the TIC as self-consistent as possible. This mainly includes ensuring various calculated parameters, which are dependent on other observed and calculated quantities, are calculated using the reported TIC values to avoid contradictory information for any given star. We provide a list of these internal consistency checks below:

\begin{enumerate}
\item \textbf{Various Columns}: $V$ and other magnitudes\\
\textit{Dependencies}: (1) \teff\, and (2) radius, if the radius is from parallax and bolometric correction\\
\textit{Issues}: The magnitudes might not be consistent with \teff\ or other quantities if those quantities are taken from an override catalog (e.g., specially curated targets list). \\
\textit{Implemented Fix}: $V$ from the cool-dwarf list does not override the default TIC $V$ for now.\\

\item \textbf{Column 65}: Effective temperature (\teff)\\
\textit{Dependencies}: $E(B-V)$ and \feh\ if from color\\	
\textit{Issues}: (1) \teff\ might not agree with \lbol\ and radius, (2) \teff\ will be different from dereddened phot-based value if from spectroscopy; (3) \teff\ will be different from dereddened phot-based value and/or from spectroscopy if taken from cool dwarf list\\
\textit{Implemented Fix}: \teff\ is set to ``NULL" if the cool dwarf list provides mass and radius but not \teff\.\\

\item \textbf{Column 67}: Surface Gravity (\logg) \\
\textit{Dependencies}: Mass and Radius if not from spectroscopy\\	
\textit{Issues}: (1) \logg\ will not in general agree with Mass and Radius if taken from spectroscopy.\\
\textit{Implemented Fix}: We always calculate \logg\ from mass (\#73) and radius (\#71) and ignore spectroscopic \logg.\\

\item \textbf{Column 71}: Radius\\
\textit{Dependencies}: (1) Parallax, reddening, \teff\, if from parallax; or (2) spectroscopic \teff\, \logg\, \feh\ if from Torres et al. (2010) relations; or (3) \teff\ if from the spline relations; or (4) cool dwarf list\\	
\textit{Issues}: No issues because radius is a primary derived quantity.\\
\textit{Implemented Fix}: The radius is calculated from parallax when available.\\

\item \textbf{Column 73}: Mass\\
\textit{Dependencies}: (1) spectroscopic \teff\, \logg\, \feh\ if from Torres et al. (2010) relations; (2) \teff\ if from spline relations; or (3) cool dwarf list\\	
\textit{Issues}: No issues because mass is a primary derived quantity.\\
\textit{Implemented Fix}: No need to fix.\\

\item \textbf{Column 75}: Density ($\rho$) \\
\textit{Dependencies}: Mass and Radius if not from literature (e.g., transit based analysis) \\	
\textit{Issues}: $\rho$ from literature (e.g., transit analysis) might not agree with calculated mass and radius\\
\textit{Implemented Fix}: We calculate $\rho$ from mass (\#73) and radius (\#71) if not from a transit analysis.\\

\item \textbf{Column 77}: Luminosity Class\\
\textit{Dependencies}: Temperature and Radius (i.e., HR diagram position) \\	
\textit{Issues}: Luminosity class might be different from spectroscopic catalogs.\\
\textit{Implemented Fix}: Currently the Luminosity Class is determined to be either a ``dwarf" or ``giant," based on the RPM$_J$ cut.\\

\item \textbf{Column 78}: Luminosity (\lbol) \\
\textit{Dependencies}: (1) Radius and \teff\ if from Stefan Boltzmann, or (2) parallax, \teff\ via BC, reddening. \\	
\textit{Issues}: Parallax based luminosity might not agree with \teff\ and radius.\\
\textit{Implemented Fix}: We always calculated \lbol\ from Stefan-Boltzmann. This will naturally include parallax based radii if radius is from parallax.\\

\item \textbf{Column 80}: Distance\\
\textit{Dependencies}: (1) Parallax, or (2) invert Stefan-Boltzmann\\	
\textit{Issues}: Distance could be inconsistent with parallax. \\
\textit{Implemented Fix}: We always calculated distance from parallax if available and less than 20\% error. Otherwise use method 2 with $V$ if available. Distance error set as ``NULL" when derived by inverting Stefan-Boltzmann relation.\\

\item \textbf{Column 82}: $E(B-V)$\\
\textit{Dependencies}: Color-color diagram reddening vector\\	
\textit{Issues}: Reddening will be inconsistent with that implied by Stefan-Boltzmann law if radius is obtained some way other than via parallax \\
\textit{Implemented Fix}: This is a second-order effect and not fixed but is documented for completeness. \\

\end{enumerate}

\section{The use of specially curated catalogs\label{sec:spec_cat}}

The TIC uses a variety of uniform relations to calculate stellar parameters, such as \teff\ and radius. Unfortunately, the majority of these relations are appropriate only for dwarf stars and may fail to accurately reproduce the appropriate physical parameters when applied to stars not considered a typical dwarf star (in the temperature range $3840\;{\rm K} \lesssim \teff \lesssim 10,000$\;K). The TESS Target Selection Working Group (TSWG) therefore has tasked the creation of specially curated catalogs, listed below, to create an accurate list of stellar parameters for specific categories of targets, such as cool dwarfs. These catalogs also aim to include high priority targets that may not be included in the CTL due to their unusual nature or poorly calculated default parameters for reasons described above. 

\noindent{The specially curated target lists currently incorporated into TIC-7 are:}

\begin{enumerate}
\item Cool Dwarfs: This list is meant to identify all dwarf stars with $\teff \lesssim 3840$\;K where the default TIC parameter calculations do a poor job of characterizing. The list contains a number of late K-dwarfs as well.
\item Bright Stars: This list is meant to identify all bright stars in the sky, $T<6$.
\item Known Planet Hosts: This list is meant to identify all stars with known exoplanet hosts.
\item Hot Sub-dwarfs: This list is meant to identify nearby bright subdwarfs which may be useful for 2-minute astroseismology studies.
\end{enumerate}



\subsection{High Level Descriptions of the Specially Curated Target Lists in TIC-7}

\subsubsection{The Cool Dwarf List \label{subsubsec:cool}}

The cool-dwarf list is a carefully vetted list of stellar parameters for K-dwarf and M-dwarf stars ($\teff<4000$~K). We provide a basic overview of the catalog here but direct the reader to \citet{Muirhead:2018} for a more detailed discussion.

The catalog itself is created using the SUPERBLINK catalog, cross matched with 2MASS and APASS. Dwarf stars are separated from giant stars using parallax measurements, when available, or the reduced proper motion criteria of \citet{Gaidos:2014}. \teff\ is calculated from $r-J$, $r-z$, and $V-J$ color with an additional $J-H$ term to account for systematic effects of [Fe/H]. For stars with trigonometric parallax, masses and radii are calculated using the $M_{K_S}-R_{*}$ relation from \citet{Benedict:2016} and \citet{Mann:2015} respectively. For stars without trigonometric parallax, radius and mass are calculated using the \teff-$R_{*}$ relation from \citet{Mann:2015} and a newly developed relation between $\teff$ and $M_{*}$.

The cool-dwarf list is treated like an override catalog and the values provided replace default calculated values for $T$, \teff\, stellar radius and stellar mass. The catalog provides a value for $V$ from the SUPERBLINK catalog but at this time these values do \textit{not} override the $V$ column in the TIC.

\subsubsection{The Bright Star List}

The bright star list is a catalog of all ($\sim10,000$) bright ($T<6$) sources in the TIC. These objects are included in the CTL regardless of whether the star passes the RPM$_J$ cut. We compared the list of bright TIC objects to the Yale Bright Star catalog (hereafter, YBC) \citep{Hoffleit:1991} and found $\sim80\%$ of the YBC is in our Bright Star list. Nearly $\sim 90\%$ of objects in the YBC but missing from the bright star list include stars where $V > 6$ or $V$ is not provided. 

These stars have contamination calculated in the normal way. However, because many of these stars may not be typical dwarf stars (with $3850<\teff<10000$), parameters such as radius and mass are not calculated unless the star already appears in the cool dwarf list or were previously identified as an RPM$_J$ dwarf. Additionally, in TIC-7, these stars have an arbitrary priority value of 1, as their priorities could not be effectively calculated using the schema described in \S~\ref{subsec:priority}. We emphasize that this priority value of 1 is for identification purposes only and does not represent the true final prioritization of these targets. 

\subsubsection{The Hot Subdwarf List \label{subsubsec:hotsb}}

The hot-subdwarf list identifies all known evolved compact stars down to $V\sim16$ that, for most of them, belong to the Extreme Horizontal Branch (EHB) and beyond (spectral types sdB, sdOB, and sdO; see \citet{Heber:2016}). The list is constructed from the \citet{Geier:2017} catalog, completed with newly identified bright objects of the kind from recent or still ongoing ground-based surveys carried out at the Steward Observatory (Green, priv. comm.), the NOT (Telting, priv. comm.), and the SAAO (Kilkenny, priv. comm.). $T_{{\rm eff}}$ and $\log g$ values are evaluated either from asteroseismology \citep{Fontaine:2012}, from spectroscopy \citep{Geier:2017}, or are given a representative value based on spectral classification if nothing else is available. Stellar mass is from asteroseismology when available \citep{Fontaine:2012} or is set to the canonical value of $0.47$ $M_{\odot}$ otherwise, considering that typical hot subdwarfs have a narrow mass distribution around that value. Radius is computed from $\log g$ and mass. \tmag is estimated from ($V$-\tmag) and ($J$-\tmag) color indices calculated from hot-subdwarf NLTE model atmospheres (interpolation in a $T_{{\rm eff}}-\log g$ grid; Fontaine et al., priv. comm.). If both $V$ and $J$ measurements are available, Tess mag is an average of the two, otherwise transformation is from $V$-band only. $V$ value is, by order of availability, from APASS\_V, or from GSC\_V. $J$, $H$, $K$ values are, by order of availability, from 2MASS, or from UKIDSS. More information about each target can be found by any registered user on the TASC Working Group 8 wiki page at the URL \url{https://tasoc.dk/wg8/}.

\subsubsection{The Known Exoplanet Host List \label{subsubsec:planets}}

The list of known exoplanet host stars is an evolving list since the number of known exoplanets is constantly being updated. The most frequently updated source of information regarding exoplanetary systems is the NASA Exoplanet Science Institute \footnote{\url{https://exoplanetarchive.ipac.caltech.edu/}}, which is described in more detail by \citet{Akeson:2013}. We extracted a subset of the data from this source for each star, including stellar coordinates, the number of planets, and fundamental stellar properties including effective temperature, mass, radius, $\log g$, $V$~mag, and $v \sin i$. The stellar properties were checked with respect to the literature and updated where necessary. The final vetted list was then cross-matched with the TIC in order to include alias information and ancillary data, such as TESS magnitudes, resulting in a first version of the list containing $\sim$2,800 exoplanet host stars. 

\section{CTL Filtergraph Portal\label{sec:appendix_filtergraph}}

Table~\ref{tbl:filtergraph} summarizes the contents of the enhanced CTL provided via the Filtergraph data visualization portal service at the URL \url{filtergraph.vanderbilt.edu/tess_ctl}.

\begin{footnotesize} 
\begin{center}
\begin{longtable}[c]{|c|l|}
 \multicolumn{2}{c}{Descriptions of CTL Contents}\\
 \hline
 Column name & Brief description \\
 \hline
 Right\_Ascension    &  Right Ascension of the star, equinox J2000.0, epoch 2000.0 (degrees)\\
 Declination   &  Declination of the star, equinox J2000.0, epoch 2000.0 (degrees) \\
 Tess\_mag & Calculated TESS magnitude \\
 Teff & Adopted `best' value from (in order of preference): (1) cool dwarf list; (2) spectroscopic;\\
  & (3) de-reddened photometric value; (4) non-dereddened photometric value. See {\tt Teff\_Src} column. \\
 Priority & Priority based on \tmag, radius, and flux contamination with boosts and de-boosts\\
 Radius & Stellar radius derived from photometry (\rsun) \\
 Mass & Stellar mass derived from photometry (\msun) \\
 ContamRatio & Ratio of contaminating flux to flux from the star\\
 Galactic\_Long & Longitude in the Galactic coordinate frame (degrees)\\
 Galactic\_Lat & Latitude in the Galactic coordinate frame (degrees)\\
 Ecliptic\_Long & Longitude in the Ecliptic coordinate frame (degrees)\\
 Ecliptic\_Lat & Latitude in the Ecliptic coordinate frame (degrees)\\
 Parallax & The parallax of the star provided by either TGAS/Gaia or Hipparcos (mas)\\
 Distance & The distance of the star provided (pc)\\
 Total\_Proper\_Motion & Total proper motion of the star (mas/yr)\\
 V\_mag & Adopted $V$ magnitude\\
 J\_mag & 2MASS $J$ magnitude\\
 H\_mag & 2MASS $H$ magnitude\\
 $K_S$\_mag & 2MASS $K_S$ magnitude\\
 G\_mag & \textit{Gaia} magnitude\\
 u\_mag & SDSS u magnitude\\
 g\_mag & SDSS g magnitude\\
 r\_mag & SDSS r magnitude\\
 i\_mag & SDSS i magnitude\\
 z\_mag & SDSS z magnitude\\
 W1\_mag & ALLWISE W1 magnitude\\
 W2\_mag & ALLWISE W2 magnitude\\
 W3\_mag & ALLWISE W3 magnitude\\
 W4\_mag & ALLWISE W4 magnitude\\
 Hipparcos\_Number & {\it Hipparcos\/} ID\\
 Tycho2\_ID & {\it Tycho-2\/} ID\\
 2MASS\_ID & 2MASS ID\\
 TICID & ID for the star in the TESS Input Catalog\\
 Special\_Lists & Identifies whether a star in a special list: The Cool Dwarf list, \\
 & the Bright Star list, Hot Subdwarf, Known Exoplanet Hosts or if the star is in multiple lists .\\
 Priority\_TIC4 & Priority based on the TIC-4 schema. \\
 Priority\_TIC5 & Priority based on the TIC-5 schema. \\
 Priority\_TIC6 & Priority based on the TIC-6 schema. \\
 Priority\_Non\_Contam & Priority without neighbor contamination. \\
 Priority\_No\_Boost &  Priority without sector boosting. \\
 Teff\_Src & Source of the effective temperature (see {\tt Teff} column)\\
 StarChar\_Src & Identifies stellar parameters adopted from: {\tt splin} (unified empirical spline relation); {\tt tplx} (TGAS parallax); \\
 & {\tt hplx} (HIP parallax); {\tt spectorr} (spectroscopic); {\tt cdwarf} (cool dwarf list)\\
 & {\tt hotsubdwarf} (hot subdwarf list)\\
 Radius\_Err & Uncertainty in the radius (solar). \\
 Mass\_Err & Uncertainty in the mass (solar). \\
 Logg & Surface gravity (cgs). \\
 Logg\_Err & Uncertainty in the surface gravity (cgs). \\
 Rho & Density (solar). \\
 Rho\_Err & Uncertainty in the density (solar). \\
 Lum & Luminosity (solar). \\
 Noise\_Star & Uncertainty from the star counts. \\
 Noise\_Sky & Uncertainty from the sky counts. \\
 Noise\_Contaminates & Uncertainty from the neighboring star counts. \\
 Noise\_Readout & Uncertainty in the detector readout. \\
 Noise\_Systematics & Uncertainty floor. \\
 \hline
 \caption{A basic description of all quantities found on the Filtergraph portal. \label{tbl:filtergraph}}
\end{longtable}
\end{center}
\end{footnotesize}

\end{document}